\documentclass[nameyear]{econsocart}
\RequirePackage[colorlinks,citecolor=blue,linkcolor=blue,urlcolor=blue,pagebackref]{hyperref}

\startlocaldefs

\newtheorem{proposition}{Proposition}
\newtheorem{algorithm}{Algorithm}

\endlocaldefs

\begin{document}

\begin{frontmatter}

\title{Identification at the Zero Lower Bound}
\runtitle{Identification at the Zero Lower Bound}

\begin{aug}
%
%
%
\author[id=au1,addressref={add1}]{\fnms{Sophocles}~\snm{Mavroeidis}\ead[label=e1]{sophocles.mavroeidis@economics.ox.ac.uk}}
\address[id=add1]{%
\orgdiv{Department of Economics},
\orgname{University of Oxford}}

\end{aug}

\support{This research is funded by the European Research Council via Consolidator grant
number 647152. I would like to thank Guido Ascari, Sergio de Ferra, James Duffy, Andrea Ferrero, Jim Hamilton, Daisuke Ikeda, Federica Romei, Frank Schorheide, Francesco Zanetti and seminar participants at the Chicago Fed, New York Fed, Board of Governors, Bank of Japan, Hitotsubashi University, Keio University, UCL, University of Cambridge, University of Oxford, University of Warwick, DNB, University of Pennsylvania, Pennsylvania State University, NBER Summer Institute, IAAE conference, Netherlands Econometrics Study Group Meeting, World Congress and North American Winter meetings of the Econometric society for useful comments and discussion, as well as Mishel Ghassibe, Lukas Freund, Shangshang Li, and Patrick Vu for research assistance.}
\begin{abstract}
I show that the Zero Lower Bound (ZLB) on interest rates can be used to
identify the causal effects of monetary policy. Identification depends on the
extent to which the ZLB limits the efficacy of monetary policy. I propose a simple way to test the efficacy of unconventional policies, modelled via a `shadow rate'. I apply this method to
U.S. monetary policy using a three-equation structural vector autoregressive model of inflation, unemployment and the federal funds rate. I reject the null hypothesis that
unconventional monetary policy has no effect at the ZLB, but find some
evidence that it is not as effective as conventional monetary policy.
\end{abstract}


\end{frontmatter}

\section{Introduction}

The zero lower bound (ZLB) on nominal interest rates has arguably been a
challenge for policy makers and researchers of monetary policy. Policy makers
have had to resort to so-called unconventional policies, such as quantitative
easing or forward guidance, which had previously been largely untested.
Researchers have to use new theoretical and empirical methodologies to analyze
macroeconomic models when the ZLB\ binds. So, the ZLB is generally viewed as a
problem or at least a nuisance. This paper proposes to turn this problem on
its head to solve another long-standing question in macroeconomics:\ the
identification of the causal effects of monetary policy on the economy.

The intuition is as follows. If the ZLB limits the ability of policy makers to
react to macroeconomic shocks, as argued, for example, by
\cite{EggertssonWoodford2003}, the response of the economy to shocks will change
when the policy instrument hits the ZLB. Because the difference in the
behavior of macroeconomic variables across the ZLB and non-ZLB regimes is only
due to the impact of monetary policy, the switch across regimes provides
information about the causal effects of policy. In the extreme case that
monetary policy completely shuts down during the ZLB regime, either because
policy makers do not use alternative (unconventional) policy instruments, or
because such instruments turn out to be completely ineffective, the only
difference in the behavior of the economy across regimes is due to the impact
of (conventional) policy during the unconstrained regime. Therefore, the ZLB
identifies the causal effect of policy during the unconstrained regime. If
monetary policy remains partially effective during the ZLB regime, e.g.,
through the use of unconventional policy instruments, then the difference in
the behavior of the economy across regimes will depend on the difference in
the effectiveness of conventional and unconventional policies. In this case,
we obtain only partial identification of the causal effects of monetary
policy, but we can still get informative bounds on the relative efficacy of
unconventional policy. In the other extreme case that unconventional policy is
as effective as conventional policy, there is no difference in the behavior of
the economy across regimes, and we have no additional information to identify
the causal effects of policy. However, we can still test this so-called ZLB
irrelevance hypothesis \citep{DebortoliGaliGambetti2019} by testing whether
the reaction of the economy to shocks is the same across the two regimes.

There are similarities between identification via occasionally binding
constraints and identification through heteroskedasticity \citep{Rigobon03},
or more generally, identification via structural change
\citep{MagnussonMavroeidis2014}. That literature showed that the switch
between different regimes generates variation in the data that identifies
parameters that are constant across regimes. For example, an exogenous shift
in a policy reaction function or in the volatility of shocks identifies the
transmission mechanism, provided the latter is unaffected by the policy
shift. When the switch from one regime to another is exogenous,
regime indicators are valid instruments, and the methodology in
\cite{MagnussonMavroeidis2014} is
applicable. However,
regimes induced by occasionally binding constraints are not exogenous --
whether the ZLB binds or not clearly depends on the structural shocks, so
regime indicators cannot be used as instruments in the usual way, and a new
methodology is needed to analyze these models.

In this paper, I show how to control for the endogeneity in regime selection
and obtain identification in structural vector autoregressions
(SVARs).\footnote{There is a related literature on Dynamic Stochastic General
Equilibrium (DSGE) models with a ZLB, see, e.g.,
\cite{FernandezGordonGuerronRubio2015}, \cite{GuerrieriIacoviello2015},
\cite{AruobaCubaBordaSchorfheide2017}, \cite{KulishMorleyRobinson2017} and
\cite{AruobaCubaBordaHigaFloresSchorfheideVillalvazo2020}. The papers in this
literature do not point out the implications of the ZLB for identification of
monetary policy shocks.} The methodology is parametric and likelihood-based,
and the analysis is similar to the well-known Tobit model \citep{Tobin1958}.
More specifically, the methodological framework builds on the early
microeconometrics literature on simultaneous equations models with censored
dependent variables, see \cite{Amemiya1974}, \cite{Lee1976},
\cite{BlundellSmith1994}, and the more recent literature on dynamic Tobit
models, see \cite{Lee1999}, and particle filtering, see
\cite{PittShephard1999}.

A further contribution of this paper is a general methodology to estimate
reduced-form VARs with a variable subject to an occasionally binding
constraint. This is a necessary starting point for SVAR analysis that uses any
of the existing popular identification schemes, such as short- or long-run
restrictions, sign restrictions, or external instruments. In the absence of
any constraints, reduced-form VARs can be estimated consistently by Ordinary
Least Squares (OLS), which is Gaussian Maximum Likelihood, or its
corresponding Bayesian counterpart, and inference is fairly well-established.
However, it is well-known that OLS estimation is inconsistent when the data is
subject to censoring or truncation, see, e.g., \cite{Gree93} for a textbook
treatment. So, it is not possible to estimate a VAR consistently by OLS using
any sample that includes the ZLB, or even using (truncated) subsamples when
the ZLB is not binding (because of selection bias), as was pointed out by
\cite{HayashiKoeda2019}. It is not possible to impose the ZLB constraint using
Markov switching models with exogenous regimes, as in
\cite{LiuTheodoridisMumtazZanetti2019}, because exogenous Markov-switching
cannot guarantee that the constraint will be respected with probability one,
and also does not account for the fact that the switch from one regime to the
other depends on the structural shocks. Finally, it is not possible to perform
consistent estimation and valid inference on the VAR (i.e., error bands with
correct coverage on impulse responses), using externally obtained measures of
the shadow rate, such as the one proposed by \cite{WuXia2016}, as any such measures are subject to large and persistent estimation error that is not accounted for if they are treated as known in subsequent analysis. See also \cite{Rossi2019} for a
comprehensive discussion of the challenges posed by the ZLB for the estimation
of structural VARs. 

The methodology developed in this paper allows for the presence of a shadow rate, estimates of which can be
obtained, but more importantly, it fully accounts for the impact of sampling
uncertainty in the estimation of the shadow rate on inference about the
structural parameters such as impulse responses. Therefore, the paper fills an
important gap in the literature, as it provides the requisite methodology to
implement any of the existing identification schemes. \cite{HayashiKoeda2019}
develop a VAR model with endogenous regime switching in which the policy
variables that are subject to a lower bound are modelled using Tobit regressions. A key difference of
their methodology from the one developed here is that they impose recursive
identification of monetary policy shocks, which the present paper shows to be
an overidentifying, and hence testable, restriction. Moreover, their model
does not include shadow rates. A more recent
paper by \cite{AruobaSchorfheideVillalvazo2020} also studies SVARs with
occasionally binding constraints, but does not focus on the implications of
these constraints for identification. 

Identification of the causal effects of policy by the ZLB does not require
that the policy reaction function be stable across regimes. However, inference
on the efficacy of unconventional policy, or equivalently, the causal effects
of shocks to the shadow rate over the ZLB period, obviously depends on whether
or not the reaction function remains the same across regimes. For example, an
attenuation of the causal effects of policy over the ZLB period may indicate
that unconventional policy is only partially effective, but it is also
consistent with unconventional policy being less active (during ZLB\ regimes)
than conventional policy (during non-ZLB regimes). This is a fundamental
identification problem that is difficult to overcome without additional
information, such as measures of unconventional policy stance, or additional
identifying assumptions, such as parametric restrictions or external
instruments. This can be done using the methodology developed in this paper.

The structure of the paper is as follows. Section \ref{s: examples} presents
the main identification results of the paper in the context of a static
bivariate simultaneous equations model with a limited dependent variable
subject to a lower bound. Section \ref{s: SVAR} generalizes the analysis to a
SVAR\ with an occasionally binding constraint and discusses identification,
estimation and inference. Section \ref{s: application} provides an application
to a three-equation SVAR in inflation, unemployment and the Federal funds rate
from \cite{StockWatson01}. Using a sample of post-1960 quarterly US data, I
find some evidence that the ZLB\ is empirically relevant, and that
unconventional policy is only partially effective. Proofs and simulation results are given in the Appendix at the end. 

\section{Simultaneous equations model\label{s: examples}}

I first illustrate the idea using a simple bivariate simultaneous equations
model (SEM), which is both analytically tractable and provides a direct link
to the related microeconometrics literature. To make the connection to the
leading application, I will motivate this using a very stylized economy
without dynamics in which the only outcome variable is inflation $\pi_{t}$ and
the (conventional) policy instrument is the short-term nominal interest rate,
$r_{t}$. In addition to the traditional interest rate channel, the model
allows for an `unconventional monetary policy' channel that can be used when
the conventional policy instrument hits the ZLB. An example of such a policy
is quantitative easing (QE), in the form of long-term asset purchases by the
central bank. Here I\ discuss a simple model of QE.\footnote{The more general
SVAR\ model of the next section can also incorporate forward guidance in the
form of \cite{ReifschneiderWilliams2000} and \cite{DebortoliGaliGambetti2019},
as shown in \cite{IkedaLiMavroeidisZanetti2020}.}

Abstracting from dynamics and other variables, the equation that links
inflation to monetary policy is given by
\begin{equation}
\pi_{t}=c+\beta\left(  r_{t}-r^{n}\right) +\varphi b_{L,t} +\varepsilon
_{1t}, \label{eq: outcome}%
\end{equation}
where $c$ is a constant, $r^{n}$ is the neutral rate, $b_{L,t}$ is the
amount of long-term bonds held by the private sector in log-deviation from its
steady state, and $\varepsilon_{1t}$ is an exogenous structural shock
unrelated to monetary policy. Equation (\ref{eq: outcome}) can be obtained
from a model of bond-market segmentation, as in \cite{ChenCurdiaFerrero2012},
where a fraction of households is constrained to invest only in long-term
bonds, see the Appendix for more details. In such a model, the
parameter $\varphi$ that determines the effectiveness of QE is proportional to
the fraction of constrained households and the elasticity of the term premium
with respect to asset holdings, both of which are assumed to be outside the
control of the central bank.

The nominal interest rate is set by a Taylor rule subject to the
ZLB\ constraint, namely,
\begin{subequations}
\label{eq: Taylor}%
\begin{align}
r_{t}  &  =\max\left(  r_{t}^{\ast},0\right)  ,\label{eq: Taylor a}\\
r_{t}^{\ast}  &  =r^{n}+\gamma\pi_{t}+\varepsilon_{2t}, \label{eq: Taylor b}%
\end{align}
where $r_{t}^{\ast}$ represents the desired target policy rate, and
$\varepsilon_{2t}$ is a monetary policy shock. When $r_{t}^{\ast}$ is
negative, it is unobserved. The unobserved $r_{t}^{\ast}$ will be referred to
as the `shadow rate', and it represents the desired policy stance prescribed
by the Taylor rule in the absence of a binding ZLB constraint.

Suppose that QE is activated only when the conventional policy instrument
$r_{t}$ hits the ZLB,\footnote{The assumption that QE is only active during
the ZLB regime is only made for simplicity, as it is inconsequential for the
resulting functional form of the transmission equation. We can let QE be
active all the time, and even allow for a different rule for QE above and
below the ZLB, i.e., $b_{L,t}=\alpha\min\left(  r_{t}^{\ast
},0\right)  +\alpha_{1}\max\left(  r_{t}^{\ast},0\right)$. Then, substituting back into
(\ref{eq: outcome}) yields an equation that is isomorphic to
(\ref{eq: outcome nesting}), i.e., $\pi_{t}=\allowbreak c_{1}\allowbreak
+\allowbreak\bar{\beta}r_{t}\allowbreak+\bar{\beta}^{\ast}%
\min\left(  r_{t}^{\ast},0\right)  \allowbreak+\varepsilon_{1t},$ with
$\bar{\beta}:=\allowbreak\beta +\varphi\alpha_{1} $ and
$\bar{\beta}^{\ast}\allowbreak:=\allowbreak\varphi\alpha.$} and follows
the same policy rule (\ref{eq: Taylor b}), up to a factor of proportionality
$\alpha,$\ i.e.,%
\end{subequations}
\[
b_{L,t}=\min\left(  \alpha r_{t}^{\ast},0\right)  .
\]
Substituting for $b_{L,t}$ in eq.~(\ref{eq: outcome}) and letting $\beta^{\ast}:=\alpha\varphi$, we obtain
\begin{equation}
\pi_{t}=c+\beta\left(  r_{t}-r^{n}\right)  +\beta^{\ast}\min\left(
r_{t}^{\ast},0\right)  +\varepsilon_{1t}. \label{eq: outcome nesting}%
\end{equation}

A special case arises when QE is ineffective $\left(  \varphi=0\right)  ,$ or
the monetary authority does not pursue a QE policy $\left(  \alpha=0\right)
,$ so that eq.~(\ref{eq: outcome nesting}) becomes
\begin{equation}
\pi_{t}=c+\beta\left(  r_{t}-r^{n}\right)  +\varepsilon_{1t},
\label{eq: outcome kinked}%
\end{equation}
and monetary policy is completely inactive at the ZLB.

Another special case of the model given by equations (\ref{eq: Taylor}) and
(\ref{eq: outcome nesting}) arises when $\beta^{\ast}=\beta$ in
eq.~(\ref{eq: outcome nesting}). This happens when $\varphi\neq0$ and $\alpha$
is chosen by the monetary authority to be equal to $\beta/\varphi.$ This can be
done when policy makers know the transmission mechanism in
eq.~(\ref{eq: outcome}) and have no restrictions in setting the policy
parameter $\alpha$ so as to fully remove the impact of the ZLB on conventional
policy. In that case, the equation for the outcome variable becomes
\begin{equation}
\pi_{t}=c+\beta\left(  r_{t}^{\ast}-r^{n}\right)  +\varepsilon_{1t}.
\label{eq: outcome censored}%
\end{equation}
The model given by equations (\ref{eq: Taylor}) and
(\ref{eq: outcome censored}) is one in which monetary policy is completely
unconstrained and there is no difference in outcomes across policy regimes.
Such models have been put forward by \cite{SwansonWilliams2014},
\cite{DebortoliGaliGambetti2019} and \cite{WuZhang2019}.

The nesting model given by eq.~(\ref{eq: outcome nesting}) allows the effects
of conventional and unconventional policy to differ. This could reflect
informational as well as political or institutional constraints that prevent
policy makers from calibrating their unconventional policy response to match
exactly the policy prescribed by the Taylor rule. For instance, it may be that
policy makers do not know the effectiveness of the QE channel $\varphi$, or
that the scale of asset purchases needed to achieve the desired policy stance
during a ZLB regime is too large to be politically acceptable. Such a
consideration may be particularly pertinent, for example, in the Eurozone.
Importantly, one does not need to take a theoretical stand on this issue,
because the methodology that I develop in the paper can accommodate a wide
range of possibilities, and, as I demonstrate below, the issue can be studied empirically.

To complete the specification of the model, I\ assume that the structural
shocks $\varepsilon_{t}=\left(  \varepsilon_{1t},\varepsilon_{2t}\right)
^{\prime}$ are independently and identically distributed (\textit{i.i.d.})
Normal with covariance matrix $\Sigma=diag\left(  \sigma_{1}^{2},\sigma
_{2}^{2}\right)$.

The analysis in this paper assumes that the shadow rate $r_t^*$ is only observed above the ZLB. If $r^*_t<0$ were observed up to scale, for
instance, if we could measure QE $b_{L,t}$ from the balance sheet of the central bank, then the computation of the likelihood would be much simpler -- no filtering would be needed to deal with
lags of $r_{t}^{\ast}<0$ on the right hand side of the SVAR model introduced
in the next section, but the identification problem would remain the same.
More generally, we could assume that $r_{t}^{\ast}<0$ is observed with some
measurement error $\eta_{t}$, and include the measurement equation
$b_{L,t}=\min\left(  \alpha r_{t}^{\ast}+\eta_{t},0\right)$ in
the model together with a specification of the distribution of $\eta_{t}$. The
estimation method in this paper can then be seen as a special case
where we are entirely agnostic about
the measurement error. Adding such measures of unconventional policy is a
potentially very useful extension of the method since, if correctly
specified, they will likely improve estimation accuracy.

\smallskip

\textit{Connection to the microeconometrics literature} 
Equations (\ref{eq: Taylor}) and (\ref{eq: outcome nesting}) form a SEM with a limited dependent variable.
The special case with $\beta^{\ast}=0$ in (\ref{eq: outcome nesting}) can be referred to as a
\emph{kinked} SEM (KSEM), while the opposite case of $\beta^{\ast}=\beta$ in
(\ref{eq: outcome nesting}) can be called a \emph{censored} SEM (CSEM). Variants of the KSEM model have
been studied in the early microeconometrics literature on limited dependent
variable SEMs. \cite{Amemiya1974} and \cite{Lee1976} studied multivariate
extensions of the well-known Tobit model \citep{Tobin1958}.
\cite{NelsonOlson1978} argued that the KSEM was less suitable for
microeconometric applications than the CSEM, and the latter subsequently
became the main focus of the literature
\citep{SmithBlundell1986,BlundellSmith1989}. \cite{BlundellSmith1994} studied
the unrestricted model using external instruments, so they did not consider
the implications of the kink for identification.

One important lesson from the microeconometrics literature is that
establishing existence and uniqueness of equilibria in this class of models is
non-trivial. \cite{GourierouxLaffontMonfort1980} define a model to be
`coherent' if it has a unique solution for the endogenous variables in terms
of the exogenous variables, i.e., if there exists a unique reduced form. More
recently, the literature has distinguished between existence and uniqueness of
solutions using the terms coherency and completeness of the model,
respectively \citep{Lewbel2007}. Establishing coherency and completeness is a
necessary first step before we can study identification and
estimation.

\subsection{Identification}

Substituting for $r_{t}^{\ast}$ in (\ref{eq: outcome nesting}) using
(\ref{eq: Taylor}) and rearranging, we obtain%
\begin{align}
\pi_{t}  &  =\widetilde{c}+\widetilde{\beta}\left(  r_{t}-r^{n}\right)
+\widetilde{\varepsilon}_{1t},\,\ \text{where}\label{eq: pi tildes}\\
\widetilde{\beta}  &  =\frac{\beta-\beta^{\ast}}{1-\gamma\beta^{\ast}},\quad
\widetilde{c}=\frac{c}{1-\gamma\beta^*},\text{ and }%
\widetilde{\varepsilon}_{1t}=\frac{\varepsilon_{1t}+\beta^*
\varepsilon_{2t}}{1-\gamma\beta^*}. \label{eq: betatil}%
\end{align}
The system of equations (\ref{eq: Taylor}) and (\ref{eq: pi tildes}) is now a
KSEM, for which the necessary and sufficient condition for coherency and
completeness (existence of a unique solution) is $\widetilde{\beta}\gamma<1$
\citep{NelsonOlson1978}. Using (\ref{eq: betatil}), the coherency and
completeness condition can be expressed in terms of the structural parameters
as 
\begin{equation}
    \frac{  1-\gamma\beta}{ 1-\gamma
\beta^{\ast}}  >0.\label{eq: CC SEM}
\end{equation}
This condition evidently restricts the admissible
range of the structural parameters. It is satisfied in the present monetary
policy model, where it is natural to assume that $\beta,\beta^{\ast}\leq0$ and
$\gamma>0$. Therefore, it is possible that the coherency condition may not
provide additional information relative to what is often available from
natural sign restrictions on the parameters.

Under condition (\ref{eq: CC SEM}), the unique solution of the
model can be written as%
\begin{align}
\pi_{t}  &  =\mu_{1}+u_{1t}-\widetilde{\beta}D_{t}\left(  \mu_{2}%
+u_{2t}\right)  ,\text{ and }\label{eq: RF y1}\\
r_{t}  &  =\max\left(  \mu_{2}+u_{2t},0\right)  , \label{eq: RF y2}%
\end{align}
where $D_{t}:=1_{\left\{  r_{t}=0\right\}  }$ is an indicator (dummy) variable
that takes the value 1 when $r_{t}$ is on the boundary and zero otherwise, and%
\begin{align}
u_{1t}  &  =\frac{\widetilde{\varepsilon}_{1t}+\widetilde{\beta}%
\varepsilon_{2t}}{1-\gamma\widetilde{\beta}}=\frac{\varepsilon_{1t}%
+\beta\varepsilon_{2t}}{1-\gamma\beta},\quad u_{2t}=\frac{\gamma
\varepsilon_{1t}+\varepsilon_{2t}}{1-\gamma\beta},\label{eq: RF solution}\\
\mu_{1}  &  =\frac{c}{1-\gamma\beta},\quad\mu_{2}=\frac{\gamma c}%
{1-\gamma\beta}+r^{n}.\nonumber
\end{align}

Equations (\ref{eq: RF y1}) and (\ref{eq: RF y2}) express the endogenous
variables $\pi_{t},r_{t}$ in terms of the exogenous variables $\varepsilon
_{1t},\varepsilon_{2t},$ and correspond to the decision rules of the agents in
the model. It is clear that those decision rules differ in a world in which
the ZLB occasionally binds, which is characterized by $\widetilde{\beta}%
\neq0,$ compared to a world in which it never does (i.e., the CSEM), where
$\widetilde{\beta}=0.$ What is important for identification, however, is that
in a world in which the ZLB occasionally binds, agents' reaction to shocks
differs across regimes, and the difference depends on the parameter
$\widetilde{\beta},$ which from eq.~(\ref{eq: betatil}), depends on the
difference between the impact of conventional and unconventional
policies, $\beta$ and $\beta^{\ast},$ respectively. I will
show that this change provides information that identifies the structural
parameters: we get point identification when $\beta^{\ast}=0$ (the KSEM case),
and partial identification when $\beta^{\ast}\neq\beta.$ The identification
argument leverages the coefficient on the kink, $\widetilde{\beta},$ in the
`incidentally kinked' regression (\ref{eq: RF y1}), which is identified by a
variant of the well-known Heckit method \citep{Heckman1979}. I will sketch out
the argument below, and provide more details for the full SVAR model in the
next section.

\subsubsection{Identification of the KSEM}

Recall that in the KSEM\ model $\widetilde{\beta}=\beta.$ Consider the
estimation of $\beta$ in (\ref{eq: outcome kinked}) from a regression of
$\pi_{t}$ on $r_{t}$ using only observations above the ZLB,%
\begin{equation}
E\left(  \pi_{t}|r_{t},r_{t}>0\right)  =c+\beta\left(  r_{t}-r^{n}\right)
+\rho\left(  r_{t}-\mu_{2}+\tau\frac{\phi\left(  a\right)  }{1-\Phi\left(
a\right)  }\right)  ,\quad a=\frac{-\mu_{2}}{\tau},
\label{eq: trunc regression}%
\end{equation}
where $\rho=cov\left(  u_{1t},u_{2t}\right)  /\tau^{2}-\beta=\gamma\sigma
_{1}^{2}\left(  1-\gamma\beta\right)  /\left(  \gamma^{2}\sigma_{1}^{2}%
+\sigma_{2}^{2}\right)  $, $\tau=\sqrt{var\left(  u_{2t}\right)  },$ and
$\phi\left(  \cdot\right)  ,$ $\Phi\left(  \cdot\right)  $ are standard Normal
density and distribution functions, respectively. The coefficient $\rho$ is
the bias in the estimation of $\beta$ from the truncated regression
(\ref{eq: trunc regression}). Now, the mean of $\pi_{t}$ using the
observations at the ZLB is%
\begin{equation}
E\left(  \pi_{t}|r_{t}=0\right)  =c-\beta r^{n}+\rho\tau\frac{\phi\left(
a\right)  }{\Phi\left(  a\right)  }. \label{eq: E(pi|r=0)}%
\end{equation}
Next, observe that $\mu_{2},$ $\tau$ and hence $\phi\left(  a\right)
/\Phi\left(  a\right)  $ can be estimated from the Tobit regression
(\ref{eq: RF y2}). Therefore, we can recover the bias $\rho$ and identify
$\beta.$ A simple way to implement this is the control function approach
\citep{Heckman1978}. Let
\[
h_{t}\left(  \mu_{2},\tau\right)  :=\left(  1-D_{t}\right)  \left(  r_{t}%
-\mu_{2}\right)  -D_{t}\frac{\tau\phi\left(  a\right)  }{\Phi\left(  a\right)
},
\]
and run the regression%
\begin{equation}
E\left(  \pi_{t}|r_{t}\right)  =c_{1}+\beta r_{t}+\rho h_{t}\left(  \mu
_{2},\tau\right)  , \label{eq: cf regression}%
\end{equation}
where $c_{1}=c-\beta r^{n}$ is an unrestricted intercept. The rank condition
for the identification of $\beta$ is simply that the regressors in
(\ref{eq: cf regression}) are not perfectly collinear. This holds if and only
if $0<\Pr\left(  D_{t}=1\right)  <1.$ So, as long as some but not all the
observations are at the boundary, the model is generically identified.

\subsubsection{Partial identification of the unrestricted
SEM\label{s: partial ID}}

The discussion of the previous subsection shows that $\widetilde{\beta}$ is
identified from the kink in the reduced-form equation for $\pi_{t}$
(\ref{eq: RF y1}). It follows from eq.~(\ref{eq: betatil}) and the order condition that $\beta,\beta^*$ are not point identified. I will now demonstrate that they are partially identified. 

The assumption $cov\left(  \varepsilon_{1t},\varepsilon
_{2t}\right)  =0$ implies (see proof of Proposition 3 for a derivation)
\begin{equation}\label{eq: gamma}
\gamma=\frac{\omega_{12}-\omega_{22}\beta}{\omega_{11}-\omega_{12}\beta
}%
\end{equation}
where $\omega_{ij}:=cov\left(u_{it},u_{jt}\right)$. Substituting for $\gamma$ in (\ref{eq: betatil}) using (\ref{eq: gamma}) yields 
\begin{equation}
\widetilde{\beta}=\frac{\beta-\beta^{\ast}}{1-\frac{\omega_{12}-\omega
_{22}\beta}{\omega_{11}-\omega_{12}\beta}\beta^{\ast}}.\label{eq: ID set condition}
\end{equation}
For any given value of the reduced-form parameters $\widetilde{\beta},\Omega:=var(u_t)$, $u_t=(u_{1t},u_{2t})'$,  the identified set for $\left(  \beta,\beta^{\ast}\right)$ is a one-dimensional
manifold in $\Re^{2}$ defined by eq.~(\ref{eq: ID set condition})
intersected with the coherency condition (\ref{eq: CC SEM}). 

\begin{figure}[htb]%
\centering
\includegraphics[
height=0.7\textwidth,
width=0.9\textwidth,
trim=2cm 1.5cm 1.5cm 1cm, clip 
]%
{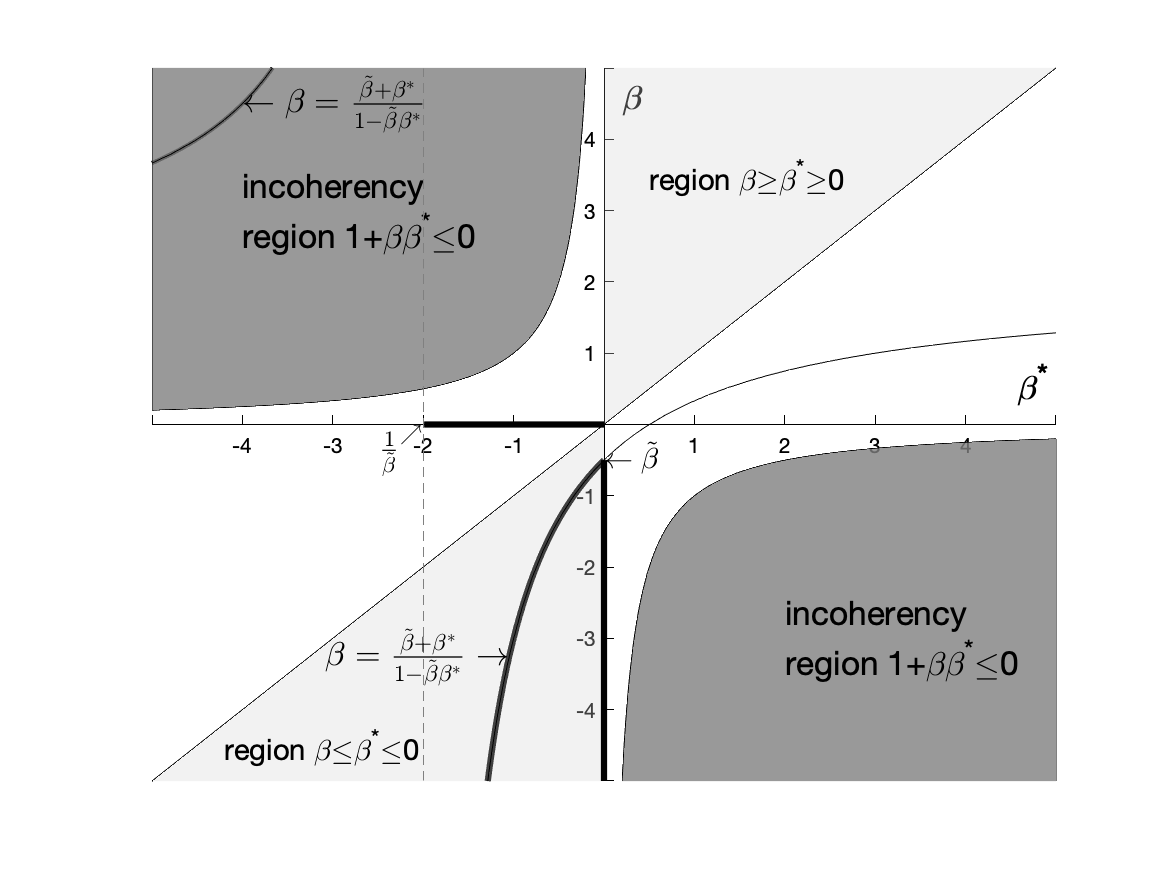}
\caption{The identified set for $\left(  \beta,\beta^{\ast}\right)  $ when
$\Omega=I$ and $\protect\widetilde{\beta}=-1/2,$ obtained by the intersection
of $\beta=\frac{\protect\widetilde{\beta}+\beta^{\ast}}%
{1-\protect\widetilde{\beta}\beta^{\ast}}$ and $1+\beta\beta^{\ast}>0$
(coherency condition). Light gray area corresponds to $\beta\leq\beta^{\ast
}\leq0$ and $\beta\geq\beta^{\ast}\geq0$. The thick part of the curve $\beta=\frac{\protect\widetilde{\beta}+\beta^{\ast}}%
{1-\protect\widetilde{\beta}\beta^{\ast}}$ indicates the identified set
obtained from the combined restrictions, and the bold intervals on the axes give the projections of the
identified set onto $\beta$ and $\beta^{\ast}$.}%
\label{fig: id set}%
\end{figure}

It is instructive to illustrate the identified set graphically at some given value of
$\widetilde{\beta}\,$and $\Omega$. Consider, for example, the case
$\Omega$ equal to the identity $I_{2}$, at which (\ref{eq: gamma}) yields $\gamma=-\beta,$ and the coherency
condition (\ref{eq: CC SEM}) reduces to $1+\beta\beta^{\ast}>0,$ and (\ref{eq: ID set condition}) yields the
function $\beta=\frac{\widetilde{\beta}+\beta^{\ast}}{1-\widetilde{\beta}%
\beta^{\ast}}$. Figure \ref{fig: id set} plots this function at
$\widetilde{\beta}=-1/2$. and highlights in dark gray the region of
incoherency defined by $1+\beta\beta^{\ast}\leq0$. The identified set is the
part of the function $\beta=\frac{\widetilde{\beta}+\beta^{\ast}%
}{1-\widetilde{\beta}\beta^{\ast}}$ that lies to the right of the pole at
$1/\widetilde{\beta},$ i.e., in the region $\beta^{\ast}>1/\widetilde{\beta
}=-2$ in this example. 

Now, consider the additional restrictions $\beta
\geq\beta^{\ast}\geq0$ or $\beta\leq\beta^{\ast}\leq0$, highlighted by the
light gray shaded areas in Figure \ref{fig: id set}. The interpretation of
those restrictions is that unconventional policy neither has the opposite
effect from conventional policy, nor is it more effective than conventional
policy. With this additional restriction, we see
that the identified set further shrinks to the part of $\beta=\frac
{\widetilde{\beta}+\beta^{\ast}}{1-\widetilde{\beta}\beta^{\ast}}$ in the
interval $(\widetilde{\beta}^{-1},0].$ The projection of the identified set
onto the $\beta$ axis yields $\beta\in(-\infty,\widetilde{\beta}],$ since
$\widetilde{\beta}<0$, while its projection onto the
$\beta^{\ast}$ axis yields $\beta^{\ast}\in$ $(\widetilde{\beta}^{-1},0]$. Because equations (\ref{eq: CC SEM}) and (\ref{eq: ID set condition})
encapsulate all the information in the reduced-form parameters about
$\beta,\beta^{\ast}$, the identified set obtained from them is
sharp.

Let us define a new parameter $\lambda$ such that $\beta^*=\lambda \beta$. The restriction indicated by the light gray areas in Figure \ref{fig: id set} corresponds to $\lambda\in\left[  0,1\right]$. If we
interpret $\lambda$ as a measure of the efficacy of unconventional policy,
this restriction implies that unconventional policy is neither counter- nor
over-productive. This reparameterization offers a convenient way to discretize the parameter space when we compute the identified set numerically, as is the case in the more general model discussed in the next section.

In this bivariate model, it is possible to characterize the identified set analytically. Here, I discuss the identified set for $\beta$ and defer the discussion of $\lambda$ to Appendix \ref{s: bounds}.

We have already established that $\beta$ is completely unidentified
when $\beta=\beta^*$ (equivalently $\lambda=1$), which corresponds to the CSEM. From the definition of
$\widetilde{\beta}$ in (\ref{eq: betatil}), it follows that
$\beta=\beta^*$ implies $\widetilde{\beta}=0.$ So, when $\widetilde{\beta}=0$,
$\beta$ is completely unidentified. It remains to see what happens when
$\widetilde{\beta}\neq0$. Let $\gamma_{0}:=\omega_{12}/\omega_{11}$, which can
be interpreted as the value the reaction function coefficient $\gamma$ in
(\ref{eq: Taylor}) would take if $\beta=0$, i.e., the value corresponding to a
Choleski identification scheme where $r_{t}$ is placed last. In Appendix \ref{s: bounds},
I prove the following bounds%
\begin{equation}%
\begin{tabular}
[c]{l}%
if $\widetilde{\beta}=0\text{ or }\widetilde{\beta}\gamma_{0}<0$, then
$\beta\in\Re;$ otherwise\\
if $\omega_{12}=\gamma_{0}=0$, then $\beta\in(-\infty,\widetilde{\beta}]$ if
$\widetilde{\beta}<0$ or $\beta\in\lbrack\widetilde{\beta},\infty)$ if
$\widetilde{\beta}>0$;\\
$\text{if }0<\widetilde{\beta}\gamma_{0}\leq1,\text{ then }\beta\in\left[
\frac{1}{\gamma_{0}},\widetilde{\beta}\right]  $ if $\widetilde{\beta}<0$ or
$\beta\in\left[  \widetilde{\beta},\frac{1}{\gamma_{0}}\right]  $ if
$\widetilde{\beta}>0$;\\
if $\widetilde{\beta}\gamma_{0}>1$, then $\lambda<0.$%
\end{tabular}
\ \ \ \ \ \ \label{eq: bounds}%
\end{equation}
We see that when $\widetilde{\beta}\neq0$ and $0\leq\widetilde{\beta}\gamma_0 \leq 1$, we
can identify both the sign of the causal effect $\beta$ of $r_{t}$ on $\pi_{t}$ and
get bounds on its magnitude. In particular, the
identified coefficient $\widetilde{\beta}$ is an attenuated measure of the
true causal effect $\beta$. Moreover, $\widetilde{\beta}\gamma_{0}>1$ implies
that $\beta^{\ast}$ has the opposite sign from $\beta$, i.e., unconventional
policy has the opposite effect of the conventional one. That could be
interpreted as saying that unconventional policy is counterproductive. 


Finally, the hypothesis that unconventional policy is as effective as conventional policy, $\beta^*
=\beta$ or $\lambda=1$, is equivalent to the null hypothesis $H_{0}:\widetilde{\beta
}=0$. The alternative that unconventional policy is less effective than
conventional policy, $\beta>\beta^{\ast}$ if $\beta>0,$ or $\beta<\beta^{\ast
}$ if $\beta<0$, corresponds to the two-sided alternative $H_{1}%
:\widetilde{\beta}\neq0$. This can be tested using a likelihood ratio test.

\section{SVAR with an occasionally binding constraint\label{s: SVAR}}

I now develop the methodology for identification and estimation of SVARs with
an occasionally binding constraint. Let $Y_{t}=\left(  Y_{1t}^{\prime}%
,Y_{2t}\right)  ^{\prime}$ be a vector of $k$ endogenous variables,
partitioned such that the first $k-1$ variables $Y_{1t}$ are unrestricted and
the $k$th variable $Y_{2t}$ is bounded from below by $b$.\footnote{The lower
bound does not need to be constant. All we need is to observe the periods in
which the economy is at the ZLB\ regime.} Define the latent process
$Y_{2t}^{\ast}$ that is only observed, and equal to $Y_{2t},$ whenever
$Y_{2t}>b$. If $Y_{2t}$ is a policy instrument, $Y_{2t}^{\ast}$ can be thought
of as the `shadow' instrument that measures the desired policy stance. The
$p$th-order SVAR model is given by the equations%
\begin{align}
A_{11}Y_{1t}+A_{12}Y_{2t}+A_{12}^{\ast}Y_{2t}^{\ast}  &  =B_{10}X_{0t}%
+\sum_{j=1}^{p}B_{1,j}Y_{t-j}+\sum_{j=1}^{p}B_{1,j}^{\ast}Y_{2,t-j}^{\ast
}+\varepsilon_{1t},\label{eq: Y1 SVAR}\\
A_{22}^{\ast}Y_{2t}^{\ast}+A_{22}Y_{2t}+A_{21}Y_{1t}  &  =B_{20}X_{0t}%
+\sum_{j=1}^{p}B_{2,j}Y_{t-j}+\sum_{j=1}^{p}B_{2,j}^{\ast}Y_{2,t-j}^{\ast
}+\varepsilon_{2t},\label{eq: Y2 SVAR}\\
Y_{2t}  &  =\max\left(  Y_{2t}^{\ast},b\right)  ,\nonumber
\end{align}
for $t\geq1$ given a set of initial values $Y_{-s},Y_{2,-s}^{\ast},$ for
$s=0,...,p-1$, and $X_{0t}$ are exogenous and predetermined variables.

Equation (\ref{eq: Y2 SVAR}) can be interpreted as a policy reaction function
because it determines the desired policy stance $Y_{2t}^{\ast}.$ Similarly,
$\varepsilon_{2t}$ is the corresponding policy shock. The above model is a
dynamic SEM. Two important differences from a standard SEM are the presence of
(i) latent lags amongst the predetermined variables on the right-hand side,
which complicates estimation; and (ii) the contemporaneous value of $Y_{2t}$
in the policy reaction function (\ref{eq: Y2 SVAR}), which allows it to vary
across ZLB and non-ZLB regimes. The presence of latent lags $Y_{2,t-j}^{\ast}$
in the policy rule (\ref{eq: Y2 SVAR}) is particularly useful because it
allows the model to incorporate forward
guidance \citep{ReifschneiderWilliams2000,DebortoliGaliGambetti2019}, see Appendix \ref{s: app FG} for details.

Collecting all the observed predetermined variables $X_{0t},Y_{t-1}%
,...,Y_{t-p}$ into a vector $X_{t},$ and the latent lags $Y_{2,t-1}^{\ast
},...,Y_{2,t-p}^{\ast}$ into $X_{t}^{\ast},$ and similarly for their
coefficients, the model can be written compactly as:
\begin{align}%
\begin{pmatrix}
A_{11} & A_{12}^{\ast} & A_{12}\\
A_{21} & A_{22}^{\ast} & A_{22}%
\end{pmatrix}%
\begin{pmatrix}
Y_{1t}\\
Y_{2t}^{\ast}\\
Y_{2t}%
\end{pmatrix}
&  =BX_{t}+B^{\ast}X_{t}^{\ast}+\varepsilon_{t},\label{eq: CKSVAR}\\
Y_{2t}  &  =\max\left\{  Y_{2t}^{\ast},b\right\}  .\nonumber
\end{align}
The vector of structural errors $\varepsilon_{t}$ is assumed to be
\textit{i.i.d.}~Normally distributed with zero mean and identity covariance.

In the previous section, we defined the KSEM as a special case of the general
model, where $Y_{2t}^{\ast}<b$ has no (contemporaneous) impact on $Y_{1t}.$ In
the dynamic setting, it feels natural to define the corresponding `kinked
SVAR' model (KSVAR) as a model in which $Y_{2t}^{\ast}$ has neither
contemporaneous nor dynamic effects. Therefore, the KSVAR obtains as a special
case of (\ref{eq: CKSVAR}) when both $A_{12}^{\ast}=0,$ and $B^{\ast}=0,$
which corresponds to a situation in which the bound is fully effective in
constraining what policy can achieve at all horizons.

The opposite extreme to the KSVAR is the censored SVAR model (CSVAR). Again,
unlike the CSEM, which only characterizes contemporaneous effects, the idea of
a CSVAR is to impose the assumption that the constraint is irrelevant at all
horizons. So, it corresponds to a fully unrestricted linear SVAR in the latent
process $\left(  Y_{1t}^{\prime},Y_{2t}^{\ast}\right)  ^{\prime}$. This is a
special case of (\ref{eq: CKSVAR}) when both $A_{12}=0$ and the elements of
$B$ corresponding to lagged $Y_{2t}$ are equal to zero. Finally, I refer to
the general model given by (\ref{eq: CKSVAR}) as the\ `censored and kinked
SVAR' (CKSVAR).

Define the $k\times k$ square matrices%
\begin{equation}
\overline{A}:=%
\begin{pmatrix}
A_{11} & A_{12}+A_{12}^{\ast}\\
A_{21} & A_{22}+A_{22}^{\ast}%
\end{pmatrix}
,\text{ \ and \ }A^{\ast}:=%
\begin{pmatrix}
A_{11} & A_{12}^{\ast}\\
A_{21} & A_{22}^{\ast}%
\end{pmatrix}
. \label{eq: Abar and A*}%
\end{equation}
$\overline{A}$ determines the impact effects of structural shocks during
periods when the constraint does not bind. $A^{\ast}$ does the same for
periods when the constraint binds.

To analyze the CKSVAR, we first need to establish existence and uniqueness of
the reduced form. This is done in the following proposition.

\begin{proposition}
\label{prop: coherency}The model given in eq.~(\ref{eq: CKSVAR}) is coherent
and complete (i.e., it has a unique solution) if and only if%
\begin{equation}
\kappa:=\frac{\overline{A}_{22}-A_{21}A_{11}^{-1}\overline{A}_{12}}%
{A_{22}^{\ast}-A_{21}A_{11}^{-1}A_{12}^{\ast}}>0. \label{eq: coherency}%
\end{equation}

\end{proposition}

Note that (\ref{eq: coherency}) does not depend on the coefficients on the
lags (whether latent or observed), so it is exactly the same as in a static
SEM. This condition is useful for inference, e.g., when constructing
confidence intervals or posteriors, because it restricts the range of
admissible values for the structural parameters. It can also be checked
empirically when the structural parameters are point-identified.

If condition (\ref{eq: coherency}) is satisfied, there exists a reduced-form
representation of the CKSVAR model (\ref{eq: CKSVAR}). For convenience of
notation, define the indicator (dummy variable) that takes the value one if
the constraint binds and zero otherwise:%
\begin{equation}
D_{t}=1_{\left\{  Y_{2t}=b\right\}  }. \label{eq: D}%
\end{equation}

\begin{proposition}
\label{prop: RF}If (\ref{eq: coherency}) holds, and for any initial values
$Y_{-s},Y_{2,-s}^{\ast},$ $s=0,...,p-1,$ the reduced-form representation of
(\ref{eq: CKSVAR}) for $t\geq1$ is given by%
\begin{align}
Y_{1t}  &  =\overline{C}_{1}X_{t}+\overline{C}_{1}^{\ast}\overline{X}%
_{t}^{\ast}+u_{1t}-\widetilde{\beta}D_{t}\left(  \overline{C}_{2}%
X_{t}+\overline{C}_{2}^{\ast}\overline{X}_{t}^{\ast}+u_{2t}-b\right)
\label{eq: RF Y1}\\
Y_{2t}  &  =\max\left(  \overline{Y}_{2t}^{\ast},b\right)  , \label{eq: RF Y2}%
\\
\overline{Y}_{2t}^{\ast}  &  =\overline{C}_{2}X_{t}+\overline{C}_{2}^{\ast
}\overline{X}_{t}^{\ast}+u_{2t},\label{eq: RF Y2*bar}\\
Y_{2t}^{\ast}  &  =\left(  1-D_{t}\right)  \overline{Y}_{2t}^{\ast}%
+D_{t}\left(  \kappa\overline{Y}_{2t}^{\ast}+\left(  1-\kappa\right)
b\right)  , \label{eq: RF Y2*}%
\end{align}
where $u_{t}=\left(  u_{1t}^{\prime},u_{2t}\right)  ^{\prime}=\overline
{A}^{-1}\varepsilon_{t},$ $\overline{C}^{\ast}=\left(  \overline{C}_{1}%
^{\ast\prime},\overline{C}_{2}^{\ast\prime}\right)  ^{\prime}=\kappa
\overline{A}^{-1}B^{\ast},$ $\overline{X}_{t}^{\ast}=\left(  \overline
{x}_{t-1},...,\overline{x}_{t-p}\right)  ^{\prime},$ $\overline{x}_{t}%
=\min\left(  \overline{Y}_{2t}^{\ast}-b,0\right)  ,$ $\overline{x}_{-s}%
=\kappa^{-1}\min\left(  Y_{2,-s}^{\ast}-b,0\right)  ,$ $s=0,...,p-1,$
\begin{equation}
\widetilde{\beta}=\left(  A_{11}-A_{12}^{\ast}A_{22}^{\ast-1}A_{21}\right)
^{-1}\left(  A_{12}^{\ast}A_{22}^{\ast-1}A_{22}-A_{12}\right)  ,
\label{eq: betatilde}%
\end{equation}
$\kappa$ is defined in (\ref{eq: coherency}) and the matrices $\overline
{C}_{1},\overline{C}_{2},$ are given in eq.~(\ref{eq: Cbar}) in the Appendix.
\end{proposition}

Note that the \textquotedblleft reduced-form\textquotedblright\ latent process
$\overline{Y}_{2t}^{\ast}$ is, in general, different from the
\textquotedblleft structural\textquotedblright\ shadow rate $Y_{2t}^{\ast}$
defined by (\ref{eq: RF Y2*}). They coincide only when $\kappa=1.$ This holds,
for example, in the CSVAR model.

Equation (\ref{eq: RF Y2}) combined with (\ref{eq: RF Y2*bar}) is a familiar
dynamic Tobit regression model with the added complexity of latent lags
included as regressors whenever $\overline{C}_{2}^{\ast}\neq0.$ Likelihood
estimation of the univariate version of this model was studied by
\cite{Lee1999}. The $k-1$ equations (\ref{eq: RF Y1}) are `incidentally
kinked' dynamic regressions, that I\ have not seen analyzed before.

\subsection{Identification}

\subsubsection{Identification of reduced-form parameters}

Let $\psi$ denote the parameters that characterize the reduced form
(\ref{eq: RF Y1})-(\ref{eq: RF Y2}): $\widetilde{\beta},\overline{C},$
$\overline{C}^{\ast}$ and $\Omega=var\left(  u_{t}\right)  .$ It is useful to
decompose $\psi$ into $\psi_{2}=\left(  \overline{C}_{2},\overline{C}%
_{2}^{\ast},\tau\right)  ^{\prime},$ where $\tau=\sqrt{var\left(
u_{2t}\right)  },$ and $\psi_{1}=(  vec\left(  \overline{C}_{1}\right)
^{\prime},vec\left(  \overline{C}_{1}^{\ast}\right)  ^{\prime}%
,\allowbreak\widetilde{\beta}^{\prime},\delta^{\prime},vech\left(  \Omega_{1.2}\right)
)  ^{\prime},$ where $\delta=\Omega_{12}/\tau^{2}$, $\Omega_{1.2}%
=\Omega_{11}-\delta\delta^{\prime}\tau^{2}$, and $\Omega_{ij}=cov\left(
u_{it},u_{jt}\right)  $.

Equation (\ref{eq: RF Y2}) is the dynamic Tobit regression model studied by
\cite{Lee1999}. So, its parameters, $\psi_{2},$ are generically identified
provided that the regressors are not perfectly collinear. This requires that
$0<\Pr\left(  D_{t}=1\right)  <1.$

Given $\psi_{2},$ the identification of the remaining parameters, $\psi_{1},$
can be characterized using a control function approach. Consider the $k-1$
regression equations%
\begin{equation}
E\left(  Y_{1t}|Y_{2t},X_{t},\overline{X}_{t}^{\ast}\right)  =\overline{C}%
_{1}X_{t}+\overline{C}_{1}^{\ast}\overline{X}_{t}^{\ast}+\widetilde{\beta
}Z_{1t}+\delta Z_{2t}, \label{eq: mom Y1}%
\end{equation}
where%
\begin{align}
Z_{1t}  &  =D_{t}\left(  b-\overline{C}_{2}X_{t}-\overline{C}_{2}^{\ast
}\overline{X}_{t}^{\ast}-\frac{\tau\phi\left(  a_{t}\right)  }{\Phi\left(
a_{t}\right)  }\right)  ,\label{eq: Z1}\\
Z_{2t}  &  =\left(  1-D_{t}\right)  \left(  Y_{2t}-\overline{C}_{2}%
X_{t}-\overline{C}_{2}^{\ast}\overline{X}_{t}^{\ast}\right)  +D_{t}\frac
{\tau\phi\left(  a_{t}\right)  }{\Phi\left(  a_{t}\right)  }, \label{eq: Z2}%
\end{align}
$a_{t}=\left(  \frac{b-\overline{C}_{2}X_{t}-\overline{C}_{2}^{\ast}%
\overline{X}_{t}^{\ast}}{\tau}\right)  $, and $\phi\left(  \cdot\right)  ,$
$\Phi\left(  \cdot\right)  $ are the standard Normal density and distribution
functions, respectively. When $\overline{C}^{\ast}$ is different from zero,
regressors $\overline{X}_{t}^{\ast}$, $Z_{1t},$ and $Z_{2t}$ in
(\ref{eq: mom Y1}) are unobserved, so we need to replace them with their
expectations conditional on $Y_{2t},Y_{t-1},...,Y_{1}.$ Then, the regressors
on the right-hand side of (\ref{eq: mom Y1}) become $\mathbf{X}_{t}:=\left(
X_{t}^{\prime},\overline{X}_{t|t}^{\ast\prime},Z_{1t|t},Z_{2t|t}\right)
^{\prime}$, where $h_{t|t}:=E\left(  h\left(  \overline{X}_{t}^{\ast}\right)
|Y_{2t},Y_{t-1},...,Y_{1}\right)  $ for any function $h\left(  \cdot\right)  $
whose expectation exists.\footnote{In the KSVAR model, we have $\overline
{C}_{1}^{\ast}=0$ and $\overline{C}_{2}^{\ast}=0,$ so $\overline{X}_{t}^{\ast
}$ drops out of (\ref{eq: mom Y1}), and the regressors $Z_{1t},Z_{2t}$ are
observed, so $Z_{jt|t}=Z_{jt},$ $j=1,2.$} The coefficients $\overline{C}%
_{1},\overline{C}_{1}^{\ast},\widetilde{\beta},$ and $\delta$ are generically
identified if the regressors $\mathbf{X}_{t}$ are not perfectly collinear.

\subsubsection{Identification of structural parameters}

From the order condition, we can easily establish that there are not enough
restrictions to identify all the structural parameters in the
CKSVAR\ (\ref{eq: CKSVAR}). Let $k_{0}=\dim\left(  X_{0t}\right)  $ denote the
number of predetermined variables other than the own lags of $Y_{t}.$ For
example, in a standard VAR without deterministic trends, we have $X_{0t}=1,$
so $k_{0}=1.$ The number of reduced-form parameters $\psi$ is $k_{0}k+k^{2}p$
(in $\overline{C}$) plus $kp$ (in $\overline{C}^{\ast}$) plus $k-1$ (in
$\widetilde{\beta}$) plus $k\left(  k+1\right)  /2$ (in $\Omega$). The number
of structural parameters in (\ref{eq: CKSVAR}) is $k_{0}k+k^{2}p$ (in $B$)
plus $kp$ (in $B^{\ast}$) plus $k^{2}$ (in $\overline{A}$) plus $k$ (in
$A_{12}^{\ast}$ and $A_{22}$). So, the CKSVAR is underidentified by $k\left(
k-1\right)  /2+1$ restrictions. Nevertheless, I will show that the impulse
responses to $\varepsilon_{2t}$ are identified. Specifically, they are
point-identified when $A_{12}^{\ast}=0,$ and partially identified when
$A_{12}^{\ast}\neq0$ but $A_{12}^{\ast}$ and $A_{12}$ have the same sign,
analogous to the bounds given in equation (\ref{eq: bounds}) in the previous section.

Because the CKSVAR is nonlinear, IRFs are obviously state-dependent, and there
are many ways one can define them, see \cite{KoopPesaranPotter1996}. The
IRF to $\varepsilon_{2t},$ according to any of the definitions proposed in the
literature, is identified if the reduced-form errors $u_{t}$ can be expressed
as a known function of $\varepsilon_{2t}$ and a process that is orthogonal to
it, i.e., $u_{t}=g\left(  \varepsilon_{2t},e_{t}\right)  ,$ where $e_{t}$ is
independent of $\varepsilon_{2t}.$ From Proposition \ref{prop: RF}, it follows
that the function $g$ is linear, and more specifically,%
\begin{align}
u_{1t}  &  =\left(  I_{k-1}-\overline{\beta}\overline{\gamma}\right)
^{-1}\left(  \bar{\varepsilon}_{1t}+\overline{\beta}\bar{\varepsilon}%
_{2t}\right)  ,\,\text{\ and}\label{eq: u1}\\
u_{2t}  &  =\left(  1-\overline{\gamma}\overline{\beta}\right)  ^{-1}\left(
\bar{\varepsilon}_{2t}+\overline{\gamma}\bar{\varepsilon}_{1t}\right)  ,
\label{eq: u2}%
\end{align}
where%
\begin{align}
\overline{\beta}  &  :=-A_{11}^{-1}\overline{A}_{12},\quad\overline{\gamma
}:=-\overline{A}_{22}^{-1}A_{21},\label{eq: beta and gamma bar}\\
\bar{\varepsilon}_{1t}  &  :=A_{11}^{-1}\varepsilon_{1t},\quad\bar
{\varepsilon}_{2t}:=\overline{A}_{22}^{-1}\varepsilon_{2t},\nonumber
\end{align}
and $\overline{A}_{22}=A_{22}^{\ast}+A_{22},$ defined in
(\ref{eq: Abar and A*}). Note that $\overline{\beta}$ can be interpreted as
the response of $Y_{1t}$ to a shock that increases $Y_{2t}$ by one unit, and
$\overline{\gamma}$ are the contemporaneous reaction function coefficients of
$Y_{2t}$ to $Y_{1t}$ when $Y_{2t}>b$ (unconstrained regime). The shock vector
$\overline{\varepsilon}_{1t}$ is not structural but it is orthogonal to
$\varepsilon_{2t},$ so it plays the role of $e_{t}$ in $u_{t}=g\left(
\varepsilon_{2t},e_{t}\right)  .$ Hence, the IRF\ is identified if and only if
$\overline{\beta},$ $\overline{\gamma},$ and $\overline{A}_{22}$ are identified.

The following proposition shows identification when $A_{12}^{\ast}=0$.

\begin{proposition}
\label{prop: point ID}When $A_{12}^{\ast}=0$ and the coherency condition
(\ref{eq: coherency}) holds, the parameters in (\ref{eq: u1})-(\ref{eq: u2})
are identified by the equations $\overline{\beta}=\widetilde{\beta},$
\begin{align}
\overline{\gamma}  &  =\left(  \Omega_{12}^{\prime}-\Omega_{22}\overline
{\beta}^{\prime}\right)  \left(  \Omega_{11}-\Omega_{12}\overline{\beta
}^{\prime}\right)  ^{-1},\text{ and}\label{eq: gammabar ident}\\
\overline{A}_{22}^{-1}  &  =\sqrt{\left(  -\overline{\gamma},1\right)
\Omega\left(  -\overline{\gamma},1\right)  ^{\prime}}.
\label{eq: A22bar ident}%
\end{align}

\end{proposition}

\textit{Remarks}
1. $\overline{\beta}=\widetilde{\beta}$ follows immediately from the
definition (\ref{eq: betatilde}) with $A_{12}^{\ast}=0$. Equations
(\ref{eq: gammabar ident}) and (\ref{eq: A22bar ident}) hold without the
restriction $A_{12}^{\ast}=0.$ They follow from the orthogonality of the
shocks $\varepsilon_{2t}$ and $\overline{\varepsilon}_{1t}.$

2. An instrumental variables interpretation of this identification result is
as follows. Define the instrument%
\[
Z_{t}:=Y_{1t}-\widetilde{\beta}Y_{2t}=A_{11}^{-1}B_{1}X_{t}+A_{11}^{-1}%
B_{1}^{\ast}X_{t}^{\ast}+A_{11}^{-1}\varepsilon_{1t},
\]
where the second equality holds when $A_{12}^{\ast}=0$. The orthogonality of
the errors $E\left(  \varepsilon_{1t}\varepsilon_{2t}\right)  =0$ implies
$E\left(  Z_{t}\varepsilon_{2t}\right)  =0.$ So, $Z_{t}$ are valid $k-1$
instruments for the $k-1$ endogenous regressors $Y_{1t}$ in the structural
equation of $Y_{2t}=\max\left(  Y_{2t}^{\ast},b\right)  $, where $Y_{2t}%
^{\ast}$ is given by (\ref{eq: Y2 SVAR}). Normalizing (\ref{eq: Y2 SVAR}) in
terms of $Y_{2t}^{\ast}$ yields the structural equation in the more familiar
form of a policy rule:%
\begin{equation}
Y_{2t}=\max\left(  \overline{\gamma}Y_{1t}+\bar{B}_{2}X_{t}+\bar{B}_{2}^{\ast
}X_{t}^{\ast}+\bar{\varepsilon}_{2t},b\right)  , \label{eq: Y2 norm}%
\end{equation}
where $\bar{B}_{2}=$ $\overline{A}_{22}^{-1}B_{2},$ $\bar{B}_{2}^{\ast}=$
$\overline{A}_{22}^{-1}B_{2}^{\ast}$. Since $A_{11}^{-1}$ is non-singular,
the\ coefficient matrix of $Z_{t}$ in the `first-stage' regressions of
$Y_{1t}$ is nonsingular, so the coefficients of (\ref{eq: Y2 norm}) are
generically identified by the rank condition. An alternative to the Tobit IV
regression model (\ref{eq: Y2 norm}) is the indirect Tobit regression approach
used in the static SEM by \cite{BlundellSmith1994}. Equation
(\ref{eq: Y2 norm}) can be written as the dynamic Tobit regression%
\begin{equation}
Y_{2t}=\max\left(  \widetilde{\gamma}Z_{t}+\tilde{B}_{2}X_{t}+\tilde{B}%
_{2}^{\ast}X_{t}^{\ast}+\tilde{\varepsilon}_{2t},b\right)  ,
\label{eq: Y2 exog}%
\end{equation}
where $\widetilde{\gamma}=\left(  1-\overline{\gamma}\overline{\beta}\right)
^{-1}\overline{\gamma},$ $\tilde{B}_{2}=\left(  1-\overline{\gamma}%
\overline{\beta}\right)  ^{-1}\bar{B}_{2},$ $\tilde{B}_{2}^{\ast}=\left(
1-\overline{\gamma}\overline{\beta}\right)  ^{-1}\bar{B}_{2}^{\ast}$ and
$\tilde{\varepsilon}_{2t}=\left(  1-\overline{\gamma}\overline{\beta}\right)
^{-1}\bar{\varepsilon}_{2t}.$ Note that the coherency condition
(\ref{eq: coherency}) becomes $\kappa=\frac{\overline{A}_{22}}{A_{22}^{\ast}%
}\left(  1-\overline{\gamma}\overline{\beta}\right)  >0,$ so $1-\overline
{\gamma}\overline{\beta}\neq0,$ which guarantees the existence of the
representation (\ref{eq: Y2 exog}). Given $\overline{\beta}=\widetilde{\beta
},$ the structural parameter $\overline{\gamma}$ can then be obtained as
$\overline{\gamma}=\widetilde{\gamma}\left(  I_{k-1}+\widetilde{\beta
}\widetilde{\gamma}\right)  ^{-1},$ and similarly for the remaining structural
parameters in (\ref{eq: Y2 norm}).

3. The parameter $A_{22}$ allows the reaction function of $Y_{2t}^{\ast}$ to
differ across the two regimes. The special case $A_{22}=0$ thus corresponds to
the restriction that the reaction function remains the same across regimes.
The parameters $A_{22}$ and $A_{22}^{\ast}$ are not separately identified.
Hence, $A_{22}^{\ast-1}$, the scale of the response to the shock
$\varepsilon_{2t}$ during periods when $Y_{2t}=b$,$\,$ is not
identified.\footnote{This is akin to the well-known property of a probit model
that the scale of the distribution of the latent process is not identifiable.}
Similarly, $\kappa=\frac{\overline{A}_{22}}{A_{22}^{\ast}}\left(
1-\overline{\gamma}\overline{\beta}\right)  $ is not identified, and
therefore, neither is the structural shadow value $Y_{2t}^{\ast}$ in eq.
(\ref{eq: RF Y2*}). Identification of these requires an additional restriction
on $A_{22}$, e.g., $A_{22}=0.$ Turning this discussion around, we see that a
change in the reaction function across regimes does not destroy the point
identification of the effects of policy during the unconstrained regime, since
the latter only requires $\overline{\beta},\overline{\gamma}$ and
$\overline{A}_{22}$, not $A_{22}^{\ast}$ or $\kappa.$ \bigskip

Next, we turn to the case $A_{12}^{\ast}\neq0,$ and derive identification
under restrictions on the sign and magnitude of $A_{12}^{\ast}$ relative to
$A_{12}$ and $A_{22}^{\ast}$ relative to $A_{22}.$ The first restriction is
motivated by a generalization of the discussion on the SEM model in Subsection
\ref{s: partial ID}. Specifically, if $\overline{A}_{12}=A_{12}+A_{12}^{\ast}$
measures the effect of conventional policy (operating in the unconstrained
regime) and $A_{12}^{\ast}$ measures the effect of unconventional policy
(operating in the constrained regime), then the assumption that $A_{12}$ and
$A_{12}^{\ast}$ have the same sign means that unconventional policy effects
are neither in the opposite direction nor larger in absolute value than
conventional policy effects. In other words, unconventional policy is neither
counterproductive nor over-productive relative to conventional policy. This
can be characterized by the specification $A_{12}^{\ast}=\Lambda\overline
{A}_{12}$ and $A_{12}=\left(  I_{k-1}-\Lambda\right)  \overline{A}_{12},$
where $\Lambda=diag\left(  \lambda_{j}\right)  ,$ $\lambda_{j}\in\left[
0,1\right]  $ for $j=1,...,k-1.$ I further impose the restriction that
$\lambda_{j}=\lambda$ for all $j,$ so that $A_{12}^{\ast}=\lambda\overline
{A}_{12}$ and $A_{12}=\left(  1-\lambda\right)  \overline{A}_{12}$ with
$\lambda\in\left[  0,1\right]  .$ This, in turn, means that $Y_{2t}$ and
$Y_{2t}^{\ast}$ enter each of the first $k-1$ structural equations for
$Y_{1t}$ only via the common linear combination $\lambda Y_{2t}^{\ast
}\allowbreak+\allowbreak\left(  1-\lambda\right)  Y_{2t},$ which can be
interpreted as a measure of the effective policy stance.

We also need to consider the impact of $A_{22}$ on identification. The
parameter $\zeta=\overline{A}_{22}/A_{22}^{\ast}$ gives the ratio of the
standard deviation of the monetary policy shock in the constrained relative to
the unconstrained regime. It is also the ratio of the reaction function
coefficients in the two regimes, e.g., $A_{22}^{\ast-1}A_{21}$ versus
$\overline{A}_{22}^{-1}A_{21}$. I will impose $\zeta>0,$ so that the sign of
the policy shock does not change across regimes. With the above
reparametrization and the definitions in (\ref{eq: beta and gamma bar}), the
identified coefficient $\widetilde{\beta}$ in (\ref{eq: betatil}) can be
written as%

\begin{equation}
\widetilde{\beta}=\left(  1-\xi\right)  \left(  I-\xi\overline{\beta}%
\overline{\gamma}\right)  ^{-1}\overline{\beta},\quad\xi:=\lambda\zeta.
\label{eq: betatil xi}%
\end{equation}
Similarly, given $\zeta>0,$ the coherency condition (\ref{eq: coherency})
reduces to $\left(  1-\overline{\gamma}\overline{\beta}\right)  \left(
1-\xi\overline{\gamma}\overline{\beta}\right)  >0.$ Notice that the parameters
$\lambda,\zeta$ only appear multiplicatively, so it suffices to consider them
together as $\xi=\lambda\zeta.$ Once $\overline{\beta}$ is known, the
remaining structural parameters needed to obtain the IRF to $\varepsilon_{2t}$
are $\overline{\gamma}$ and $\overline{A}_{22}$, and they are obtained from
Proposition \ref{prop: point ID}. So, the identified set can be characterized
by varying $\xi$ over its admissible range. Without further restrictions on
$\zeta,$ the admissible range is obviously $\xi\geq0.$ If we further assume
that $\zeta\leq1,$ i.e., that the slope of the reaction function coefficients
is no steeper in the constrained regime than in the unconstrained regime, then
$\xi\in\left[  0,1\right]  ,$ and so partial identification proceeds exactly
along the lines of the SEM in the previous section where $\lambda$ played the
role of $\xi.$ In the case $k=2,$ the bounds derived in eq.~(\ref{eq: bounds})
apply, with $\beta=\overline{\beta}$ in the notation of the present section.
However, when $k>2,$ it is difficult to obtain a simple analytical
characterization of the identified set for $\overline{\beta}.$ In any case, we
will typically wish to obtain the identified set for functions of the
structural parameters, such as the IRF. This can be done numerically by
searching over a fine discretization of the admissible range for $\xi.$ An
algorithm for doing this is provided in Appendix \ref{s: ID set}.

\subsection{Estimation}

Estimation of the CKSVAR is carried out by Maximum Likelihood (ML) using either a version of the sequential importance sampler (SIS) of \cite{Lee1999} or the fully adapted particle filter (FAPF) of \cite{MalikPitt2011} to evaluate the likelihood, except in the case of the KSVAR model for which the likelihood is available analytically. The details are given in Appendix \ref{s: likelihood}. 

Using the limit theory of \cite{Newe94Mc}, the ML estimator can be shown
to be consistent and asymptotically Normal and the LR statistic asymptotically
$\chi^{2}$ with degrees of freedom equal to the number of restrictions. Standard asymptotics arise when the probability of each regime occurring is bounded away from zero. Infrequent visits to one of the two regimes will slow down the rate of convergence of the estimator, but will not lead to a non-standard limiting distribution. Since the focus of this paper is on identification, I will not discuss primitive conditions for these results, such as geometric ergodicity, which can be shown, for example, by bounding the joint spectral radius of the companion-form representation of the model \citep{Liebscher2005}. Instead, I report Monte Carlo simulation results on the finite-sample properties of ML estimators and LR tests in Appendix
\ref{s: numerical}. They show that the Normal distribution provides a very
good approximation to the finite-sample distribution of the ML estimators. I
find some finite-sample size distortion in the LR tests of various
restrictions on the CKSVAR, but this can be addressed effectively with a
parametric bootstrap, as shown in the Appendix.

One interesting observation from the simulations is that the LR test of the
CSVAR restrictions against the CKSVAR appears to be less powerful than the
corresponding test of the KSVAR restrictions against the CKSVAR. Thus, we
expect to be able to detect deviations from KSVAR more easily than deviations
from CSVAR. In other words, finding evidence against the hypothesis that
unconventional policies are fully effective (CSVAR) may be harder than finding
evidence against the opposite hypothesis that they are completely ineffective (KSVAR).

\section{Application\label{s: application}}

I use the three-equation SVAR of \cite{StockWatson01}, consisting of
inflation, the unemployment rate and the Federal Funds rate to provide a
simple empirical illustration of the methodology developed in this paper. As
discussed in \cite{StockWatson01}, this model is far too limited to provide
credible identification of structural shocks, so the results in this section
are meant as an illustration of the new methods.

\begin{figure}[htb]%
\centering
\includegraphics[
height=4in,
width=\textwidth
]%
{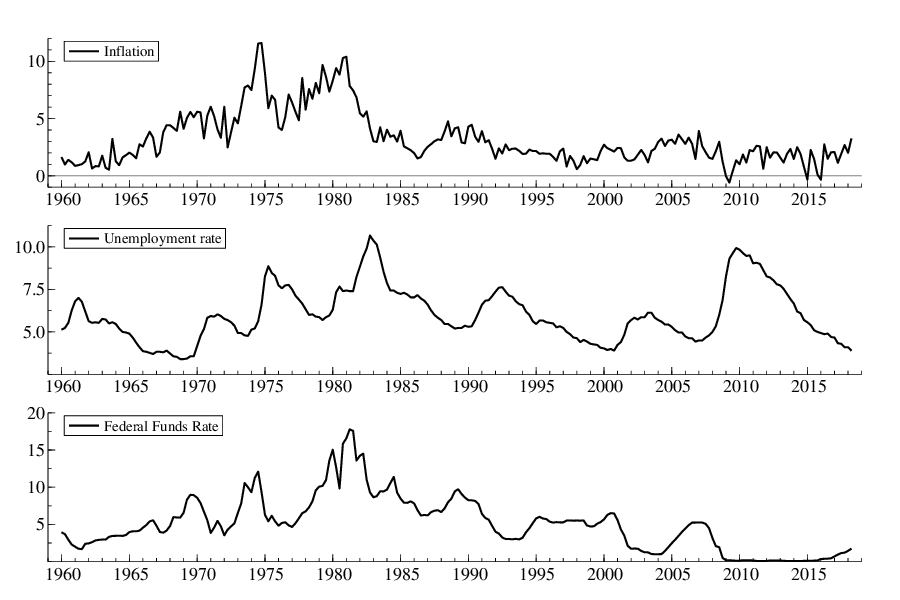}%
\caption{Data used in Stock and Watson (2001) over the extended sample 1960q1
to 2018q2.}%
\label{fig: SWdata}%
\end{figure}

The data are quarterly and are constructed exactly as in \cite{StockWatson01}%
.\footnote{The inflation data are computed as $\pi_{t}=$ $400ln(P_{t}%
/P_{t-1})$, where $P_{t}$, is the implicit GDP deflator and $u_{t}$ is the
civilian unemployment rate. Quarterly data on $u_{t}$ and $i_{t}$ are formed
by taking quarterly averages of their monthly values.} The variables are
plotted in Figure \ref{fig: SWdata} over the extended sample 1960q1 to 2018q2.
I will consider all periods in which the Fed funds rate was below 20 basis
points to be on the ZLB. This includes 28 quarters, or 11\% of the sample.%

\subsection{Tests of efficacy of unconventional policy}

I estimate three specifications of the SVAR(4) with the ZLB: the unrestricted
CKSVAR specification, as well as the restricted KSVAR and CSVAR
specifications. The maximum log-likelihood for each model is reported in Table
\ref{t: likelihood}, computed using the SIS\ algorithm in the case of CKSVAR
and CSVAR, with 1000 particles. The accuracy of the SIS algorithm was gauged
by comparing the log-likelihood to the one obtained using the resampling FAPF
algorithm. In both CKSVAR and CSVAR the difference is very small. The results
are also very similar when we increase the number of particles to 10000.
Finally, the table reports the LR tests of KSVAR and
CSVAR against CKSVAR using both asymptotic and parametric bootstrap p-values.%

\begin{table}[tbp] \centering
\caption{Estimated CKSVAR models in Inflation, Unemployment and the Federal Funds rate\tabnoteref[a]{tab1}}
\label{t: likelihood}
\begin{tabular}
[c]{lllllll}\hline
Model & log lik & (FAPF) & \# restr. & LR stat. & Asym. p-val. &
\multicolumn{1}{|l}{Boot. p-val.}\\\hline
\multicolumn{1}{r|}{CKSVAR(4)} & \multicolumn{1}{|c|}{-81.64} &
\multicolumn{1}{|c|}{-81.94} & \multicolumn{1}{|c|}{} & \multicolumn{1}{|c|}{}
& \multicolumn{1}{|c}{} & \multicolumn{1}{|c}{}\\
\multicolumn{1}{r|}{KSVAR(4)} & \multicolumn{1}{|c|}{-97.05} &
\multicolumn{1}{|c|}{-} & \multicolumn{1}{|c|}{12} &
\multicolumn{1}{|c|}{30.82} & \multicolumn{1}{|c}{0.002} &
\multicolumn{1}{|c}{0.011}\\
\multicolumn{1}{r|}{CSVAR(4)} & \multicolumn{1}{|c|}{-94.86} &
\multicolumn{1}{|c|}{-94.87} & \multicolumn{1}{|c|}{14} &
\multicolumn{1}{|c|}{26.43} & \multicolumn{1}{|c}{0.023} &
\multicolumn{1}{|c}{0.117}\\\hline
\end{tabular}
\tabnotetext[a]{tab1}{Maximized log-likelihood of various SVAR models in inflation, unemployment and Federal funds rate. CKSVAR corresponds to the unrestricted specification (\ref{eq: RF Y1})-(\ref{eq: RF Y2}); KSVAR excludes latent lags; CSVAR is a purely censored model. CKSVAR and CSVAR likelihoods computed using sequential importance sampling with 1000 particles (alternative estimates based on Fully Adapted Particle Filtering with resampling are shown in parentheses). Asymptotic p-values from $\chi^2_q$, $q=$ number of restrictions. Bootstrap p-values from parametric bootstrap with 999 replications. Sample: 1960q1-2018q2 (234 obs, 11\% at ZLB)}%
\end{table}%

The KSVAR imposes the restriction that no latent lags (i.e., lags of the
shadow rate) should appear on the right hand side of the model, i.e.,
$B^{\ast}=0$ in (\ref{eq: CKSVAR}) or $\overline{C}_{1}^{\ast}=0$ and
$\overline{C}_{2}^{\ast}=0$ in (\ref{eq: RF Y1}) and (\ref{eq: RF Y2}). This
amounts to 12 exclusion restrictions on the CKSVAR(4), four restrictions in
each of the three equations. This is necessary (but not sufficient) for the
hypothesis that unconventional policy is completely ineffective at all
horizons. It is necessary because $\overline{C}^{\ast}=\left(  \overline
{C}_{1}^{\ast\prime},\overline{C}_{2}^{\ast\prime}\right)  ^{\prime}\neq0$
would imply that unconventional policy has at least a lagged effect on
$Y_{t}.$ $\overline{C}^{\ast}=0$ is not sufficient to infer that
unconventional policy is completely ineffective because it may still have a
contemporaneous effect on $Y_{1t}$ if $A_{12}^{\ast}\neq0,$ and the latter is
not point-identified. The result of the test in Table \ref{t: likelihood}
shows that lags of the shadow rate are statistically significant at the 5\%
level, rejecting the null hypothesis that unconventional policy has no effect.

The CSVAR model imposes the restriction that only the coefficients on the lags
of the shadow rate (which is equal to the actual rate above the ZLB) are
different from zero in the model, i.e., the elements of $B$ corresponding to
lags of $Y_{2t}$ in (\ref{eq: CKSVAR}) are all zero, or equivalently, the
elements of $\overline{C}$ corresponding to lags of $Y_{2t}$ in
(\ref{eq: RF Y1}) and (\ref{eq: RF Y2}) are all equal to $\overline{C}^{\ast}%
$. In addition, it imposes the restriction that $\widetilde{\beta}=0$ in
(\ref{eq: RF Y1}), i.e., no kink in the reduced-form equations for inflation
and unemployment across regimes, yielding 14 restrictions in total. This is
necessary for the hypothesis that the ZLB\ is empirically irrelevant for
policy in that it does not limit what monetary policy can achieve. The
evidence against this hypothesis is not as strong as in the case of the KSVAR.
The asymptotic p-value is 0.023, indicating rejection of the null hypothesis
that unconventional policy is as effective as conventional policy at the 5\%
level, but the bootstrap p-value is 0.117. Note that this difference could
also be due to fact that the test of the CSVAR restrictions may be less
powerful than the test of the KSVAR restrictions, as indicated by the
simulations reported in the previous section. Thus, I\ would cautiously
conclude that the evidence on the empirical relevance of the ZLB\ is mixed.
Further evidence on the efficacy of unconventional policy will also be
provided in the next subsection.

\subsection{Impulse response functions}

Based on the evidence reported in the previous section, I estimate the IRFs
associated with the monetary policy shock using the unrestricted CKSVAR
specification, and compare them to recursive IRFs from the CSVAR specification
that place the Federal funds rate last in the causal ordering. From the
identification results in Section \ref{s: SVAR}, the CKSVAR point-identifies
the nonrecursive IRFs only under the assumption that the shadow rate has no
contemporaneous effect of $Y_{1t},$ i.e., $A_{12}^{\ast}=0$ in
(\ref{eq: Y1 SVAR}). Note that, due to the nonlinearity of the model, the IRFs
are state-dependent. I use the following definition of the IRF from
\cite{KoopPesaranPotter1996}:\footnote{The kink in the reduced-form representation of the model makes it difficult to approximate the IRFs by local projections on simple nonlinear functions of the data, such as powers or interactions with the regime indicator, see Appendix \ref{s: IRF} for further discussion of this point.}
\begin{equation}
IRF_{h,t}\left(  \varsigma,X_{t},\overline{X}_{t}^{\ast}\right)  =E\left(
Y_{t+h}|\varepsilon_{2t}=\varsigma,X_{t},\overline{X}_{t}^{\ast}\right)
-E\left(  Y_{t+h}|\varepsilon_{2t}=0,X_{t},\overline{X}_{t}^{\ast}\right)  .
\label{eq: IRF}%
\end{equation}

\begin{figure}[htb]%
\centering
\includegraphics[
height=4.0439in,
width=\textwidth
]%
{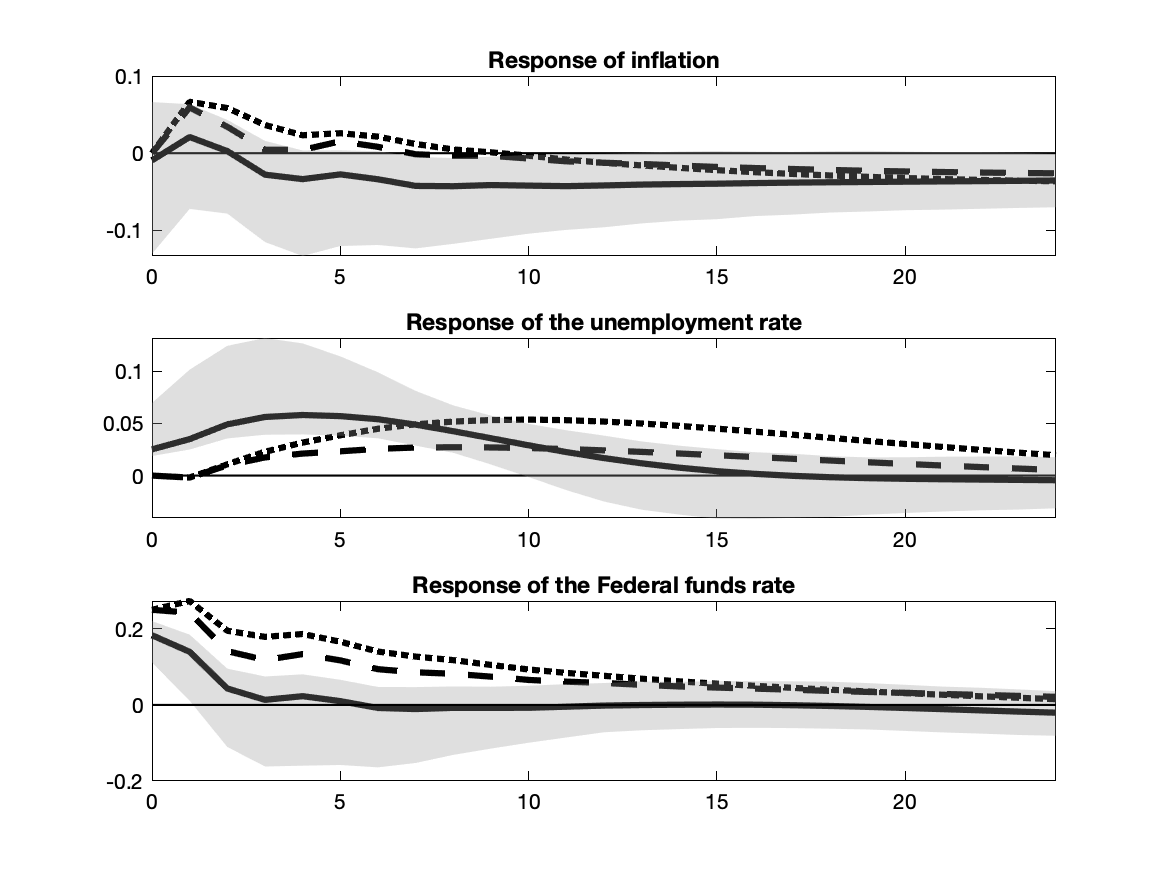}%
\caption{IRFs to a 25bp monetary policy shock in 2018q3 from a CKSVAR(4)
estimated over the period 1960q1 to 2018q2. The solid line corresponds to point estimates from
the nonrecursive identification using the ZLB under the assumption that
unconventional policy is ineffective on impact, with 90\% bootstrap error bands in gray. The dashed line corresponds to the nonlinear recursive IRF, estimated with the CSVAR(4) under the restriction that the contemporaneous impact of Fed Funds on inflation and unemployment is
zero. The dotted line corresponds to the recursive IRF from a
linear SVAR(4) estimated by OLS with Fed Funds ordered last.}%
\label{fig: IRFs}%
\end{figure}

Figure \ref{fig: IRFs} reports the nonrecursive IRFs to a 25 basis points
monetary policy shock at the end of the sample, 2018q3 (at which point $\overline{X}_{t}^{\ast}=0$ in (\ref{eq: IRF}) because interest rates had been above the ZLB over the previous four quarters), from the CKSVAR under the assumption that unconventional policy has no contemporaneous effect ($\lambda=0$). It also reports two different
estimates of recursive IRFs using the identification scheme in
\cite{StockWatson01} with interest rates placed last. The first estimate is obtained
from the CSVAR specification, and the second is a ``naive'' OLS estimate of the IRF that ignores the ZLB constraint -- a direct application of the method in \cite{StockWatson01} to the present sample. The figure also
reports 90\% bootstrap error bands for the nonrecursive IRFs.

In the nonrecursive IRF, the response of inflation to a monetary tightening is
negative on impact, albeit very small, and, with the exception of the first
quarter when it is positive, it stays negative throughout the horizon. Hence,
the incidence of a price puzzle is mitigated relative to the recursive IRFs,
according to which inflation rises for up to 6 quarters after a monetary
tightening (9 quarters in the OLS case). Note, however, that the error bands
are so wide that they cover (pointwise) most of the recursive IRF, though less
so for the OLS one. Turning to the unemployment response, we see that the
nonrecursive IRF starts significantly positive on impact (no transmission lag)
and peaks much earlier (after 4 quarters) than the recursive IRF (10
quarters). In this case, the recursive IRF\ is outside the error bands for
several quarters (more so for the naive OLS IRF). Finally, the response of the
Federal funds rate to the monetary tightening is less than one on impact and
generally significantly lower than the recursive IRFs. This is both due to the
contemporaneous feedback from inflation and unemployment, as well as the fact
that there is a considerable probability of returning to the ZLB, which
mitigates the impact of monetary tightening.%

\begin{figure}[htb]%
\centering
\includegraphics[
height=4.0439in,
width=\textwidth
]%
{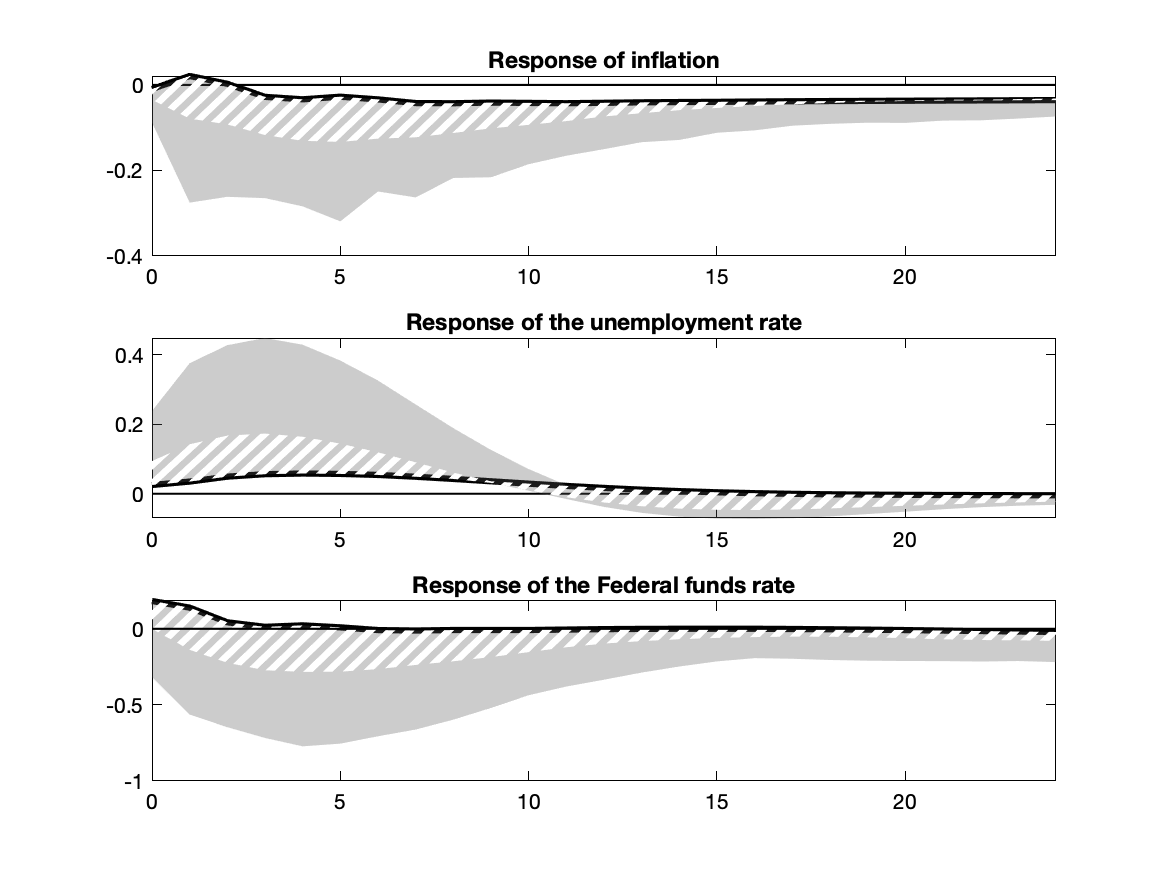}%
\caption{Identified sets of the IRFs to a 25bp monetary policy shock in 2018q3 from a
CKSVAR(4) estimated over the period 1960q1 to 2018q2. The shaded area denotes the identified set, the solid line
indicates the point-identified IRF\ under the restriction that
unconventional policy is ineffective on impact. The striped area imposes
the restriction that the response of the Fed funds rate on impact should be nonnegative.}%
\label{fig: Set ID IRFs}%
\end{figure}

Next, I turn to the identified sets of the IRFs that arise when I relax the
restriction that unconventional policy is ineffective, i.e., $\lambda$ can be
greater than zero. I consider the range of $\xi=\lambda\zeta\in\left[
0,1\right]  ,$ recalling that $\lambda$ measures the efficacy of
unconventional policy and $\zeta$ measures the ratio of the reaction function
coefficients and shock volatilities in the constrained versus the
unconstrained regimes. The shaded areas in Figure
\ref{fig: Set ID IRFs} report the identified sets without any other
restrictions. The striped areas (a subset of the aforementioned identified sets) show the tightening of the identified sets when I impose the additional sign restriction that the contemporaneous effect of the
monetary policy shock to the Fed Funds rate should be nonnegative. The bold lines show the IRFs under the (point-identifying) assumption $\lambda=0$. The latter are the same as the
nonrecursive point estimates reported in Figure \ref{fig: IRFs}.

We observe that the identified set for the IRF of inflation is bounded from
above by the limiting case $\lambda=0$. This is also true of the response of
the Fed Funds rate. The case $\lambda=0$ provides a lower bound on the effect
to unemployment only from 0 to 9 quarters. Even though the point estimate of
the unemployment response under $\lambda=0$ remains positive over all
horizons, the identified set includes negative values beyond 10 quarters
ahead. We also notice that the identified sets are fairly large, albeit still
informative. Interestingly, the identified IRF of the Fed Funds rate includes
a range of negative values on impact. These values arise because for values of
$\xi>0$, there are generally two solutions for the structural VAR parameters
$\overline{\beta},\overline{\gamma}$ in the equations
(\ref{eq: gammabar ident}), (\ref{eq: betatil xi}), with one of them inducing
such strong responses of inflation and unemployment to the interest rate that
the contemporaneous feedback in the policy rule would in fact revert the
direct positive effect of the policy shock on the interest rate. If we impose
the additional sign restriction that the contemporaneous impact of the policy
shock to the Fed Funds rate must be non-negative, then those values are ruled
out and the identified sets become considerably tighter. This is an example of
how sign restrictions can lead to tighter partial identification of the IRF.%

With an additional assumption on $\zeta,$ the method can be used to obtain an
estimate of the identified set for $\lambda,$ the measure of the efficacy of
unconventional policy. In particular, if we set $\zeta=1,$ i.e., the reaction
function remains the same across the two regimes, then the identified set for
$\lambda$ is $\left[  0.0.506\right]  .$ In other words, the identified set
excludes values of the efficacy of policy beyond 51\%, so that, roughly
speaking, unconventional policy is at most 51\% as effective as conventional
one. Note that this estimate does not account for sampling uncertainty and
relies crucially on the assumption that the reaction function remains the same
across the two regimes. This assumption could be justified by arguing that
there is no reason to believe that policy objectives may have shifted over the
ZLB period, and that any desired policy stance was feasible over that period.
The latter assumption may be questionable. For example, one can imagine that
there may be financial and political constraints on the amount of quantitative
easing policy makers could do, which may cause them to proceed more cautiously
over the ZLB period than over regular times. Within the context of our model,
this would be reflected as a flatter policy reaction function over the ZLB
period than over the non-ZLB periods, i.e., it will correspond to $\zeta<1$.
To illustrate the implications of this for the identification of $\lambda$,
suppose that $\zeta=1/2,$ i.e., the shadow rate reacts half as fast to shocks
during the ZLB period than it does in the non-ZLB period. Then, the identified
set for $\lambda$ would include 1, i.e., the data would be consistent with the
view that unconventional policy is fully effective. So, under this alternative
assumption on $\zeta$, the reason we observed a subdued response to policy
shocks over the ZLB period is because policy was less active over that period,
and policy shocks were smaller, not because unconventional policy was
partially ineffective.

As I discussed in the introduction, it is difficult to make further progress
on this issue without further information or additional assumptions. The
technical reason is that the scale of the latent regression over the censored
sample is not identified, so additional information is required to untangle
the structural parameters $\lambda$ and $\zeta$ from $\xi=\lambda\zeta$. One
possibility would be to identify $\lambda$ from the coefficients on the lags
of $Y_{2t}$ and $Y_{2t}^{\ast}$ by imposing the (overidentifying) restriction
that $Y_{2t-j}$ and $Y_{2t-j}^{\ast}$ appear in the model only via the linear
combination $Y_{2t-j}^{\text{eff}}:=\lambda Y_{2t-j}^{\ast}\allowbreak
+\allowbreak\left(  1-\lambda\right)  \allowbreak Y_{2t-j}$ for all lags
$j=1,...,p,$ where $Y_{2t}^{\text{eff}}$ can be interpreted as the effective
policy stance. Provided that the coefficients on the lags of $Y_{2t}^{\ast}$
or $Y_{2t}$ are not all zero, this restriction point identifies $\lambda,$ and
hence, partially identifies $\zeta$ from $\xi.$ One could obtain tighter
bounds by using sign restrictions \citep[see][]{IkedaLiMavroeidisZanetti2020}, or obtain point identification by using conventional identification schemes. For instance, one can identify $\overline{\beta}$ directly
using external instruments, as in \cite{GertlerKaradi2015}, and hence point
identify $\xi$ from (\ref{eq: betatil xi}).

\section{Conclusion}

This paper has shown that the ZLB can be used constructively to identify the
causal effects of monetary policy on the economy. Identification relies on the
(in)efficacy of alternative (unconventional) policies. When unconventional
policies are partially effective in mitigating the impact of the ZLB, the
causal effects of monetary policy are only partially identified. A general
method is proposed to estimate SVARs subject to an occasionally binding
constraint. The method can be used to test the efficacy of unconventional
policy, modelled via a shadow rate. Application to a core three-equation SVAR
with US data suggests that the ZLB is empirically relevant and unconventional
policy is only partially effective.

\begin{appendix}
\section{Proofs\label{s: proofs}}

\subsection{Derivation of identified set for model of Section \ref{s: examples}}\label{s: bounds}

Using the notation $\beta^* = \lambda\beta$, eq.~(\ref{eq: ID set condition}) can be expressed as
\begin{equation}
\widetilde{\beta}=g\left(  \beta\right)  \beta,\text{\quad}g\left(
\beta\right)  :=\frac{1-\lambda}{1-\lambda\frac{\beta\left(  \omega
_{12}-\omega_{22}\beta\right)  }{\omega_{11}-\omega_{12}\beta}}.
\label{eq: betatil 1}%
\end{equation}

When $\omega_{12}=0,$ we have $g\left(  \beta\right)  =$ $\frac{1-\lambda
}{1+\lambda\omega_{22}\beta^{2}/\omega_{11}}\in\left(  0,1\right)  $ for all
$\lambda\in\left(  0,1\right)  .$ Therefore, when $\widetilde{\beta}\neq0,$
the sign of $\beta$ is the same as that of $\widetilde{\beta}$ and its
magnitude is lower, as stated in (\ref{eq: bounds}).

Next, consider $\omega_{12}\neq0.$ It is easily seen that $g\left(  0\right)
=1-\lambda$ and $\lim_{\beta\rightarrow\pm\infty}\left(  \beta\right)  =0.$
Moreover,
\[
\frac{\partial g}{\partial\beta}=\lambda\left(  1-\lambda\right)  \frac
{\omega_{12}\omega_{22}\beta^{2}-2\omega_{11}\omega_{22}\beta+\omega
_{11}\omega_{12}}{\left(  \omega_{11}-\beta\omega_{12}-\beta\lambda\omega
_{12}+\beta^{2}\lambda\omega_{22}\right)  ^{2}}.
\]
For $\lambda\in\left(  0,1\right)  ,$ the above derivative function has zeros
at $
\omega_{12}\omega_{22}\beta^{2}-2\omega_{11}\omega_{22}\beta+\omega_{11}%
\omega_{12}=0$,
which occur at
\[%
\begin{array}
[c]{c}%
\beta_{1}=\frac{\omega_{11}\omega_{22}+\sqrt{\omega_{11}\omega_{22}\left(
\omega_{11}\omega_{22}-\omega_{12}^{2}\right)  }}{\omega_{12}\omega_{22}}\\
\beta_{2}=\frac{\omega_{11}\omega_{22}-\sqrt{\omega_{11}\omega_{22}\left(
\omega_{11}\omega_{22}-\omega_{12}^{2}\right)  }}{\omega_{12}\omega_{22}}%
\end{array}
,\quad\text{if }\omega_{12}\neq0.
\]
Now, because $0<\left(  \omega_{11}\omega_{22}-\omega_{12}^{2}\right)
<\omega_{11}\omega_{22}$ implies $\sqrt{\omega_{11}\omega_{22}\left(
\omega_{11}\omega_{22}-\omega_{12}^{2}\right)  }<\omega_{11}\omega_{22},$ we
have $\beta_{i}<0,$ $i=1,2,$ when $\omega_{12}<0$ and $\beta_{i}>0,$ $i=1,2,$
when $\omega_{12}>0.$

By symmetry, it suffices to consider only one of the two cases, e.g., the case
$\omega_{12}<0.$ In this case, $g^{\prime}\left(  \beta\right)  =\frac
{\partial g}{\partial\beta}<0$ for all $\beta>0$ and, since $g\left(
0\right)  =1-\lambda$ and $g\left(  \infty\right)  =0,$ it follows that
$g\left(  \beta\right)  \in\left(  0,1-\lambda\right)  $ for all $\beta>0.$
Thus, from (\ref{eq: betatil 1}) we see that $\widetilde{\beta}<0$ cannot
arise from $\beta>0$ when $\omega_{12}<0.$ In other words, observing
$\widetilde{\beta}<0$ must mean that $\beta<0.$ Moreover, since $g^{\prime
}\left(  \beta\right)  <0$ for all $\beta>\beta_{1}$ and $\beta_{1}<0,$ it
must be that $g\left(  \beta\right)  >0$ for all $\beta>\beta_{1},$ and hence,
also for $\beta_{1}<\beta\leq0.$ At $\beta<\beta_{1},$ $g^{\prime}\left(
\beta\right)  >0,$ and since $g^{\prime}\left(  \beta\right)  <0$ for all
$\beta<\beta_{2}<\beta_{1},$ and $g\left(  -\infty\right)  =0,$ it has to be
that $g\left(  \beta\right)  $ approaches zero from below as $\beta
\rightarrow-\infty,$ and therefore, $g\left(  \beta\right)  $ must cross zero
at some $\beta_{0}\in\left(  \beta_{2},\beta_{1}\right)  ,$ and $g\left(
\beta\right)  \geq0$ for all $\beta\in\left[  \beta_{0},0\right]  .$
Inspection of (\ref{eq: betatil 1}) shows that $\beta_{0}=\omega_{11}%
/\omega_{12}=1/\gamma_{0},$ which corresponds to $\gamma=-\infty$ from
(\ref{eq: gamma}). Since $g\left(  \beta\right)  \in\left[
0,1-\lambda\right]  $ for all $\beta\in\left[  \beta_{0},0\right]  ,$ and
$\lambda\in(0,1),$ it follows from (\ref{eq: betatil 1}) that $\left\vert
\widetilde{\beta}\right\vert \leq\left\vert \beta\right\vert $. In other
words, $\widetilde{\beta}$ is attenuated relative to the true $\beta.$

Finally, we notice that there is a minimum value of $\widetilde{\beta}$ that
one can observe under the restriction $\lambda\in\left[  0,1\right]  $ (at
$\lambda=1,$ $\widetilde{\beta}=0$). Given the attenuation bias and the fact
that $\widetilde{\beta}<0$ if and only if $\beta\in\left[  \beta_{0},1\right]
,$ the smallest value of $\widetilde{\beta}$ occurs when $\lambda=0$ and
$\beta=\omega_{11}/\omega_{12},$ so $\widetilde{\beta}_{\min}=\omega
_{11}/\omega_{12}=1/\gamma_{0}$. Thus, observing $\widetilde{\beta}%
<\omega_{11}/\omega_{12}$ and $\omega_{12}<0,$ or $\widetilde{\beta}%
\omega_{12}/\omega_{11}>1,$ violates the identifying restriction that
$\lambda\geq0$, for only with a $\lambda<0$ can we get $g\left(  \beta\right)
>1$ when $\beta<0$ and hence $\widetilde{\beta}<\beta<0$.

\subsubsection{Bounds on $\lambda$}

The bounds on $\lambda$ are obtained by finding all the values of $\lambda$
for which equation (\ref{eq: betatil 1}) has a solution for $\beta$. This
equation implies%
\[
\beta^{2}\left(  \left(  1-\lambda\right)  \omega_{12}+\widetilde{\beta
}\lambda\omega_{22}\right)  -\beta\left(  \left(  1-\lambda\right)
\omega_{11}+\widetilde{\beta}\left(  1+\lambda\right)  \omega_{12}\right)
+\widetilde{\beta}\omega_{11}=0,
\]
whose discriminant is the following quadratic function of lambda:
\begin{align*}
D\left(  \lambda\right)   &  =\left(  \left(  1-\lambda\right)  \omega
_{11}+\widetilde{\beta}\left(  1+\lambda\right)  \omega_{12}\right)
^{2}-4\widetilde{\beta}\omega_{11}\left(  \left(  1-\lambda\right)
\omega_{12}+\widetilde{\beta}\lambda\omega_{22}\right).
\end{align*}
Hence, the identified set for $\lambda$ corresponds to $S_{\lambda}=\left\{
\lambda:D\left(  \lambda\right)  \geq0\right\}  $. This set is non-empty
because $D\left(  0\right)  \geq0.$ It can be computed analytically and can
take the following three shapes: (i) $S_{\lambda}=\Re$ if $D\left(
\lambda\right)  \allowbreak\geq\allowbreak0$ for all $\lambda\in\Re$; (ii)
$S_{\lambda}\allowbreak=\allowbreak(-\infty,\lambda_{lo}]\allowbreak
\cup\allowbreak\lbrack\lambda_{up},\infty)$ if $\omega_{11}\allowbreak
-\allowbreak\widetilde{\beta}\omega_{12}\allowbreak\neq\allowbreak0,$ where
$\lambda_{lo}<\lambda_{up}$ are the roots of $D\left(  \lambda\right)  =0$;
and (iii) $S_{\lambda}\allowbreak=\allowbreak(-\infty,\lambda_{lo}]$\ if
$\omega_{11}\allowbreak-\allowbreak\widetilde{\beta}\omega_{12}\allowbreak
=\allowbreak0,$ because $\omega_{11}^{2}\allowbreak-\allowbreak\left(
\frac{\omega_{11}}{\omega_{12}}\right)  ^{2}\allowbreak\omega_{12}%
^{2}\allowbreak-\allowbreak2\left(  \frac{\omega_{11}}{\omega_{12}}\right)
\allowbreak\omega_{11}\omega_{12}\allowbreak+\allowbreak2\left(  \frac
{\omega_{11}}{\omega_{12}}\right)  ^{2}\allowbreak\omega_{11}\omega
_{22}\allowbreak=\allowbreak2\omega_{11}^{2}\allowbreak\frac{\omega_{11}%
\omega_{22}-\omega_{12}^{2}}{\omega_{12}^{2}}\allowbreak>\allowbreak0$. If we
also impose the restriction $\lambda\in\left[  0,1\right]  ,$ then the
identified set is $S_{\lambda}\cap\left[  0,1\right]  $.

\subsection{Proof of Proposition \ref{prop: coherency}}

Define $\overline{A}_{i2}:=A_{i2}^{\ast}+A_{i2},$ $i=1,2$ as the right blocks
of $\overline{A}$ that was defined in (\ref{eq: Abar and A*}). Applying
\cite[Theorem 1]{GourierouxLaffontMonfort1980}, coherency holds if and only if
$\det\overline{A}$ and $\det A^{\ast}$ have the same sign. Without loss of
generality, we can assume that $A_{11}$ is nonsingular (this can always be
achieved by reordering the variables in $Y_{t}$). From (\ref{eq: Abar and A*}%
), we have $\det A^{\ast}=\det A_{11}\det\left(  A_{22}^{\ast}-A_{21}%
A_{11}^{-1}A_{12}^{\ast}\right)  $ and $\det\overline{A}=\det A_{11}%
\det\left(  \overline{A}_{22}-A_{21}A_{11}^{-1}\overline{A}_{12}\right)  $
\citep[p. 50 (6)]{Lutkepohl96}. The coherency condition can be written as
$\det\overline{A}/\det A^{\ast}>0,$ which, given that $\left(  \overline
{A}_{22}-A_{21}A_{11}^{-1}\overline{A}_{12}\right)  $ and $\left(
A_{22}^{\ast}-A_{21}A_{11}^{-1}A_{12}^{\ast}\right)  $ are scalars, yields
(\ref{eq: coherency}).

\subsection{Proof of Proposition \ref{prop: RF}}

Define $\overline{A}_{i2}:=A_{i2}^{\ast}+A_{i2},$ $i=1,2$ as the right blocks
of $\overline{A}$ that was defined in (\ref{eq: Abar and A*}). Also let
$Y_{t}^{\ast}:=\left(  Y_{1t}^{\prime},Y_{2t}^{\ast}\right)  ^{\prime}.$ When
the coherency condition (\ref{eq: coherency}) holds, the solution of
(\ref{eq: CKSVAR}) exists and is unique. It can be expressed as%
\begin{equation}
Y_{t}^{\ast}=\left\{
\begin{array}
[c]{ll}%
CX_{t}+C^{\ast}X_{t}^{\ast}+u_{t}, & \text{if }D_{t}=0\\
\widetilde{C}X_{t}+\widetilde{C}^{\ast}X_{t}^{\ast}+\widetilde{c}%
b+\widetilde{u}_{t}, & \text{if }D_{t}=1
\end{array}
\right.  \label{eq: Y*}%
\end{equation}
where%
\begin{equation}
C=\overline{A}^{-1}B,\quad C^{\ast}=\overline{A}^{-1}B^{\ast},\quad
u_{t}=\overline{A}^{-1}\varepsilon_{t} \label{eq: C}%
\end{equation}
and%
\begin{equation}
\widetilde{C}=A^{\ast-1}B,\quad\widetilde{C}^{\ast}=A^{\ast-1}B^{\ast}%
,\quad\widetilde{c}=-A^{\ast-1}\binom{A_{12}}{A_{22}}b,\quad\widetilde{u}%
_{t}=A^{\ast-1}\varepsilon_{t}. \label{eq: Ctilde}%
\end{equation}

Using the partitioned inverse formula, we obtain%
\begin{align*}
\widetilde{C}_{1}  &  =\left(  A_{11}-A_{12}^{\ast}A_{22}^{\ast-1}%
A_{21}\right)  ^{-1}\left(  B_{1}-A_{12}^{\ast}A_{22}^{\ast-1}B_{2}\right) \\
\widetilde{C}_{2}  &  =\left(  A_{22}^{\ast}-A_{21}A_{11}^{-1}A_{12}^{\ast
}\right)  ^{-1}\left(  B_{2}-A_{21}A_{11}^{-1}B_{1}\right)
\end{align*}
and%
\begin{align*}
C_{1}  &  =\left(  A_{11}-\overline{A}_{12}\overline{A}_{22}^{-1}%
A_{21}\right)  ^{-1}\left(  B_{1}-\overline{A}_{12}\overline{A}_{22}^{-1}%
B_{2}\right) \\
C_{2}  &  =\left(  \overline{A}_{22}-A_{21}A_{11}^{-1}\overline{A}%
_{12}\right)  ^{-1}\left(  B_{2}-A_{21}A_{11}^{-1}B_{1}\right)  .
\end{align*}
Solving the latter for $B_{1}$ and $B_{2}$ yields%
\[
B_{1}=A_{11}C_{1}+\overline{A}_{12}C_{2},\text{ and }B_{2}=\overline{A}%
_{22}C_{2}+A_{21}C_{1}\text{.}%
\]
Thus,%
\begin{align*}
\widetilde{C}_{1}  &  =C_{1}+\left(  A_{11}-A_{12}^{\ast}A_{22}^{\ast-1}%
A_{21}\right)  ^{-1}\left(  \overline{A}_{12}-A_{12}^{\ast}A_{22}^{\ast
-1}\overline{A}_{22}\right)  C_{2}=C_{1}-\widetilde{\beta}C_{2},\text{ and}\\
\widetilde{C}_{2}  &  =\left(  A_{22}^{\ast}-A_{21}A_{11}^{-1}A_{12}^{\ast
}\right)  ^{-1}\left(  A_{22}^{\ast}-A_{21}A_{11}^{-1}A_{12}^{\ast}%
+A_{22}-A_{21}A_{11}^{-1}A_{12}\right)  C_{2}=\kappa C_{2},
\end{align*}
where $\kappa$ is given in (\ref{eq: coherency}). The exact same derivations
apply to $\widetilde{C}^{\ast},$ i.e.,
\[
\widetilde{C}_{1}^{\ast}=C_{1}^{\ast}-\widetilde{\beta}C_{2}^{\ast},\text{
\ and \ }\widetilde{C}_{2}^{\ast}=\kappa C_{2}^{\ast}.
\]

Next,
\begin{align*}
\widetilde{c}_{1}  &  =\left(  A_{11}-A_{12}^{\ast}A_{22}^{\ast-1}%
A_{21}\right)  ^{-1}\left(  A_{12}^{\ast}A_{22}^{\ast-1}A_{22}-A_{12}\right)
b=\widetilde{\beta}b,\text{ and}\\
\widetilde{c}_{2}  &  =-\frac{A_{22}-A_{21}A_{11}^{-1}A_{12}}{A_{22}^{\ast
}-A_{21}A_{11}^{-1}A_{12}^{\ast}}b=\left(  1-\kappa\right)  b.
\end{align*}
Finally, $\widetilde{u}_{t}=A^{\ast-1}\overline{A}u_{t}=\left(  \widetilde{u}%
_{1t}^{\prime},\widetilde{u}_{2t}\right)  ^{\prime}$, where%
\[
\widetilde{u}_{1t}=u_{1t}-\widetilde{\beta}u_{2t},\text{ and \ }%
\widetilde{u}_{2t}=\kappa u_{2t}.
\]
Substituting back into (\ref{eq: Y*}), the reduced-form model\ for $Y_{1t}$
becomes%
\begin{align}
Y_{1t}  &  =\left(  1-D_{t}\right)  \left(  C_{1}X_{t}+C_{1}^{\ast}X_{t}%
^{\ast}+u_{1t}\right) \nonumber\\
&  +D_{t}\left(  \left(  C_{1}-\widetilde{\beta}C_{2}\right)  X_{t}+\left(
C_{1}^{\ast}-\widetilde{\beta}C_{2}^{\ast}\right)  X_{t}^{\ast}+u_{1t}%
-\widetilde{\beta}u_{2t}\right)  , \label{eq: Y1 1}%
\end{align}
and for $Y_{2t}^{\ast}$ it is%
\begin{equation}
Y_{2t}^{\ast}=C_{2}X_{t}+C_{2}^{\ast}X_{t}^{\ast}+u_{2t}-\left(
1-\kappa\right)  D_{t}\left(  C_{2}X_{t}+C_{2}^{\ast}X_{t}^{\ast}%
+u_{2t}-b\right)  . \label{eq: RF Y2* p}%
\end{equation}

Next, define%
\begin{equation}
\widetilde{Y}_{2t}^{\ast}:=C_{2}X_{t}+C_{2}^{\ast}X_{t}^{\ast}+u_{2t},
\label{eq: Y2*til 0}%
\end{equation}
and rewrite (\ref{eq: RF Y2* p}) as
\begin{align}
Y_{2t}^{\ast}  &  =\widetilde{Y}_{2t}^{\ast}-\left(  1-\kappa\right)
D_{t}\left(  \widetilde{Y}_{2t}^{\ast}-b\right) \nonumber\\
&  =\left(  1-D_{t}\right)  \widetilde{Y}_{2t}^{\ast}+D_{t}\left(
\kappa\widetilde{Y}_{2t}^{\ast}+\left(  1-\kappa\right)  b\right)  .
\label{eq: RF Y2* pp}%
\end{align}
Let $q=\dim X_{t}$ denote the number of elements of $X_{t}$ and define, for
each $i=1,2,$%
\begin{equation}
\overline{C}_{ij}=\left\{
\begin{array}
[c]{ll}%
C_{ij}, & j\in\left\{  1,q\right\}  :X_{tj}\neq Y_{2,t-s}\text{ for all }%
s\in\left\{  1,p\right\} \\
C_{ij}+C_{is}^{\ast}, & j\in\left\{  1,q\right\}  :X_{tj}=Y_{2,t-s},\text{ for
some }s\in\left\{  1,p\right\}  .
\end{array}
\right.  \label{eq: Cbar}%
\end{equation}
In other words, $\overline{C}$ contains the original coefficients on all the
regressors other than the lags of $Y_{2t},$ while the coefficients on the lags
of $Y_{2t}$ are augmented by the corresponding coefficients of the lags of
$Y_{2t}^{\ast}.$ For example, if $p=1$ and there are no other exogenous
regressors $X_{0t},$ then, for $i=1,2,$%
\[
C_{i}X_{t}+C_{i}^{\ast}X_{t}^{\ast}=C_{i1}Y_{1t-1}+C_{i2}Y_{2t-1}+C_{i}^{\ast
}Y_{2t-1}^{\ast},
\]
so $\overline{C}_{i}=\left(  C_{i1},C_{i2}+C_{i}^{\ast}\right)  $. Using
(\ref{eq: Cbar}), we can rewrite (\ref{eq: Y2*til 0}) as
\begin{equation}
\widetilde{Y}_{2t}^{\ast}=\overline{C}_{2}X_{t}+C_{2}^{\ast}\min\left(
X_{t}^{\ast}-b,0\right)  +u_{2t}. \label{eq: Y2*til 1}%
\end{equation}
Now, observe that%
\[
\min\left(  Y_{2t}^{\ast}-b,0\right)  =D_{t}\left(  Y_{2t}^{\ast}-b\right)
=\kappa D_{t}\left(  \widetilde{Y}_{2t}^{\ast}-b\right)  =\kappa\min\left(
\widetilde{Y}_{2t}^{\ast}-b,0\right)
\]
So, letting $\widetilde{X}_{t}^{\ast}$ denote the lags of $\widetilde{Y}%
_{2t}^{\ast},$ we have $\min\left(  X_{t}^{\ast}-b,0\right)  =\kappa
\min\left(  \widetilde{X}_{t}^{\ast}-b,0\right)  ,$ and consequently,
\[
C^{\ast}\min\left(  X_{t}^{\ast}-b,0\right)  =\overline{C}^{\ast}\min\left(
\widetilde{X}_{t}^{\ast}-b,0\right)  ,
\]
where $\overline{C}^{\ast}=\kappa C^{\ast}.$ Now, from (\ref{eq: Y2*til 1}) we
have%
\[
\widetilde{Y}_{2t}^{\ast}=\overline{C}_{2}X_{t}+\overline{C}_{2}^{\ast}%
\min\left(  \widetilde{X}_{t}^{\ast}-b,0\right)  +u_{2t}.
\]
Recall the definition of $\overline{Y}_{2t}^{\ast}$ in (\ref{eq: RF Y2*bar}):%
\[
\overline{Y}_{2t}^{\ast}:=\overline{C}_{2}X_{t}+\overline{C}_{2}^{\ast
}\overline{X}_{t}^{\ast}+u_{2t},
\]
where $\overline{X}_{t}^{\ast}:=\left(  \overline{x}_{t-1},...,\overline
{x}_{t-p}\right)  ^{\prime},$ and $\overline{x}_{t}\allowbreak:=\allowbreak
\min\left(  \overline{Y}_{2t}^{\ast}\allowbreak-b,0\right)  ,$ with initial
conditions $\overline{x}_{-s}\allowbreak=\kappa^{-1}\allowbreak\min\left(
Y_{2,-s}^{\ast}\allowbreak-b,0\right)  ,$ $s=0,...,\allowbreak p-1.$ It
follows that $\min\left(  \widetilde{X}_{t}^{\ast}-b,0\right)  \allowbreak
=\overline{X}_{t}^{\ast}$ for all $t\geq1,$ so that $\widetilde{Y}_{2t}^{\ast
}=\overline{Y}_{2t}^{\ast}.$ Substituting $\overline{Y}_{2}^{\ast}$ for
$\widetilde{Y}_{2}^{\ast}$ in (\ref{eq: RF Y2* pp}), we get (\ref{eq: RF Y2*}%
). Using the reparametrization (\ref{eq: Cbar}) and the relationship between
$X_{t}^{\ast}$ and $\overline{X}_{t}^{\ast}$ in (\ref{eq: Y1 1}), we obtain
(\ref{eq: RF Y1}).

Finally, from eq.~(\ref{eq: RF Y2* p}), it follows that the event
$Y_{2t}^{\ast}<b$ is equivalent to
\[
b+\kappa\left(  C_{2}X_{t}+C_{2}^{\ast}X_{t}^{\ast}+u_{2t}-b\right)  <b,
\]
which, since $\kappa>0$ by the coherency condition (\ref{eq: coherency}), is
equivalent to%
\begin{equation}
u_{2t}<b-C_{2}X_{t}-C_{2}^{\ast}X_{t}^{\ast}. \label{eq: u2 inequality}%
\end{equation}
Using the definition (\ref{eq: RF Y2*bar}), and (\ref{eq: Cbar}), the
inequality (\ref{eq: u2 inequality}) can be written as $\overline{Y}%
_{2t}^{\ast}<b,$ which establishes (\ref{eq: RF Y2}).

\textbf{Comment:} Note that $\kappa$ appears in the reduced form only
multiplicatively with $C^{\ast},$ so $\kappa$ and $C^{\ast}$ are not
separately identified, only $\overline{C}^{\ast}=\kappa C^{\ast}$ is. The
reparametrization from $C$ to $\overline{C}$ is convenient because
$\overline{C}$ is identified independently of $\kappa,$ while $C,C^{\ast}$ and
$\kappa$ are not separately identified.

\subsection{Proof of Proposition \ref{prop: point ID}}

We solve $u_{t}=\overline{A}^{-1}\varepsilon_{t}$ using the partitioned
inverse formula to get%
\begin{align}
u_{1t}  &  =\left(  A_{11}-\overline{A}_{12}\overline{A}_{22}^{-1}%
A_{21}\right)  ^{-1}\left(  \varepsilon_{1t}-\overline{A}_{12}\overline
{A}_{22}^{-1}\varepsilon_{2t}\right) \label{eq: u1 0}\\
u_{2t}  &  =\left(  \overline{A}_{22}-A_{21}A_{11}^{-1}\overline{A}%
_{12}\right)  ^{-1}\left(  \varepsilon_{2t}-A_{21}A_{11}^{-1}\varepsilon
_{1t}\right)  . \label{eq: u2 0}%
\end{align}
Using the definitions
\begin{align*}
\overline{\beta}  &  :=-A_{11}^{-1}\overline{A}_{12},\quad\overline{\gamma
}:=-\overline{A}_{22}^{-1}A_{21},\\
\bar{\varepsilon}_{1t}  &  :=A_{11}^{-1}\varepsilon_{1t},\quad\bar
{\varepsilon}_{2t}:=\overline{A}_{22}^{-1}\varepsilon_{2t},
\end{align*}
we can rewrite (\ref{eq: u1 0})-(\ref{eq: u2 0}) as (\ref{eq: u1}%
)-(\ref{eq: u2}).

Note that
\begin{align*}
\bar{\varepsilon}_{1t}  &  =A_{11}^{-1}\left(  A_{11}u_{1t}+\overline{A}%
_{12}u_{2t}\right)  =u_{1t}-\bar{\beta}u_{2t},\\
\bar{\varepsilon}_{2t}  &  =\overline{A}_{22}^{-1}\left(  A_{21}%
u_{1t}+\overline{A}_{22}u_{2t}\right)  =-\overline{\gamma}u_{1t}+u_{2t},
\end{align*}
so,%
\begin{align*}
var\left(  \bar{\varepsilon}_{1t}\right)   &  =\left(  I_{k-1},-\bar{\beta
}\right)  \Omega\left(  I_{k-1},-\bar{\beta}\right)  ^{\prime},\\
var\left(  \bar{\varepsilon}_{2t}\right)   &  =\left(  -\bar{\gamma},1\right)
\Omega\left(  -\bar{\gamma},1\right)  ^{\prime},
\end{align*}
and
\begin{align*}
cov\left(  \bar{\varepsilon}_{1t},\bar{\varepsilon}_{2t}\right)   &  =\left(
I_{k-1},-\overline{\beta}\right)
\begin{pmatrix}
\Omega_{11} & \Omega_{12}\\
\Omega_{12}^{\prime} & \Omega_{22}%
\end{pmatrix}
\left(  -\overline{\gamma},1\right)  ^{\prime}\\
&  =-\left(  \Omega_{11}-\overline{\beta}\Omega_{12}^{\prime}\right)
\overline{\gamma}^{\prime}+\Omega_{12}-\overline{\beta}\Omega_{22}=0.
\end{align*}
The last equation identifies $\overline{\gamma}$ given $\overline{\beta}$.
Specifically,%
\[
\overline{\gamma}^{\prime}=\left(  \Omega_{11}-\overline{\beta}\Omega
_{12}^{\prime}\right)  ^{-1}\left(  \Omega_{12}-\overline{\beta}\Omega
_{22}\right)  .
\]

\section{Numerical results\label{s: numerical}}

This section provides Monte-Carlo evidence on the finite-sample properties of
the proposed estimators and tests. The data generating process (DGP) is a
trivariate VAR(1), given by equations (\ref{eq: Y1 SVAR}) and
(\ref{eq: Y2 SVAR}). I\ consider three different DGPs corresponding to the
CKSVAR, KSVAR and CSVAR models, respectively. In all three DGPs, the following
parameters are set to the same values: the contemporaneous coefficients are
$A_{11}=I_{2},$ $A_{12}=$ $A_{12}^{\ast}=0_{2\times1},$ $A_{22}^{\ast}=1$ and
$A_{22}=0;$ the intercepts are set to zero, $B_{10}=0_{2\times1}$ and
$B_{20}=0;$ the coefficients on the lags are $B_{1,1}=\left(  \rho
I_{2},0\right)  ,$ $B_{1,1}^{\ast}=0_{2\times1},$ $B_{2,1}=\left(
0_{1\times2},B_{22,1}\right)  $, with $\rho=0.5.$ Finally, each of the three
DGPs is determined as follows. DGP1:\ $B_{22,1}=B_{2,1}^{\ast}=0$ (both KSVAR
and CSVAR restrictions hold, since lags of $Y_{2,t}$ and $Y_{2,t}^{\ast}$ all
have zero coefficients); DGP2: $B_{22,1}=\rho,\,B_{2,1}^{\ast}=0\ $(KSVAR
restrictions hold but CSVAR restrictions do not); DGP3: $B_{22,1}%
=0,\,B_{2,1}^{\ast}=\rho$ (CSVAR restrictions hold but KSVAR restrictions do
not). The setting of the autoregressive coefficient $\rho=0.5$ leads to a
lower degree of persistence than is typically observed in macro data (e.g., in
the Stock and Watson, 2001, application, the three largest roots are 0.97,
0.97 and 0.8), because I want to avoid confounding any possible finite-sample
issues arising from the ZLB with well-known problems of bias and size
distortion due to strong persistence (near unit roots) in the data. Finally,
the bound on $Y_{2t}$ is set to $b=0,$ the sample size is $T=250,$ the initial
conditions are set to 0 and the number of Monte Carlo replications is 1000. In
all cases, the CKSVAR and CSVAR likelihoods are computed using SIS with
$R=1000$ particles. The notation for the reported parameters is given in Table
\ref{t: notation}.%

\begin{table}[tbp] \centering
\caption{Parameter notation in reported simulation results}\label{t: notation}%
\begin{tabular}
[c]{l|l}\hline
\textbf{Mnemonic} & \textbf{Description}\\\hline
$\tau$ & st. dev. of reduced form error $u_{2t}$ in $Y_{2t}$ (constrained
variable)\\
Eq.~3 & reduced form equation for $Y_{2t}$\\
Eq.~j & red. form equation for $Y_{1j,t},$ $j=1,2$ (unconstrained variables)\\
$\widetilde{\beta}_{j}$ & coefficient on kink in eq.~j\\
eq.~i Y1j\_1 & coefficient of $Y_{1j,t-1}$ in eq.~i\\
eq.~i Y2\_1 & coefficient of $Y_{2,t-1}$ in eq.~i\\
eq.~i lY2\_1 & coefficient of $\min\left(  Y_{2,t-1}^{\ast}-b,0\right)  $ in
eq.~i\\
$\delta_{j}$ & coefficient of regression of $u_{1j,t}$ (red. form error in Eq.
j) on $u_{2t}$\\
Ch\_ij & (i,j) element of Choleski factor of $\Omega_{1.2}$\\\hline
\end{tabular}
\end{table}%

\begin{figure}[htb]%
\centering
\includegraphics[
height=0.6\textwidth,
width=\textwidth
]%
{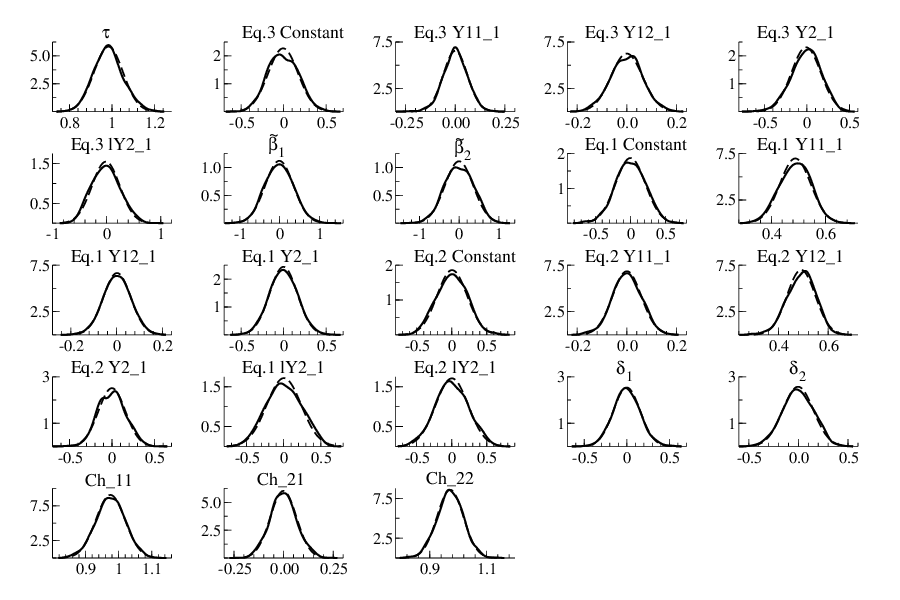}%
\caption{Sampling densities of ML estimators of reduced-form coefficients of CKSVAR(1) under
DGP1 (solid lines) and approximating Normal densities (dashed lines). $T=250$, 1000 Monte Carlo replications. Parameter names described in
Table \ref{t: notation}.}%
\label{fig: ML cksvar}%
\end{figure}

Figure \ref{fig: ML cksvar} reports the sampling distribution of the ML estimators of the reduced-form
parameters in Proposition \ref{prop: RF} for the CKSVAR model under DGP1. The results for the KSVAR and CSVAR
models, which are also correctly specified under DGP1, are omitted because they are entirely analogous.
The sampling densities appear to be very close to the superimposed Normal
approximations, indicating that the Normal asymptotic approximation is fairly accurate.%

Table \ref{t: ML cksvar} reports
moments of the sampling distributions of the above mentioned estimators for the CKSVAR model. Again, the results for the KSVAR and CSVAR models are entirely analogous and are therefore omitted. We
notice no discernible biases. Additional simulation results with $T=100$ and $T=1000$ given in the Tables \ref{t: CKSVAR DGP1}, \ref{t: KSVAR DGP1} and \ref{t: CSVAR DGP1}
indicate that the RMSE declines at rate $\sqrt{T}$ in accordance with
asymptotic theory. It is noteworthy that the
estimators of $\widetilde{\beta}$ have substantially
larger RMSE than the estimators of the other parameters. 

\begin{table}[tbp] \centering
\caption{Moments of sampling distribution of ML estimators of the parameters of CKSVAR(1)\tabnoteref[a]{tab2}}\label{t: ML cksvar}%
\centering%
\begin{tabular}
[c]{r|r|r|r|r|r}\hline
ML-CKSVAR & true & mean & bias & sd & RMSE\\\hline
$\tau$ & 1.000 & 0.983 & -0.017 & 0.068 & 0.070\\
Eq.3 Constant & 0.000 & -0.006 & -0.006 & 0.175 & 0.176\\
Eq.3 Y11\_1 & 0.000 & 0.000 & 0.000 & 0.061 & 0.061\\
Eq.3 Y12\_1 & 0.000 & -0.000 & -0.000 & 0.064 & 0.064\\
Eq.3 Y2\_1 & 0.000 & -0.010 & -0.010 & 0.173 & 0.173\\
Eq.3 lY2\_1 & 0.000 & -0.013 & -0.013 & 0.258 & 0.258\\
$\tilde{\beta}_{1}$ & -0.000 & -0.008 & -0.008 & 0.356 & 0.356\\
$\tilde{\beta}_{2}$ & 0.000 & -0.004 & -0.004 & 0.359 & 0.359\\
Eq.1 Constant & 0.000 & 0.002 & 0.002 & 0.213 & 0.213\\
Eq.1 Y11\_1 & 0.500 & 0.489 & -0.011 & 0.057 & 0.058\\
Eq.1 Y12\_1 & 0.000 & 0.002 & 0.002 & 0.060 & 0.060\\
Eq.1 Y2\_1 & 0.000 & -0.002 & -0.002 & 0.163 & 0.163\\
Eq.2 Constant & 0.000 & 0.005 & 0.005 & 0.214 & 0.214\\
Eq.2 Y11\_1 & 0.000 & 0.000 & 0.000 & 0.058 & 0.058\\
Eq.2 Y12\_1 & 0.500 & 0.491 & -0.009 & 0.057 & 0.058\\
Eq.2 Y2\_1 & 0.000 & -0.001 & -0.001 & 0.159 & 0.159\\
Eq.1 lY2\_1 & 0.000 & 0.005 & 0.005 & 0.232 & 0.232\\
Eq.2 lY2\_1 & 0.000 & 0.002 & 0.002 & 0.233 & 0.233\\
$\delta_{1}$ & 0.000 & -0.001 & -0.001 & 0.157 & 0.157\\
$\delta_{2}$ & 0.000 & -0.004 & -0.004 & 0.155 & 0.155\\
Ch\_11 & 1.000 & 0.975 & -0.025 & 0.044 & 0.051\\
Ch\_21 & 0.000 & -0.001 & -0.001 & 0.066 & 0.066\\
Ch\_22 & 1.000 & 0.972 & -0.028 & 0.046 & 0.054\\\hline
\end{tabular}
\tabnotetext[a]{tab2}{Computed under DGP1 with $T=250$ using 1000 MC replications. Parameter names described in Table \ref{t: notation}.}
\end{table}%

\begin{table}[htb]
\centering%
\caption{Bias, standard deviation and Root Mean Square Error of Maximum Likelihood estimator of parameters of CKSVAR(1) model\tabnoteref[a]{tab4}}%
\label{t: CKSVAR DGP1}%
\begin{tabular}
[c]{r|r|r|r|r|rr|r|r|r}\hline
ML-CKSVAR & \multicolumn{3}{|c|}{$T=100$} & \multicolumn{3}{|c|}{$T=250$} &
\multicolumn{3}{c}{$T=1000$}\\\hline
Parameter & bias & sd & RMSE & bias & sd & RMSE & bias & sd & RMSE\\\hline
$\tau$ & \multicolumn{1}{|r|}{-0.048} & 0.111 & 0.121 & -0.017 & 0.068 &
\multicolumn{1}{|r|}{0.070} & -0.003 & 0.035 & 0.035\\
Eq.3 Constant & \multicolumn{1}{|r|}{0.001} & 0.293 & 0.293 & -0.006 & 0.175 &
\multicolumn{1}{|r|}{0.176} & 0.004 & 0.085 & 0.085\\
Eq.3 Y11\_1 & \multicolumn{1}{|r|}{-0.001} & 0.111 & 0.111 & 0.000 & 0.061 &
\multicolumn{1}{|r|}{0.061} & 0.000 & 0.032 & 0.032\\
Eq.3 Y12\_1 & \multicolumn{1}{|r|}{-0.004} & 0.109 & 0.109 & -0.000 & 0.064 &
\multicolumn{1}{|r|}{0.064} & -0.000 & 0.031 & 0.031\\
Eq.3 Y2\_1 & \multicolumn{1}{|r|}{-0.031} & 0.289 & 0.291 & -0.010 & 0.173 &
\multicolumn{1}{|r|}{0.173} & -0.003 & 0.082 & 0.082\\
Eq.3 lY2\_1 & \multicolumn{1}{|r|}{-0.022} & 0.435 & 0.436 & -0.013 & 0.258 &
\multicolumn{1}{|r|}{0.258} & 0.003 & 0.122 & 0.122\\
$\tilde{\beta}_{1}$ & \multicolumn{1}{|r|}{0.012} & 0.586 & 0.586 & -0.008 &
0.356 & \multicolumn{1}{|r|}{0.356} & 0.000 & 0.176 & 0.176\\
$\tilde{\beta}_{2}$ & \multicolumn{1}{|r|}{-0.011} & 0.606 & 0.606 & -0.004 &
0.359 & \multicolumn{1}{|r|}{0.359} & -0.003 & 0.168 & 0.168\\
Eq.1 Constant & \multicolumn{1}{|r|}{0.000} & 0.360 & 0.360 & 0.002 & 0.213 &
\multicolumn{1}{|r|}{0.213} & 0.007 & 0.105 & 0.105\\
Eq.1 Y11\_1 & \multicolumn{1}{|r|}{-0.034} & 0.098 & 0.104 & -0.011 & 0.057 &
\multicolumn{1}{|r|}{0.058} & -0.002 & 0.029 & 0.029\\
Eq.1 Y12\_1 & \multicolumn{1}{|r|}{-0.004} & 0.108 & 0.108 & 0.002 & 0.060 &
\multicolumn{1}{|r|}{0.060} & 0.001 & 0.028 & 0.028\\
Eq.1 Y2\_1 & \multicolumn{1}{|r|}{-0.004} & 0.281 & 0.281 & -0.002 & 0.163 &
\multicolumn{1}{|r|}{0.163} & -0.006 & 0.079 & 0.079\\
Eq.2 Constant & \multicolumn{1}{|r|}{0.006} & 0.356 & 0.356 & 0.005 & 0.214 &
\multicolumn{1}{|r|}{0.214} & 0.002 & 0.102 & 0.103\\
Eq.2 Y11\_1 & \multicolumn{1}{|r|}{0.004} & 0.103 & 0.103 & 0.000 & 0.058 &
\multicolumn{1}{|r|}{0.058} & 0.000 & 0.028 & 0.028\\
Eq.2 Y12\_1 & \multicolumn{1}{|r|}{-0.029} & 0.103 & 0.107 & -0.009 & 0.057 &
\multicolumn{1}{|r|}{0.058} & -0.002 & 0.029 & 0.029\\
Eq.2 Y2\_1 & \multicolumn{1}{|r|}{0.000} & 0.269 & 0.269 & -0.001 & 0.159 &
\multicolumn{1}{|r|}{0.159} & 0.001 & 0.078 & 0.078\\
Eq.1 lY2\_1 & \multicolumn{1}{|r|}{0.009} & 0.403 & 0.403 & 0.005 & 0.232 &
\multicolumn{1}{|r|}{0.232} & 0.010 & 0.112 & 0.113\\
Eq.2 lY2\_1 & \multicolumn{1}{|r|}{-0.003} & 0.408 & 0.408 & 0.002 & 0.233 &
\multicolumn{1}{|r|}{0.233} & 0.004 & 0.115 & 0.115\\
$\delta_{1}$ & \multicolumn{1}{|r|}{0.005} & 0.260 & 0.260 & -0.001 & 0.157 &
\multicolumn{1}{|r|}{0.157} & 0.001 & 0.075 & 0.075\\
$\delta_{2}$ & \multicolumn{1}{|r|}{-0.005} & 0.260 & 0.260 & -0.004 & 0.155 &
\multicolumn{1}{|r|}{0.155} & -0.001 & 0.073 & 0.073\\
Ch\_11 & \multicolumn{1}{|r|}{-0.067} & 0.073 & 0.099 & -0.025 & 0.044 &
\multicolumn{1}{|r|}{0.051} & -0.006 & 0.024 & 0.024\\
Ch\_21 & \multicolumn{1}{|r|}{-0.005} & 0.111 & 0.111 & -0.001 & 0.066 &
\multicolumn{1}{|r|}{0.066} & -0.000 & 0.031 & 0.031\\
Ch\_22 & \multicolumn{1}{|r|}{-0.075} & 0.073 & 0.105 & -0.028 & 0.046 &
\multicolumn{1}{|r|}{0.054} & -0.008 & 0.023 & 0.024\\\hline
\end{tabular}
\tabnotetext[a]{tab4}{Computed under DGP1 with $R=1000$ particles using 1000 MC replications. Parameter names described in Table \ref{t: notation}.}%
\end{table}

\begin{table}[htb]
\centering%
\caption{Bias, standard deviation and Root Mean Square Error of Maximum Likelihood estimator of parameters of KSVAR(1) model\tabnoteref[a]{tab5}}%
\label{t: KSVAR DGP1}%
\begin{tabular}
[c]{r|r|r|r|r|r|r|r|r|r}\hline
ML-KSVAR & \multicolumn{3}{|c|}{$T=100$} & \multicolumn{3}{c|}{$T=250$} &
\multicolumn{3}{c}{$T=1000$}\\\hline
Parameter & bias & sd & RMSE & bias & sd & RMSE & bias & sd & RMSE\\\hline
$\tau$ & -0.024 & 0.111 & 0.113 & -0.008 & 0.068 & 0.069 & -0.001 & 0.035 &
0.035\\
Eq.3 Constant & 0.011 & 0.145 & 0.145 & 0.001 & 0.092 & 0.092 & 0.003 &
0.046 & 0.046\\
Eq.3 Y11\_1 & -0.001 & 0.103 & 0.103 & 0.001 & 0.060 & 0.060 & -0.000 &
0.031 & 0.031\\
Eq.3 Y12\_1 & -0.004 & 0.102 & 0.102 & -0.000 & 0.062 & 0.062 & -0.000 &
0.030 & 0.030\\
Eq.3 Y2\_1 & -0.048 & 0.199 & 0.204 & -0.019 & 0.122 & 0.124 & -0.003 &
0.060 & 0.060\\
$\tilde{\beta}_{1}$ & -0.003 & 0.571 & 0.571 & -0.013 & 0.349 & 0.349 &
-0.001 & 0.174 & 0.174\\
$\tilde{\beta}_{2}$ & -0.003 & 0.584 & 0.584 & -0.001 & 0.348 & 0.348 &
-0.004 & 0.168 & 0.168\\
Eq.1 Constant & 0.002 & 0.264 & 0.264 & 0.001 & 0.165 & 0.165 & 0.001 &
0.080 & 0.080\\
Eq.1 Y11\_1 & -0.033 & 0.093 & 0.099 & -0.012 & 0.056 & 0.057 & -0.002 &
0.028 & 0.028\\
Eq.1 Y12\_1 & -0.002 & 0.100 & 0.100 & 0.002 & 0.058 & 0.058 & 0.001 & 0.027 &
0.027\\
Eq.1 Y2\_1 & -0.002 & 0.197 & 0.197 & -0.000 & 0.117 & 0.117 & -0.001 &
0.057 & 0.057\\
Eq.2 Constant & 0.004 & 0.258 & 0.258 & 0.003 & 0.158 & 0.158 & 0.001 &
0.078 & 0.078\\
Eq.2 Y11\_1 & 0.006 & 0.096 & 0.096 & 0.001 & 0.057 & 0.057 & 0.000 & 0.027 &
0.027\\
Eq.2 Y12\_1 & -0.028 & 0.094 & 0.098 & -0.008 & 0.055 & 0.056 & -0.002 &
0.028 & 0.029\\
Eq.2 Y2\_1 & -0.001 & 0.189 & 0.189 & -0.000 & 0.113 & 0.113 & 0.003 & 0.054 &
0.055\\
$\delta_{1}$ & -0.000 & 0.252 & 0.252 & -0.003 & 0.156 & 0.156 & 0.001 &
0.075 & 0.075\\
$\delta_{2}$ & -0.003 & 0.253 & 0.253 & -0.003 & 0.152 & 0.152 & -0.001 &
0.073 & 0.073\\
Ch\_11 & -0.048 & 0.070 & 0.085 & -0.018 & 0.044 & 0.047 & -0.005 & 0.023 &
0.024\\
Ch\_21 & -0.003 & 0.108 & 0.108 & -0.000 & 0.065 & 0.065 & -0.000 & 0.031 &
0.031\\
Ch\_22 & -0.054 & 0.070 & 0.088 & -0.020 & 0.045 & 0.050 & -0.007 & 0.023 &
0.024\\\hline
\end{tabular}
\tabnotetext[a]{tab5}{Computed under DGP1 with $R=1000$ particles using 1000 MC replications. Parameter names described in Table \ref{t: notation}.}
\end{table}

\begin{table}[ptb]
\caption{Bias, standard deviation and Root Mean Square Error of Maximum Likelihood estimator of parameters of CSVAR(1) model\tabnoteref[a]{tab6}}
\label{t: CSVAR DGP1}%
\centering%
\begin{tabular}
[c]{r|r|r|r|r|r|r|r|r|r}\hline
ML-CSVAR & \multicolumn{3}{|c}{$T=100$} & \multicolumn{3}{c|}{$T=250$} &
\multicolumn{3}{c}{$T=1000$}\\\hline
Parameter & bias & sd & RMSE & bias & sd & RMSE & bias & sd & RMSE\\\hline
$\tau$ & -0.026 & 0.111 & 0.114 & -0.008 & 0.068 & 0.069 & -0.001 & 0.035 &
0.035\\
Eq.3 Constant & -0.009 & 0.135 & 0.135 & -0.006 & 0.081 & 0.081 & 0.001 &
0.040 & 0.040\\
Eq.3 Y11\_1 & -0.001 & 0.104 & 0.104 & 0.001 & 0.060 & 0.060 & -0.000 &
0.031 & 0.031\\
Eq.3 Y12\_1 & -0.004 & 0.103 & 0.103 & -0.000 & 0.061 & 0.061 & -0.000 &
0.030 & 0.030\\
Eq.3 Y2\_1 & -0.027 & 0.125 & 0.128 & -0.011 & 0.078 & 0.079 & -0.001 &
0.038 & 0.038\\
Eq.1 Constant & -0.002 & 0.109 & 0.110 & -0.003 & 0.065 & 0.065 & 0.000 &
0.031 & 0.031\\
Eq.1 Y11\_1 & -0.033 & 0.090 & 0.096 & -0.012 & 0.054 & 0.056 & -0.002 &
0.028 & 0.028\\
Eq.1 Y12\_1 & -0.002 & 0.096 & 0.096 & 0.002 & 0.057 & 0.057 & 0.001 & 0.027 &
0.027\\
Eq.1 Y2\_1 & -0.000 & 0.118 & 0.118 & 0.001 & 0.072 & 0.072 & 0.000 & 0.036 &
0.036\\
Eq.2 Constant & 0.003 & 0.106 & 0.106 & 0.002 & 0.063 & 0.063 & -0.000 &
0.031 & 0.031\\
Eq.2 Y11\_1 & 0.005 & 0.091 & 0.092 & 0.001 & 0.056 & 0.056 & 0.000 & 0.027 &
0.027\\
Eq.2 Y12\_1 & -0.026 & 0.090 & 0.094 & -0.008 & 0.054 & 0.055 & -0.002 &
0.028 & 0.028\\
Eq.2 Y2\_1 & -0.001 & 0.115 & 0.115 & 0.000 & 0.070 & 0.070 & 0.002 & 0.035 &
0.035\\
$\delta_{1}$ & 0.001 & 0.114 & 0.114 & 0.002 & 0.071 & 0.071 & 0.001 & 0.035 &
0.035\\
$\delta_{2}$ & -0.001 & 0.116 & 0.116 & -0.002 & 0.070 & 0.070 & -0.000 &
0.034 & 0.034\\
Ch\_11 & -0.033 & 0.069 & 0.077 & -0.012 & 0.043 & 0.044 & -0.003 & 0.023 &
0.023\\
Ch\_21 & -0.005 & 0.103 & 0.104 & -0.001 & 0.064 & 0.064 & -0.000 & 0.031 &
0.031\\
Ch\_22 & -0.037 & 0.068 & 0.077 & -0.014 & 0.044 & 0.046 & -0.005 & 0.023 &
0.024\\\hline
\end{tabular}
\tabnotetext[a]{tab6}{Computed under DGP1 with $R=1000$ particles using 1000 MC replications. Parameter names described in Table \ref{t: notation}.}
\end{table}

Next, I turn to the properties of the LR test of KSVAR against CKSVAR and
CSVAR against CKSVAR. The former hypothesis involves three restrictions
(exclusion of the latent lag $Y_{2,t-1}^{\ast}$ from each of the three
equations), so the LR statistic is asymptotically distributed as $\chi_{3}%
^{2}$ under the null. The latter hypothesis involves five restrictions
(exclusion of the observed lag $Y_{2,t-1}$ from each of the three equations,
plus $\widetilde{\beta}=0$), and the LR statistic is asymptotically
distributed as $\chi_{5}^{2}.$ Table \ref{t: LR sim} reports the rejection
frequencies of the LR tests for each of the two hypotheses in each of the
three DGPs at three significance levels: 10\%, 5\% and 1\%. In addition to the
asymptotic tests, I\ also report the rejection frequency of the tests using
parametric bootstrap critical values. The parametric bootstrap uses draws of
Normal errors and the estimated reduced-form parameters to generate the
bootstrap samples of $Y_{t}$ and $Y_{2t}^{\ast}$. The Monte Carlo rejection
frequencies are computed using the ``warp-speed'' method method of
\cite{GiacominiPolitisWhite2013}. Note that both null hypotheses hold under
DGP1, but only the KSVAR is valid under DGP2 and only the CSVAR is valid under
DGP3. For convenience, I indicate the rejection frequencies under the
alternative in bold in the table.%

\begin{table}[htbp] \centering
\caption{Rejection frequencies of LR tests of $H_0$ against $H_1$ across different DGPs\tabnoteref[a]{tab3}\label{t: LR sim}}
\begin{tabular}
[c]{ll|c|c|c|c|c|c}\hline
&  & \multicolumn{3}{|c|}{$H_{0}:$ KSVAR, $H_{1}:$ CKSVAR} &
\multicolumn{3}{|l}{$H_{0}:$ CSVAR, $H_{1}:$ CKSVAR}\\\hline
& Sign. Level & 10\% & 5\% & 1\% & 10\% & 5\% & 1\%\\\hline
DGP1 & asymptotic & 0.173 & 0.093 & 0.020 & 0.155 & 0.084 & 0.022\\
& bootstrap & 0.107 & 0.045 & 0.005 & 0.109 & 0.052 & 0.012\\\hline
DGP2 & asymptotic & 0.149 & 0.080 & 0.018 & \textbf{0.319} & \textbf{0.206} &
\textbf{0.073}\\
& bootstrap & 0.117 & 0.050 & 0.011 & \textbf{0.242} & \textbf{0.141} &
\textbf{0.041}\\\hline
DGP3 & asymptotic & \textbf{0.587} & \textbf{0.454} & \textbf{0.244} & 0.141 &
0.067 & 0.019\\
& bootstrap & \textbf{0.471} & \textbf{0.365} & \textbf{0.194} & 0.103 &
0.053 & 0.018\\\hline
\end{tabular}
\tabnotetext[a]{tab3}{Computed using 1000 Monte Carlo replications, $T=250$. The asymptotic tests use $\chi^2_3$ and $\chi^2_5$ critical values for KSVAR and CSVAR resp. The bootstrap rej. frequencies were computed uisng the warp-speed method of \cite{GiacominiPolitisWhite2013}. Bold numbers indicate that the rejecction frequencies were computed under $H_1$ (power).}%
\end{table}%

\begin{figure}[ptb]%
\centering
\includegraphics[
height=4in,
width=5in
]%
{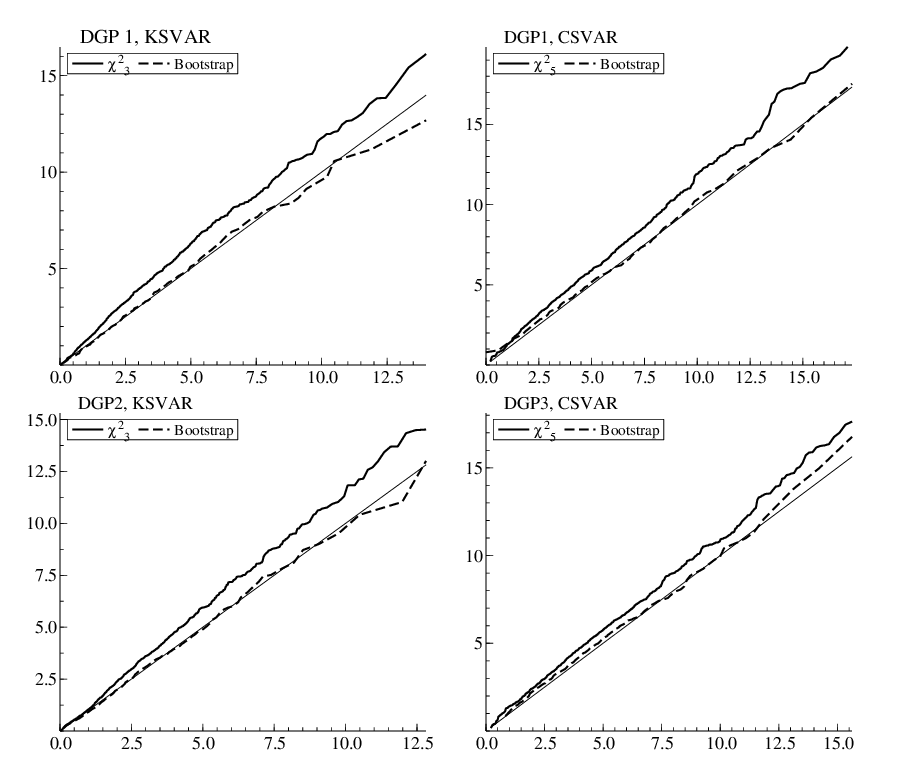}%
\caption{QQ plots of the sampling distribution under the null hypothesis of LR statistics of KSVAR against
CKSVAR (left) and CSVAR against CKSVAR (right). Solid (dashed) lines plot quantiles against asymptotic $\chi^2$ (bootstrap) approximation. Computed for $T=250$ using 1000 Monte Carlo replications.}%
\label{fig: QQ plots}%
\end{figure}

There is evidence that the LR tests reject too often under $H_{0}$ relative to
their nominal level when we use asymptotic critical values. Moreover, the size
distortions are very similar across null hypotheses and DGPs. Unreported
results show that size distortion eventually disappears as the sample gets
large, but this level of overrejection is unsatisfactory at $T=250,$ which is
a typical sample size in macroeconomic applications. The parametric bootstrap
appears to do a remarkably good job at correcting the size of the tests. In
all cases considered, the parametric bootstrap rejection frequency is not
significantly different from the nominal level when the null hypothesis holds
(all but the numbers in bold in the Table). To shed further light on this
issue, Figure \ref{fig: QQ plots} reports the QQ plots of the
sampling distributions of the two LR statistics against their asymptotic and parametric
bootstrap approximations for all three DGPs under the null hypothesis. The sampling distributions of the LR statistics stochastically dominate their asymptotic approximations, but the bootstrap
approximations are quite accurate.

Finally, the rejection frequencies highlighted in bold in Table
\ref{t: LR sim} correspond to the power of the tests against two very similar
deviations from the null hypothesis. The numbers on the left under DGP3 show
the power of the test to reject the KSVAR specification under the alternative
at which the coefficient on the latent lag $B_{2,1}^{\ast}=0.5.$ Similarly,
the bold numbers on the right give the power of rejecting CSVAR against the
alternative where the coefficient on the observed lag $B_{22,1}=0.5.$ Since
the lower bound is set to zero, and the sample contains about 50\% of
observations at the ZLB, the two deviations from the null are of equal
magnitude. Yet, we notice the LR\ test is significantly more powerful against
the KSVAR than against the CSVAR. This could be because CSVAR\ imposes more
restrictions than the KSVAR, so one would expect it to have lower power than
the KSVAR against similar deviations from the null.%

\subsection*{Alternative DGP}

\begin{figure}[htb]%
\centering
\includegraphics[
height=0.6\textwidth,
width=\textwidth
]%
{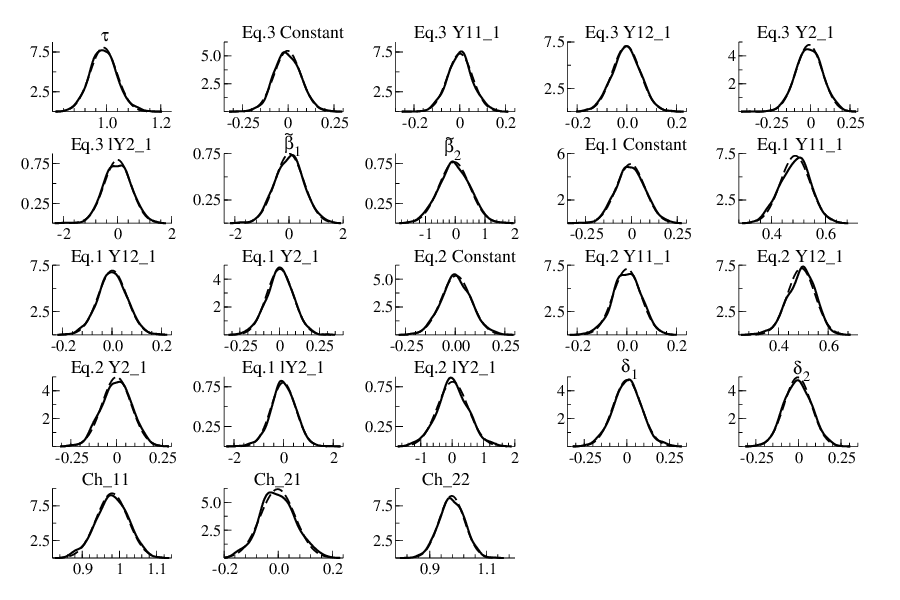}%
\caption{Sampling densities of ML estimators of reduced-form coefficients of CKSVAR(1) under
DGP1 with $b$ chosen such that $\Pr\left(  Y_{2t}=b\right)  =0.11$ (solid lines) and approximating Normal densities (dashed lines). $T=250$, 1000 Monte Carlo replications. Parameter names described in
Table \ref{t: notation}.}%
\label{fig: ML cksvar'}%
\end{figure}

The DGPs in the previous simulations have the property that the frequency of
the ZLB regime is around 50\%. I reran those simulations with a slight
modification to the DGPs to match the frequency in the sample of the empirical
application in the paper. Specifically, I reduce the lower bound $b$ to a
level that makes the frequency of the ZLB regime equal to 11\%. The results
are given in Figure \ref{fig: ML cksvar'} and Table
\ref{t: ML cksvar'}. The results are very similar to the
ones reported in Figure \ref{fig: ML cksvar} and
Table \ref{t: ML cksvar} above: the Normal approximation of the sampling distribution of the MLE
appears to be very good, and the bias is negligible. The only difference is that the standard deviation of $\widetilde{\beta}$ is larger.%

\begin{table}[ptb]
\centering
\caption{Moments of sampling distribution of ML estimators of the parameters of CKSVAR(1)\tabnoteref[a]{tab7}}\label{t: ML cksvar'}%
\begin{tabular}
[c]{r|r|r|r|r|r}\hline
ML-CKSVAR & true & mean & bias & sd & RMSE\\\hline
$\tau$ & 1.000 & 0.988 & -0.012 & 0.050 & 0.051\\
Eq.3 Constant & 0.000 & -0.004 & -0.004 & 0.073 & 0.073\\
Eq.3 Y11\_1 & 0.000 & 0.001 & 0.001 & 0.055 & 0.055\\
Eq.3 Y12\_1 & 0.000 & -0.002 & -0.002 & 0.056 & 0.056\\
Eq.3 Y2\_1 & 0.000 & -0.008 & -0.008 & 0.083 & 0.083\\
Eq.3 lY2\_1 & 0.000 & 0.003 & 0.003 & 0.501 & 0.501\\
$\tilde{\beta}_{1}$ & 0.000 & 0.013 & 0.013 & 0.533 & 0.533\\
$\tilde{\beta}_{2}$ & 0.000 & -0.030 & -0.030 & 0.518 & 0.519\\
Eq.1 Constant & 0.000 & -0.005 & -0.005 & 0.078 & 0.078\\
Eq.1 Y11\_1 & 0.500 & 0.488 & -0.012 & 0.055 & 0.056\\
Eq.1 Y12\_1 & 0.000 & 0.001 & 0.001 & 0.058 & 0.058\\
Eq.1 Y2\_1 & 0.000 & 0.001 & 0.001 & 0.084 & 0.084\\
Eq.2 Constant & 0.000 & 0.005 & 0.005 & 0.075 & 0.075\\
Eq.2 Y11\_1 & 0.000 & 0.001 & 0.001 & 0.056 & 0.056\\
Eq.2 Y12\_1 & 0.500 & 0.491 & -0.009 & 0.056 & 0.056\\
Eq.2 Y2\_1 & 0.000 & -0.000 & -0.000 & 0.080 & 0.080\\
Eq.1 lY2\_1 & 0.000 & -0.014 & -0.014 & 0.497 & 0.497\\
Eq.2 lY2\_1 & 0.000 & 0.018 & 0.018 & 0.491 & 0.491\\
$\delta_{1}$ & 0.000 & 0.002 & 0.002 & 0.084 & 0.084\\
$\delta_{2}$ & 0.000 & -0.004 & -0.004 & 0.080 & 0.080\\
Ch\_11 & 1.000 & 0.980 & -0.020 & 0.043 & 0.047\\
Ch\_21 & 0.000 & -0.002 & -0.002 & 0.065 & 0.065\\
Ch\_22 & 1.000 & 0.978 & -0.022 & 0.045 & 0.050\\\hline
\end{tabular}
\tabnotetext[a]{tab7}{Computed under DGP1 with $b$ chosen such that $\Pr\left(  Y_{2t}=b\right)
=0.11$, $T=250$ using 1000 MC replications. Parameter names described in Table \ref{t: notation}.}
\end{table}

\section{A simple model of QE\label{s: QE}}

This is a simplified version of the New Keynesian model of bond market
segmentation that appears in \cite{IkedaLiMavroeidisZanetti2020} and is based
on \cite{ChenCurdiaFerrero2012}. The economy consists of two types of
households. A fraction $\omega_{r}$ of type `r' households can only trade
long-term government bonds. The remaining $1-\omega_{r}$ households of type
`u' can purchase both short-term and long-term government bonds, the latter
subject to a trading cost $\zeta_{t}$. This trading cost gives rise to a term
premium, i.e., a spread between long-term and short-term yields, that the
central bank can manipulate by purchasing long-term bonds. The term premium
affects aggregate demand through the consumption decisions of constrained
households. This generates an unconventional monetary policy channel.

The transmission mechanism of monetary policy is obtained from the equilibrium
conditions of households and firms in the economy. Households choose
consumption to maximize an isoelastic utility function and firms set prices
subject to Calvo frictions. These give rise to an Euler equation for output
and a Phillips curve, respectively. Equation (\ref{eq: outcome}) in the paper
can be derived by combining those two equations. I will derive the Euler
equation in some detail in order to illustrate the origins of the QE channel.
The Phillips curve derivation is standard and is therefore omitted.

Up to a loglinear approximation, the relevant first-order conditions of the
households' optimization problem can be written as%
\begin{align}
0  &  =E_{t}\left[  -\frac{1}{\sigma}\left(  \hat{c}_{t+1}^{u}-\hat{c}_{t}%
^{u}\right)  +\hat{r}_{t}-\pi_{t+1}\right]  ,\label{l_HHu2}\\
\frac{\zeta}{1+\zeta}\hat{\zeta}_{t}  &  =E_{t}\left[  -\frac{1}{\sigma
}\left(  \hat{c}_{t+1}^{u}-\hat{c}_{t}^{u}\right)  +\hat{R}_{L,t+1}-\pi
_{t+1}\right]  ,\label{l_HHu3}\\
0  &  =E_{t}\left[  -\frac{1}{\sigma}\left(  \hat{c}_{t+1}^{r}-\hat{c}_{t}%
^{r}\right)  +\hat{R}_{L,t+1}-\pi_{t+1}\right]  , \label{l_HHr2}%
\end{align}
where $\sigma$ is the elasticity of intertemporal substitution, $\zeta$ is the
steady state value of $\zeta_{t},$ hatted variables denote log-deviations from
steady state, $c_{t}^{j}$ is consumption of household $j\in\left\{
u,r\right\}  ,$ $r_{t}$ is the short-term nominal interest rate, and $R_{L,t}$
is the gross yield on long-term government bonds from period $t-1$ to
$t$.\footnote{I do not put a hat over $\pi_{t}$ because I assume a zero
inflation target for simplicity.} Goods market clearing yields
\begin{equation}
\hat{y}_{t}=\omega_{r}\hat{c}_{t}^{r}+\left(  1-\omega_{r}\right)  \hat{c}%
_{t}^{u}, \label{eq: good mkt clearing}%
\end{equation}
where $y_{t}$ is output, and I\ have assumed, for simplicity, that in steady
state $c^{u}=c^{r}$, which implies $c^{u}=c^{r}=y.$ Multiplying (\ref{l_HHu2})
and (\ref{l_HHr2}) by $\left(  1-\omega_{r}\right)  $ and $\omega_{r},$
respectively, and adding them yields%
\begin{equation}
\hat{y}_{t}=E_{t}\hat{y}_{t+1}-\sigma E_{t}\left[  \left(  1-\omega
_{r}\right)  \hat{r}_{t}+\omega_{r}\hat{R}_{L,t+1}-\pi_{t+1}\right]  .
\label{eq: EE with RL}%
\end{equation}
Subtracting (\ref{l_HHu2}) from (\ref{l_HHu3}) yields%
\begin{equation}
E_{t}\left(  \hat{R}_{L,t+1}\right)  =\hat{r}_{t}+\frac{\zeta}{1+\zeta}%
\hat{\zeta}_{t}, \label{eq: yield spread}%
\end{equation}
which establishes that the term premium between long and short yields is
proportional to $\hat{\zeta}_{t}.$ Substituting for $E_{t}\left(  \hat
{R}_{L,t+1}\right)  $ in (\ref{eq: EE with RL}) using (\ref{eq: yield spread})
yields%
\begin{equation}
\hat{y}_{t}=E_{t}\hat{y}_{t+1}-\sigma\left(  \hat{r}_{t}+\omega_{r}\frac
{\zeta}{1+\zeta}\hat{\zeta}_{t}\right)  +\sigma E_{t}\left(  \pi_{t+1}\right)
\label{eq: EE with zeta}%
\end{equation}
Next, assume that the cost of trading long-term bonds depends on their supply,
$b_{L,t},$ i.e.,%
\[
\hat{\zeta}_{t}=\rho_{\zeta}\hat{b}_{L,t},\quad\rho_{\zeta}\geq0.
\]
Substituting for $\hat{\zeta}_{t}$ in (\ref{eq: EE with zeta}) yields the
Euler equation%
\begin{equation}
\hat{y}_{t}=E_{t}\hat{y}_{t+1}-\sigma\left(  \hat{r}_{t}+\omega_{r}\frac
{\zeta}{1+\zeta}\rho_{\zeta}\hat{b}_{L,t}\right)  +\sigma E_{t}\left(
\pi_{t+1}\right)  . \label{eq: EE}%
\end{equation}

The second equation is a standard New Keynesian Phillips curve that links
inflation to output:%
\begin{equation}
\pi_{t}=\delta E_{t}\left(  \pi_{t+1}\right)  +\varpi\hat{y}_{t}%
+\varepsilon_{1t} \label{eq: NKPC}%
\end{equation}
where $\delta$ is the average discount factor of the two households,
$\varpi\geq0$ is a parameter that depends on the degree of price stickiness
(the Calvo parameter), and $\varepsilon_{1t}$ is proportional to an
\textit{i.i.d.}~technology shock. Substituting for $\hat{y}_{t}$ in
(\ref{eq: NKPC}) using (\ref{eq: EE}) yields%
\begin{equation}
\pi_{t}=\left(  \delta+\sigma\varpi\right)  E_{t}\left(  \pi_{t+1}\right)
+\varpi E_{t}\left(  \hat{y}_{t+1}\right)  -\varpi\sigma\left(  \hat{r}%
_{t}+\omega_{r}\frac{\zeta}{1+\zeta}\rho_{\zeta}\hat{b}_{L,t}\right)
+\varepsilon_{1t}. \label{eq: aux}%
\end{equation}

Finally, at an equilibrium in which inflation and output depend only on the
exogenous shocks $\varepsilon_{t}=\left(  \varepsilon_{1t},\varepsilon
_{2t}\right)  ^{\prime},$ which are the only state variables in the system,
and when the shocks have no memory, $E_{t}\left(  \pi_{t+1}\right)  $ and
$E_{t}\left(  \hat{y}_{t+1}\right)  $ will be equal to the corresponding
unconditional expectations, which are constants.\footnote{Such an equilibrium
always exists if the volatility of the shocks is not too large, see
\cite{Mendes2011}.} Therefore, (\ref{eq: aux}) reduces to
eq.~(\ref{eq: outcome}) in the paper by setting $c=\left(  \delta+\sigma
\varpi\right)  E\left(  \pi_{t+1}\right)  +\varpi E\left(  \hat{y}%
_{t+1}\right)  ,$ $\beta=-\varpi\sigma$, $\hat{r}_{t}=r_{t}-r^{n},$ where
$r^{n}$ is the discount rate of the unconstrained households, and
$\varphi=\omega_{r}\frac{\zeta}{1+\zeta}\rho_{\zeta}\beta$, and dropping the hat from $b_{L,t}$ for simplicity. The parameter $\varphi$ depends on the fraction of constrained households, $\omega_{r}$, and the
sensitivity of the term premium to long-term asset holdings, $\frac{\zeta
}{1+\zeta}\rho_{\zeta}$. 

\section{Forward guidance rules}\label{s: app FG}

\cite{DebortoliGaliGambetti2019} discuss the following two inertial policy
rules
\begin{equation}
r_{t}=\max\left\{  0,\phi_{r}r_{t-1}+\left(  1-\phi_{r}\right)  \left(
\rho+\phi_{\pi} \pi_{t}+\phi_{y}\Delta y_{t}\right)  \right\}
\label{eq: conv taylor}%
\end{equation}
and
\begin{subequations}
\label{eq: fg taylor}%
\begin{align}
r_{t}  &  =\max\left(  0,r_{t}^{\ast}\right) \label{eq: fg taylor a}\\
r_{t}^{\ast}  &  =\phi_{r}r_{t-1}^{\ast}+\left(  1-\phi_{r}\right)  \left(
\rho+\phi_{\pi} \pi_{t}+\phi_{y}\Delta y_{t}\right)  , \label{eq: fg taylor b}%
\end{align}
where I have set the inflation target to zero, and $\Delta y_{t}$ is output
growth. Both of these rules are nested within equation (\ref{eq: Y2 SVAR}) of
the CKSVAR, with $Y_{1t}=\left(  \pi_{t},\Delta y_{t}\right)  ^{\prime}\,$and
$Y_{2t}=r_{t}.$ Rule (\ref{eq: conv taylor}) sets the coefficients on
$Y_{2,t-1}$ and $Y_{2,t-1}^{\ast}$ as $B_{22}=\phi_{r}$ and $B_{22}^{\ast}=0,$
respectively, while rule (\ref{eq: fg taylor}) sets them as $B_{22}=0$ and
$B_{22}^{\ast}=\phi_{r}.$ \cite{DebortoliGaliGambetti2019} argue rule
(\ref{eq: fg taylor}) is consistent with forward guidance, because it will
tend to keep interest rates at zero for longer than rule
(\ref{eq: conv taylor}). It also ensures policy reaction is the same across
regimes, and so it is consistent with the ZLB irrelevance hypothesis that the
paper puts forward.

\cite{ReifschneiderWilliams2000} propose a slightly more elaborate policy rule
for forward guidance:
\end{subequations}
\begin{subequations}
\label{eq: FG rule}%
\begin{align}
r_{t}^{\ast}  &  =r_{t}^{Taylor}-\alpha Z_{t},\quad Z_{t}=Z_{t-1}+d_{t},\quad
d_{t}:=r_{t}-r_{t}^{Taylor},\label{eq: FG rule a}\\
r_{t}  &  =\max\left(  r_{t}^{\ast},0\right)  ,\nonumber\\
r_{t}^{Taylor}  &  =\rho+\phi_{\pi}\pi_{t}+\phi_{y}y_{t} \label{eq: FG rule b}%
\end{align}
where $y_{t}$ is the output gap, and the inflation target is zero.
Differencing (\ref{eq: FG rule a}) yields%
\end{subequations}
\[
r_{t}^{\ast}=r_{t-1}^{\ast}+\Delta r_{t}^{Taylor}-\alpha\left(  r_{t}%
-r_{t}^{Taylor}\right)  .
\]
Substituting for $r_{t}^{Taylor}$ using (\ref{eq: FG rule b}) yields%
\begin{align*}
r_{t}^{\ast}  &  =r_{t-1}^{\ast}+\phi_{\pi}\Delta\pi_{t}+\phi_{y}\Delta
y_{t}-\alpha r_{t}+\alpha\left(  \rho+\phi_{\pi}\pi_{t}+\phi_{y}y_{t}\right)
\\
&  =\alpha\rho-\alpha r_{t}+\left(  1+\alpha\right)  \left(  \phi_{\pi}\pi
_{t}+\phi_{y}y_{t}\right)  -\left(  \phi_{\pi}\pi_{t-1}+\phi_{y}%
y_{t-1}\right)  +r_{t-1}^{\ast}.
\end{align*}
This is again nested within equation (\ref{eq: Y2 SVAR}) of the CKSVAR with
$Y_{1t}=\left(  \pi_{t},y_{t}\right)  ^{\prime},$ $Y_{2t}=r_{t},$
$Y_{2t}^{\ast}=r_{t}^{\ast},$ $X_{1t}=\left(  1,\pi_{t-1},y_{t-1}\right)
^{\prime},$ $X_{2t}=r_{t-1},$ $X_{2t}^{\ast}=r_{t-1}^{\ast},$ and parameters
$A_{21}=-\left(  1+\alpha\right)  \left(  \phi_{\pi},\phi_{y}\right)  ,$
$A_{22}=\alpha,$ $A_{22}^{\ast}=1,$ $B_{21}=\left(  \alpha\rho,-\phi_{\pi
},-\phi_{y}\right)  ,$ $B_{22}=0$ and $B_{22}^{\ast}=1.$

More examples of forward guidance policy rules that are nested within the
CKSVAR are discussed in \cite{IkedaLiMavroeidisZanetti2020}.

\section{Computational details\label{s: computation}}

\subsection{Likelihood\label{s: likelihood}}

To compute the likelihood, we need to obtain the prediction error densities.
The first step is to write the model in state-space form. Define%
\[
s_{t}=%
\begin{pmatrix}
\mathbf{y}_{t}\\
\vdots\\
\mathbf{y}_{t-p+1}%
\end{pmatrix}
,\quad\underset{\left(  k+1\right)  \times1}{\mathbf{y}_{t}}=%
\begin{pmatrix}
Y_{t}\\
\overline{Y}_{2t}^{\ast}%
\end{pmatrix}
,
\]
and write the state transition equation as%
\begin{equation}
s_{t}=F\left(  s_{t-1},u_{t};\psi\right)  =%
\begin{pmatrix}
F_{1}\left(  s_{t-1},u_{t};\psi\right) \\
\mathbf{y}_{t-1}\\
\vdots\\
\mathbf{y}_{t-p+1}%
\end{pmatrix}
, \label{eq: state transition}%
\end{equation}
\[
F_{1}\left(  s_{t-1},u_{t};\psi\right)  =%
\begin{pmatrix}
\overline{C}_{1}X_{t}+\overline{C}_{1}^{\ast}\overline{X}_{t}^{\ast}%
+u_{1t}-\widetilde{\beta}D_{t}\left(  \overline{C}_{2}X_{t}+\overline{C}%
_{2}^{\ast}\overline{X}_{t}^{\ast}+u_{2t}-b\right) \\
\max\left(  b,\overline{C}_{2}X_{t}+\overline{C}_{2}^{\ast}\overline{X}%
_{t}^{\ast}+u_{2t}\right) \\
\overline{C}_{2}X_{t}+\overline{C}_{2}^{\ast}\overline{X}_{t}^{\ast}+u_{2t}%
\end{pmatrix}
,
\]
and the observation equation as%
\begin{equation}
Y_{t}=%
\begin{pmatrix}
I_{k} & 0_{k\times1+\left(  p-1\right)  \left(  k+1\right)  }%
\end{pmatrix}
s_{t}. \label{eq: obs equation}%
\end{equation}

Next, I will derive the predictive density and mass functions. With Gaussian
errors, the joint predictive density of $Y_{t}$ corresponding to the
observations with $D_{t}=0$ is:%
\begin{multline}
f_{0}\left(  Y_{t}|s_{t-1},\psi\right)  =\left\vert \Omega\right\vert
^{-1/2}\exp\left\{  -\frac{1}{2}tr\left(  \left(  Y_{t}-\overline{C}%
X_{t}-\overline{C}^{\ast}\overline{X}_{t}^{\ast}\right) \right.\right. \\\left.\left. \left(
Y_{t}-\overline{C}X_{t}-\overline{C}^{\ast}\overline{X}_{t}^{\ast}\right)
^{\prime}\Omega^{-1}\right)  \right\}  . \label{eq: f0}%
\end{multline}
At $D_{t}=1,$ the predictive density of $Y_{1t}$ can be written as:%
\begin{align}
f_{1}\left(  Y_{1t}|s_{t-1},\psi\right)   &  :=\left\vert \Xi_{1}\right\vert
^{-\frac{1}{2}}\exp\left[  -\frac{1}{2}\left(  Y_{1t}-\mu_{1t}\right)
^{\prime}\Xi_{1}^{-1}\left(  Y_{1t}-\mu_{1t}\right)  \right] \label{eq: f1}\\
\mu_{1t}  &  :=\widetilde{\beta}b+\left(  \overline{C}_{1}-\widetilde{\beta
}\overline{C}_{2}\right)  X_{t}+\left(  \overline{C}_{1}^{\ast}%
-\widetilde{\beta}\overline{C}_{2}^{\ast}\right)  \overline{X}_{t}^{\ast
}\label{eq: mu1}\\
\Xi_{1}  &  :=\Omega_{1.2}+\widetilde{\delta}\widetilde{\delta}^{\prime}%
\tau^{2}=%
\begin{pmatrix}
I_{k-1} & -\widetilde{\beta}%
\end{pmatrix}
\Omega\binom{I_{k-1}}{-\widetilde{\beta}^{\prime}},\quad\widetilde{\delta
}=\Omega_{12}\omega_{22}^{-1}-\widetilde{\beta}, \label{eq: Xi1}%
\end{align}
where $\Omega_{1.2}=\Omega_{11}-\Omega_{12}\omega_{22}^{-1}\Omega_{21},$ and
$\tau=\sqrt{\omega_{22}}$. Next,%
\begin{align}
u_{2t}|Y_{1t},s_{t-1}  &  \sim N\left(  \mu_{2t},\tau_{2}^{2}\right)  ,\text{
with}\label{eq: dens u2|Y1}\\
\mu_{2t}  &  :=\tau^{2}\widetilde{\delta}^{\prime}\Xi_{1}^{-1}\left(
Y_{1t}-\mu_{1t}\right)  ,\quad\tau_{2}=\tau\sqrt{\left(  1-\tau^{2}%
\widetilde{\delta}^{\prime}\Xi_{1}^{-1}\widetilde{\delta}\right)  }.
\label{eq: mu2, tau2}%
\end{align}
Hence,
\begin{equation}
\Pr\left(  D_{t}=1|Y_{1t},s_{t-1},\psi\right)  =\Phi\left(  \frac
{b-\overline{C}_{2}X_{t}-\overline{C}_{2}^{\ast}\overline{X}_{t}^{\ast}%
-\mu_{2t}}{\tau_{2}}\right)  . \label{eq: prob D=1}%
\end{equation}

In the case of the KSVAR model, there are no latent lags ($\overline{C}^{\ast
}=0,$ $\overline{C}=C$), so the log-likelihood is available analytically:%
\begin{multline}
\log L\left(  \psi\right)  =\sum_{t=1}^{T}  \left(  1-D_{t}\right)  \log
f_{0}\left(  Y_{t}|s_{t-1},\psi\right) \\ . +\sum_{t=1}^{T}D_{t}\log\left(  f_{1}\left(
Y_{1t}|s_{t-1},\psi\right)  \Phi\left(  \frac{b-C_{2}X_{t}-\mu_{2t}}{\tau_{2}%
}\right)  \right)   \label{eq: lik KSVAR}%
\end{multline}
where $f_{0}\left(  Y_{t}|s_{t-1},\theta\right)  $ and $f_{1}\left(
Y_{1t}|s_{t-1},\theta\right)  $ are given by (\ref{eq: f0}) and (\ref{eq: f1}%
), resp., with $\overline{C}^{\ast}=0$.

The likelihood for the unrestricted CKSVAR ($\overline{C}^{\ast}\neq0$) can be computed approximately by simulation (particle filtering). I provide two different simulation algorithms. The first
is a sequential importance sampler (SIS), proposed originally by
\cite{Lee1999} for the univariate dynamic Tobit model. It is extended here to
the CKSVAR model. The second algorithm is a fully adapted particle filter
(FAPF), which is a sequential importance resampling algorithm designed to
address the sample degeneracy problem. It is proposed by \cite{MalikPitt2011}
and is a special case of the auxiliary particle filter developed by
\cite{PittShephard1999}.

Both algorithms require sampling from the predictive density of $\overline
{Y}_{2t}^{\ast}$ conditional on $Y_{1t},D_{t}=1$ and $s_{t-1}.$ From
(\ref{eq: RF Y2*bar}) and (\ref{eq: dens u2|Y1}), we see that this is a
truncated Normal with original mean $\mu_{2t}^{\ast}=\overline{C}_{2}%
X_{t}+\overline{C}_{2}^{\ast}\overline{X}_{t}^{\ast}+\mu_{2t}$ and standard
deviation $\tau_{2},$ where $\mu_{2t},\tau_{2}$ are given in
(\ref{eq: mu2, tau2}), i.e.,%
\begin{equation}
f_{2}\left(  Y_{2t}^{\ast}|Y_{1t},D_{t}=1,s_{t-1},\psi\right)  =TN\left(
\mu_{2t}^{\ast},\tau_{2},\overline{Y}_{2t}^{\ast}<b\right)  \label{eq: f2}%
\end{equation}
Draws from this truncated distribution can be obtained using, for instance,
the procedure in \cite{Lee1999}. Let $\xi_{t}^{\left(  j\right)  }\sim
U\left[  0,1\right]  $ be \textit{i.i.d.} uniform random draws, $j=1,...,M$.
Then, a draw from $\overline{Y}_{2t}^{\ast}|Y_{1t},s_{t-1},\overline{Y}%
_{2t}^{\ast}<b$ is given by%
\begin{equation}
\overline{Y}_{2t}^{\ast\left(  j\right)  }=\mu_{2t}^{\ast}+\tau_{2}\Phi
^{-1}\left[  \xi_{t}^{\left(  j\right)  }\Phi\left(  \frac{b-\mu_{2t}^{\ast}%
}{\tau_{2}}\right)  \right]  . \label{eq: trunc norm draw}%
\end{equation}

\begin{algorithm}
[SIS]\label{algo: SIS}Sequential Importance Sampler

\begin{enumerate}
\item \emph{Initialization.} For $j=1:M,$ set $W_{0}^{j}=1$ and $s_{0}%
^{j}=\left(  \mathbf{y}_{0}^{j},\ldots,\mathbf{y}_{-p+1}^{j}\right)  ,$ with
$\mathbf{y}_{-s}^{j}=\left(  Y_{0}^{\prime},Y_{2,0}\right)  ^{\prime},$ for
$s=0,...,p-1.$ (in other words, initialize $\overline{Y}_{2,-s}^{\ast}$ at the
observed values of $Y_{2,-s}$).

\item \emph{Recursion.} For $t=1:T$:

\begin{enumerate}
\item For $j=1:M,$ compute the incremental weights
\[
w_{t-1|t}^{j}=p\left(  Y_{t}|s_{t-1}^{j},\psi\right)  =\left\{
\begin{array}
[c]{ll}%
f_{0}\left(  Y_{t}|s_{t-1}^{j},\psi\right)  , & \text{if }D_{t}=0\\
f_{1}\left(  Y_{1t}|s_{t-1}^{j},\psi\right)  \Pr\left(  D_{t}=1|Y_{1t}%
,s_{t-1}^{j},\psi\right)  , & \text{if }D_{t}=1
\end{array}
\right.
\]
where $f_{0},f_{1},$ and $\Pr\left(  D_{t}=1|Y_{1t},s_{t-1};\psi\right)  $ are
given by (\ref{eq: f0}), (\ref{eq: f1}), and (\ref{eq: prob D=1}), resp., and%
\[
S_{t}=\frac{1}{M}\sum_{j=1}^{M}w_{t-1|t}^{j}W_{t-1}^{j}%
\]

\item Sample $s_{t}^{j}$ randomly from $p\left(  s_{t}|s_{t-1}^{j}%
,Y_{t}\right)  $. That is, $s_{t}^{j}=\left(  \mathbf{y}_{t}^{j}%
,\mathbf{y}_{t-1}^{j},\ldots,\mathbf{y}_{t-p}^{j}\right)  $ where
$\mathbf{y}_{t}^{j}=\left(  Y_{t}^{\prime},\overline{Y}_{2t}^{\ast\left(
j\right)  }\right)  \ $and $\overline{Y}_{2t}^{\ast\left(  j\right)  }$ is a
draw from $f_{2}\left(  Y_{2t}^{\ast}|Y_{1t},D_{t}=1,s_{t-1}^{j},\psi\right)
$ using (\ref{eq: trunc norm draw}).

\item Update the weights:%
\[
W_{t}^{j}=\frac{w_{t-1|t}^{j}W_{t-1}^{j}}{S_{t}}.
\]

\end{enumerate}

\item Likelihood approximation%
\[
\log\widehat{p}(Y_{T}|\psi)=\sum_{t=1}^{T}\log S_{t}%
\]

\end{enumerate}
\end{algorithm}

If the draws $\xi_{t}^{\left(  j\right)  }$ are kept fixed across different
values of $\psi$, the simulated likelihood in step 3 is smooth. Note that when
$k=1$ and $Y_{t}=Y_{2t}$ (no $Y_{1t}$ variables), the model reduces to a
univariate dynamic Tobit model, and Algorithm \ref{algo: SIS} reduces exactly
to the sequential importance sampler proposed by \cite{Lee1999}. A possible
weakness of this algorithm is sample degeneracy, which arises when all but a
few weights $W_{t}^{J}$ are zero. To gauge possible sample degeneracy, we can
look at the effective sample size (ESS), as recommended by
\cite{HerbstSchorfheide2015}%
\begin{equation}
ESS_{t}=\frac{M}{\frac{1}{M}\sum_{j=1}^{M}\left(  W_{t}^{j}\right)  ^{2}}.
\label{eq: ESS}%
\end{equation}

Next, I turn to the FAPF algorithm.

\begin{algorithm}
[FAPF]\label{algo: FAPF}Fully Adapted Particle Filter

\begin{enumerate}
\item \emph{Initialization.} For $j=1:M,$ set $s_{0}^{j}=\left(
\mathbf{y}_{0}^{j},\ldots,\mathbf{y}_{-p+1}^{j}\right)  ,$ with $\mathbf{y}%
_{-s}^{j}=\left(  Y_{0}^{\prime},Y_{2,0}\right)  ^{\prime},$ for
$s=0,...,p-1.$ (in other words, initialize $\overline{Y}_{2,-s}^{\ast}$ at the
observed values of $Y_{2,-s}$).

\item \emph{Recursion.} For $t=1:T$:

\begin{enumerate}
\item For $j=1:M,$ compute
\[
w_{t-1|t}^{j}=p\left(  Y_{t}|s_{t-1}^{j},\psi\right)  =\left\{
\begin{array}
[c]{ll}%
f_{0}\left(  Y_{t}|s_{t-1}^{j},\psi\right)  , & \text{if }D_{t}=0\\
f_{1}\left(  Y_{1t}|s_{t-1}^{j},\psi\right)  \Pr\left(  D_{t}=1|Y_{1t}%
,s_{t-1}^{j},\psi\right)  , & \text{if }D_{t}=1
\end{array}
\right.
\]
where $f_{0},f_{1},$ and $\Pr\left(  D_{t}=1|Y_{1t},s_{t-1};\psi\right)  $ are
given by (\ref{eq: f0}), (\ref{eq: f1}), and (\ref{eq: prob D=1}), resp., and%
\[
\pi_{t-1|t}^{j}=\frac{w_{t-1|t}^{j}}{\sum_{j=1}^{M}w_{t-1|t}^{j}}.
\]

\item For $j=1:M$, sample $k_{j}$ randomly from the multinomial distribution
$\left\{  j,\pi_{t-1|t}^{j}\right\}  .$ Then, set $\tilde{s}_{t-1}^{j}%
=s_{t-1}^{k_{j}}$ (this applies only to the elements in $s_{t-1}^{j}$ that
correspond to $X_{t}^{\ast j},$ since all the other elements are observed and
constant across all $j$. That is, $\tilde{s}_{t-1}^{j}=\left(  \mathbf{\tilde
{y}}_{t-1}^{j},\allowbreak\ldots\allowbreak,\mathbf{\tilde{y}}_{t-p}^{j}\right)  ,$ $\mathbf{\tilde
{y}}_{t-s}^{j}=\left(  Y_{t-1}^{\prime},\overline{Y}_{2,t-s}^{\ast\left(
k_{j}\right)  }\right)  ,$ $s=1,...,p.$)

\item For $j=1:M$, sample $s_{t}^{j}$ randomly from $p\left(  s_{t}|\tilde
{s}_{t-1}^{j},Y_{t}\right)  $. That is, $s_{t}^{j}=\left(  \mathbf{y}_{t}%
^{j},\allowbreak\ldots,\allowbreak\mathbf{\tilde{y}}_{t-p}^{j}\right)
$ where $\mathbf{y}_{t}^{j}=\left(  Y_{t}^{\prime},\overline{Y}_{2t}%
^{\ast\left(  j\right)  }\right)  \ $and $\overline{Y}_{2t}^{\ast\left(
j\right)  }$ is a draw from $f_{2}\left(  Y_{2t}^{\ast}|Y_{1t},D_{t}%
=1,\tilde{s}_{t-1}^{j},\psi\right)  $ using (\ref{eq: trunc norm draw}).
\end{enumerate}

\item Likelihood approximation%
\[
\ln\hat{p}\left(  Y_{T}|\psi\right)  =\sum_{t=1}^{T}\ln\left(  \frac{1}{M}%
\sum_{j=1}^{M}w_{t-1|t}^{j}\right)
\]

\end{enumerate}
\end{algorithm}

Many of the generic particle filtering algorithms used in the macro
literature, described in \cite{HerbstSchorfheide2015}, are inapplicable in a
censoring context because of the absence of measurement error in the
observation equation. It is, of course, possible to introduce a small
measurement error in $Y_{2t},$ so that the constraint $Y_{2t}\geq b$ is not
fully respected, but there is no reason to expect other particle filters
discussed in \cite{HerbstSchorfheide2015} to estimate the likelihood more
accurately than the FAPF algorithm described above.

Moments or quantiles of the filtering or smoothing distribution of any
function $h\left(  \cdot\right)  $ of the latent states $s_{t}$ can be
computed using the drawn sample of particles. When we use Algorithm
\ref{algo: FAPF}, simple average or quantiles of $h\left(  s_{t}^{j}\right)  $
produce the requisite average or quantiles of $h\left(  s_{t}\right)  $
conditional on $Y_{1},\ldots,Y_{t}$ (the filtering density). For particles
generated using Algorithm \ref{algo: SIS}, we need to take weighted averages
using the importance sampling weights $W_{t}.$ Smoothing estimates of
$h\left(  s_{t}^{j}\right)  $ can be obtained using weights $W_{T}.$

\subsection{Computation of the identified set\label{s: ID set}}

Substitute for $\overline{\gamma}$ in (\ref{eq: betatil xi}) using Proposition
\ref{prop: point ID} to get%
\begin{equation}
\widetilde{\beta}=\left(  1-\xi\right)  \left(  I-\xi\overline{\beta}\left(
\Omega_{12}^{\prime}-\Omega_{22}\overline{\beta}^{\prime}\right)  \left(
\Omega_{11}-\Omega_{12}\overline{\beta}^{\prime}\right)  ^{-1}\right)
^{-1}\overline{\beta}. \label{eq: betatil to betabar}%
\end{equation}
For each value of $\xi\in\lbrack0,1),$ the above equation defines a
correspondence from $\Re^{k-1}$ to $\Re^{k-1}$. The range of $\overline{\beta
}$ can then be obtained numerically by solving (\ref{eq: betatil to betabar})
for $\overline{\beta}$ as a function of the reduced-form parameters and $\xi$
for each value of $\xi,$ and gathering all the solutions in the set. 

Rearranging (\ref{eq: betatil to betabar}) yields%
\begin{equation}
\widetilde{\beta}=\xi\overline{\beta}\left(  \Omega_{12}^{\prime}-\Omega
_{22}\overline{\beta}^{\prime}\right)  \left(  \Omega_{11}-\Omega
_{12}\overline{\beta}^{\prime}\right)  ^{-1}\widetilde{\beta}+\left(
1-\xi\right)  \overline{\beta}. \label{eq: betatil to betabar 1}
\end{equation}
Note that
\[
\left(  \Omega_{11}-\Omega_{12}\overline{\beta}^{\prime}\right)  ^{-1}%
=\Omega_{11}^{-1}+\Omega_{11}^{-1}\Omega_{12}\left(  1-\overline{\beta
}^{\prime}\Omega_{11}^{-1}\Omega_{12}\right)  ^{-1}\overline{\beta}^{\prime
}\Omega_{11}^{-1}.
\]
Hence, 
\begin{align*}
&  \left(  \Omega_{12}^{\prime}-\Omega_{22}\overline{\beta}^{\prime}\right)
\left(  \Omega_{11}-\Omega_{12}\overline{\beta}^{\prime}\right)  ^{-1}\\
&  =\left(  \Omega_{12}^{\prime}-\Omega_{22}\overline{\beta}^{\prime}\right)
\Omega_{11}^{-1}+\frac{\left(  \Omega_{12}^{\prime}\Omega_{11}^{-1}%
\Omega_{12}-\Omega_{22}\overline{\beta}^{\prime}\Omega_{11}^{-1}\Omega
_{12}\right)  \overline{\beta}^{\prime}\Omega_{11}^{-1}}{1-\overline{\beta
}^{\prime}\Omega_{11}^{-1}\Omega_{12}}\\
&  =\frac{\left(  \Omega_{12}^{\prime}-\Omega_{22}\overline{\beta}^{\prime
}\right)  \Omega_{11}^{-1}+\Omega_{12}^{\prime}\Omega_{11}^{-1}\left(
\Omega_{12}\overline{\beta}^{\prime}\Omega_{11}^{-1}-\overline{\beta}^{\prime
}\Omega_{11}^{-1}\Omega_{12}I_{k-1}\right)  }{1-\overline{\beta}^{\prime
}\Omega_{11}^{-1}\Omega_{12}}.
\end{align*}
Substituting this back into (\ref{eq: betatil to betabar 1}), we get%
\[
\widetilde{\beta}=\xi\overline{\beta}\frac{\left(  \Omega_{12}^{\prime}%
-\Omega_{22}\overline{\beta}^{\prime}\right)  \Omega_{11}^{-1}+\Omega
_{12}^{\prime}\Omega_{11}^{-1}\left(  \Omega_{12}\overline{\beta}^{\prime
}\Omega_{11}^{-1}-\overline{\beta}^{\prime}\Omega_{11}^{-1}\Omega_{12}%
I_{k-1}\right)  }{1-\overline{\beta}^{\prime}\Omega_{11}^{-1}\Omega_{12}%
}\widetilde{\beta}+\left(  1-\xi\right)  \overline{\beta}.%
\]
Multiplying both sides by $1-\overline{\beta}^{\prime}\Omega_{11}^{-1}\Omega_{12}$ yields
\begin{align*}
\widetilde{\beta}-\overline{\beta}^{\prime}\Omega_{11}^{-1}\Omega
_{12}\widetilde{\beta}  &  =\xi\overline{\beta}\Omega_{12}^{\prime
}\widetilde{\beta}-\xi\overline{\beta}\Omega_{22}\overline{\beta}^{\prime
}\Omega_{11}^{-1}\widetilde{\beta}+\xi\overline{\beta}\Omega_{12}^{\prime
}\Omega_{11}^{-1}\Omega_{12}\overline{\beta}^{\prime}\Omega_{11}%
^{-1}\widetilde{\beta}\\
&  -\xi\overline{\beta}\overline{\beta}^{\prime}\Omega_{11}^{-1}\Omega
_{12}\Omega_{12}^{\prime}\Omega_{11}^{-1}\widetilde{\beta}+\left(
1-\xi\right)  \overline{\beta}\left(  1-\overline{\beta}^{\prime}\Omega
_{11}^{-1}\Omega_{12}\right).
\end{align*}
Rearranging, we have
\begin{align*}
\widetilde{\beta}  &  =\widetilde{\beta}\Omega_{12}^{\prime}\Omega_{11}%
^{-1}\overline{\beta}+\xi\Omega_{12}^{\prime}\widetilde{\beta}\overline{\beta
}+\left(  1-\xi\right)  \overline{\beta}-\left(  1-\xi\right)  \overline
{\beta}\overline{\beta}^{\prime}\Omega_{11}^{-1}\Omega_{12}\\
&  +\overline{\beta}\overline{\beta}^{\prime}\Omega_{11}^{-1}\widetilde{\beta
}\xi\Omega_{12}^{\prime}\Omega_{11}^{-1}\Omega_{12}-\overline{\beta}%
\overline{\beta}^{\prime}\Omega_{11}^{-1}\widetilde{\beta}\xi\Omega
_{22}-\overline{\beta}\overline{\beta}^{\prime}\Omega_{11}^{-1}\Omega
_{12}\Omega_{12}^{\prime}\Omega_{11}^{-1}\widetilde{\beta}\xi
\\  &  =\left(  \widetilde{\beta}\Omega_{12}^{\prime}%
\Omega_{11}^{-1}+\left(  \xi\Omega_{12}^{\prime}\widetilde{\beta}%
+1-\xi\right)  I_{k-1}\right)  \overline{\beta}\\
&  +\overline{\beta}\overline{\beta}^{\prime}\Omega_{11}^{-1}\left(  \left(
\left(  \Omega_{12}^{\prime}\Omega_{11}^{-1}\Omega_{12}-\Omega_{22}\right)
I_{k-1}-\Omega_{12}\Omega_{12}^{\prime}\Omega_{11}^{-1}\right)
\widetilde{\beta}\xi-\left(  1-\xi\right)  \Omega_{12}\right).
\end{align*}
This can be written as
\begin{equation}
\widetilde{\beta}-\tilde{A}\overline{\beta}+\overline{\beta}\overline{\beta
}^{\prime}\tilde{b}=0, \label{eq: quadratic}%
\end{equation}
where
\begin{align*}
\tilde{b}  &  :=-\Omega_{11}^{-1}\left(  \left(  \left(  \Omega_{12}^{\prime}%
\Omega_{11}^{-1}\Omega_{12}-\Omega_{22}\right)  I_{k-1}-\Omega_{12}\Omega
_{12}^{\prime}\Omega_{11}^{-1}\right)  \widetilde{\beta}\xi-\left(
1-\xi\right)  \Omega_{12}\right) , \text{ and}\\
\bar{A}  &  :=\widetilde{\beta}\Omega_{12}^{\prime}\Omega_{11}^{-1}+\left(
\xi\Omega_{12}^{\prime}\widetilde{\beta}+1-\xi\right)  I_{k-1}.
\end{align*}

Define%
\[
z:=\tilde{b}^{\prime}x\text{ \ \ and \ \ }w:=\tilde{b}_{\perp}^{\prime}x,
\]
where $\tilde{b}_{\perp}^{\prime}\tilde{b}_{\perp}=1$ and $\tilde{b}_{\perp
}^{\prime}\tilde{b}=0$. Hence, rewrite (\ref{eq: quadratic}) as%
\[
\widetilde{\beta}-\tilde{A}\tilde{b}\left(  \tilde{b}^{\prime}\tilde
{b}\right)  ^{-1}z-\tilde{A}\tilde{b}_{\perp}w+\tilde{b}\left(  \tilde
{b}^{\prime}\tilde{b}\right)  ^{-1}z^{2}+\tilde{b}_{\perp}wz=0.
\]
Premultiply by $\tilde{b}_{\perp}^{\prime}$ to get
\[
\tilde{b}_{\perp}^{\prime}\widetilde{\beta}-\tilde{b}_{\perp}^{\prime}%
\tilde{A}\tilde{b}\left(  \tilde{b}^{\prime}\tilde{b}\right)  ^{-1}z-\tilde
{b}_{\perp}^{\prime}\tilde{A}\tilde{b}_{\perp}w+wz=0.
\]
Solve that for $w$ to get%
\begin{align*}
w  &  =\left(  \tilde{b}_{\perp}^{\prime}\tilde{A}\tilde{b}_{\perp}-z\right)
^{-1}\left(  \tilde{b}_{\perp}^{\prime}\widetilde{\beta}-\tilde{b}_{\perp
}^{\prime}\tilde{A}\tilde{b}\left(  \tilde{b}^{\prime}\tilde{b}\right)
^{-1}z\right) \\
&  =C_{0}\left(  z\right)  ^{-1}c_{1}\left(  z\right)  ,
\end{align*}
with%
\begin{align*}
C_{0}\left(  z\right)   &  :=\left(  \tilde{b}_{\perp}^{\prime}\tilde{A}%
\tilde{b}_{\perp}-z\right)  ,\text{ \ and}\\
c_{1}\left(  z\right)   &  :=\tilde{b}_{\perp}^{\prime}\widetilde{\beta
}-\tilde{b}_{\perp}^{\prime}\tilde{A}\tilde{b}\left(  \tilde{b}^{\prime}%
\tilde{b}\right)  ^{-1}z,
\end{align*}
provided that $\det\left(  C_{0}\left(  z\right)  \right)  \neq0.$

Next, premultiply (\ref{eq: quadratic}) by $\tilde{b}^{\prime}$ and substitute
for $w$ to get%
\begin{equation}
\tilde{b}^{\prime}\widetilde{\beta}-\tilde{b}^{\prime}\tilde{A}\tilde
{b}\left(  \tilde{b}^{\prime}\tilde{b}\right)  ^{-1}z-\tilde{b}^{\prime}%
\tilde{A}\tilde{b}_{\perp}C_{0}\left(  z\right)  ^{-1}c_{1}\left(  z\right)
+z^{2}=0. \label{eq: quadratic 1}%
\end{equation}
Now, notice that $C_{0}\left(  z\right)  ^{-1}=C_{0}\left(  z\right)
^{adj}/\det\left(  C_{0}\left(  z\right)  \right)  ,$ where $C^{adj}$ is the
adjoint of a square matrix $C.$ Moreover, since $C_{0}\left(  z\right)  $ is
of dimension $k-2$ and its elements are linear in $z,$ $\det\left(
C_{0}\left(  z\right)  \right)  $ is a polynomial in $z$ of order at most
$k-2,$ and the elements of $C_{0}\left(  z\right)  ^{adj}$ are polynomials in
$z$ of order at most $k-3.$ For $k=2,$ $w$ is empty, so (\ref{eq: quadratic})
is simply a quadratic in $z.$ When $k>2,$ $\det\left(  C_{0}\left(  z\right)
\right)  $ is nonzero and we can multiply (\ref{eq: quadratic 1}) by it to get%
\begin{align}
0  &  =\tilde{b}^{\prime}\widetilde{\beta}\det\left(  C_{0}\left(  z\right)
\right)  +\tilde{b}^{\prime}\tilde{A}\tilde{b}\left(  \tilde{b}^{\prime}%
\tilde{b}\right)  ^{-1}z\det\left(  C_{0}\left(  z\right)  \right)
\nonumber\\
&  -\tilde{b}^{\prime}\tilde{A}\tilde{b}_{\perp}C_{0}\left(  z\right)
^{adj}c_{1}\left(  z\right)  +\det\left(  C_{0}\left(  z\right)  \right)
z^{2},\label{eq: quadratic 2}
\end{align}
This is a polynomial equation of order $k$ and has at most $k$ solutions,
denoted $z_{i},$ say. Then, the solutions for $\overline{\beta}$ are given
by%
\begin{align}
\overline{\beta}_{i}  &  =\left[  \tilde{b}\left(  \tilde{b}^{\prime}\tilde{b}\right)
^{-1},\tilde{b}_{\perp}\right]
\begin{pmatrix}
z_{i}\\
C_{0}\left(  z_{i}\right)  ^{-1}c_{1}\left(  z_{i}\right)
\end{pmatrix}
\nonumber\\
&  =\tilde{b}\left(  \tilde{b}^{\prime}\tilde{b}\right)  ^{-1}z_{i}+\tilde
{b}_{\perp}C_{0}\left(  z_{i}\right)  ^{-1}c_{1}\left(  z_{i}\right).\label{eq: betabar solution}
\end{align}
Below I give some special cases.

\textit{Case $k=2$:} In this case, $w$ is empty, $\overline{\beta}$ is a scalar, and the equation
(\ref{eq: quadratic}) is a quadratic
\[
\widetilde{\beta}-\tilde{A}\overline{\beta}+\overline{\beta}^{2}\tilde{b}=0.
\]
If $\tilde{A}^{2}-4\widetilde{\beta}\tilde{b}>0,$ the two real solutions are$
\overline{\beta}_{1,2}=\frac{\tilde{A}\pm\sqrt{\tilde{A}^{2}-4\widetilde{\beta}\tilde{b}}%
}{2\tilde{b}}.$
\smallskip

\textit{Case $k=3$:} In this case, $w$ is a scalar, and the equation (\ref{eq: quadratic 2}) can be
written as a cubic in $z,$ i.e.,%
\begin{equation}
C_{0}\left(  z\right)  \tilde{b}^{\prime}\widetilde{\beta}-C_{0}\left(
z\right)  \tilde{b}^{\prime}\tilde{A}\tilde{b}\left(  \tilde{b}^{\prime}%
\tilde{b}\right)  ^{-1}z-\tilde{b}^{\prime}\tilde{A}\tilde{b}_{\perp}%
c_{1}\left(  z\right)  +C_{0}\left(  z\right)  z^{2}=0, \label{eq: cubic}%
\end{equation}
since $C_{0}\left(  z\right)  $ is a scalar linear function of $z.$ It can be
shown that one of the roots of (\ref{eq: cubic}) satisfies $\Omega
_{12}^{\prime}\Omega_{11}^{-1}\overline{\beta}=1,$ which implies $\det\left(
\Omega_{11}-\Omega_{12}\overline{\beta}^{\prime}\right)  =0,$ and hence
violates the equation for $\overline{\gamma}=\left(  \Omega_{12}^{\prime
}-\Omega_{22}\overline{\beta}^{\prime}\right)  \left(  \Omega_{11}-\Omega
_{12}\overline{\beta}^{\prime}\right)  ^{-1},$ so it is not a valid solution.
The root in question is%
\[
z_{1}=\frac{\Omega_{12}^{\prime}\Omega_{11}^{-1}\left(  \tilde{b}_{\perp
}\tilde{b}^{\prime}\tilde{A}\tilde{b}-\tilde{b}\tilde{b}^{\prime}\tilde
{A}\tilde{b}_{\perp}\right)  }{\Omega_{12}^{\prime}\Omega_{11}^{-1}\tilde
{b}_{\perp}\left(  \tilde{b}^{\prime}\tilde{b}\right)  }%
\]
We can then factor out a term $z-z_{1}$ from (\ref{eq: cubic}), and obtain the
remaining two roots from a quadratic equation. Therefore, there will be zero
or two solutions for $\overline{\beta}$, as in the case $k=2$.\smallskip

An algorithm for obtaining the identified set of the IRF (\ref{eq: IRF}) is as follows.
\begin{algorithm}
[ID set]\label{algo: ID set}Discretize the space $(0,1)$ into $R$ equidistant
points. \newline For each $r=1:R,$ set $\xi_{r}=\frac{r}{R+1}$ and solve
equation (\ref{eq: quadratic 1}).

\begin{enumerate}
\item If no solution exists, proceed to the next $r.$

\item If $0<q_{r}\leq k$ solutions exist, denote them $z_{i,r},$ and, for each
$i=1:q_{r}$,

\begin{enumerate}
\item derive $\overline{\beta}_{i,r}$ from (\ref{eq: betabar solution}),
$\overline{\gamma}_{i,r}=\left(  \Omega_{12}^{\prime}-\Omega_{22}%
\overline{\beta}_{i,r}^{\prime}\right)  \left(  \Omega_{11}-\Omega
_{12}\overline{\beta}_{i,r}^{\prime}\right)  ^{-1}$,\newline$\overline
{A}_{22,i,r}^{-1}\allowbreak=\allowbreak\sqrt{\left(  -\overline{\gamma}%
_{i,r},1\right)  \Omega\left(  -\overline{\gamma}_{i,r},1\right)  ^{\prime}}$,
and $\Xi_{1,i.r}=\left(  I_{k-1},-\overline{\beta}_{i,r}\right)  \Omega\left(
I_{k-1},-\overline{\beta}_{i,r}\right)  ^{\prime};$

\item for $j=1:M,$

\begin{enumerate}
\item draw independently $\bar{\varepsilon}_{1t,i,r}^{j}\sim N\left(
0,\Xi_{1,i.r}\right)  $ and $u_{t+h}^{j}\sim N\left(  0,\Omega\right)  $ for
$h=1,...,H;$

\item for any scalar $\varsigma$, set%
\begin{align*}
u_{1t,i,r}^{j}\left(  \varsigma\right)   &  =\left(  I_{k-1}-\overline{\beta
}_{i,r}\overline{\gamma}_{i,r}\right)  ^{-1}\left(  \bar{\varepsilon}%
_{1t,i,r}^{j}-\overline{\beta}_{i,r}\varsigma\right) \\
u_{2t,i,r}^{j}\left(  \varsigma\right)   &  =\left(  1-\overline{\gamma}%
_{i,r}\overline{\beta}_{i,r}\right)  ^{-1}\left(  \varsigma-\overline{\gamma
}_{i,r}\bar{\varepsilon}_{1t,i,r}^{j}\right)  ,
\end{align*}
and compute $Y_{t,i,r}^{j}\left(  \varsigma\right)  $ using (\ref{eq: RF Y1}%
)-(\ref{eq: RF Y2}) with $u_{t,i,r}^{j}\left(  \varsigma\right)  $ in place of
$u_{t},$ and iterate forward to obtain $Y_{t+h,i,r}^{j}\left(  \varsigma
\right)  $ using $u_{t+h}^{j}$ computed in step i. Set $\varsigma=1$ for a
one-unit (e.g., 100 basis points) impulse to the policy shock $\overline
{\varepsilon}_{2t},$ or $\varsigma=\overline{A}_{22,i,r}^{-1}$ for a
one-standard deviation impulse.
\end{enumerate}

\item compute%
\[
\widehat{IRF}_{h,t,i,r}\left(  \varsigma\right)  =\frac{1}{M}\sum_{j=1}%
^{M}\left(  Y_{t+h,i,r}^{j}\left(  \varsigma\right)  -Y_{t+h,i,r}^{j}\left(
0\right)  \right)  .
\]

\end{enumerate}
\end{enumerate}

The identified set is given by the collection of $\widehat{IRF}_{h,t,i,r}%
\left(  \varsigma\right)  $ over $i=1:q_{r},$ $r=1:R,$ and the single
point-identified IRF\ at $\xi=0.$
\end{algorithm}

\subsection{IRFs and local projections\label{s: IRF}}

I will briefly discuss the difficulty in getting a local projection-like representation of the IRF in a
dynamic Tobit model, which is a univariate CKSVAR(1). The model is given by
the equations%
\begin{align*}
y_{t}^{\ast}  &  =\rho y_{t-1}+\rho^{\ast}\min\left(  y_{t-1}-b,0\right)
+u_{t}\\
&  =\rho y_{t-1}+\rho^{\ast}D_{t-1}\left(  y_{t-1}^{\ast}-b\right)
+u_{t},\quad D_{t}=1\left\{  y_{t}^{\ast}<b\right\}  ,\\
y_{t}  &  =\max\left(  y_{t}^{\ast},b\right)  =\left(  1-D_{t}\right)
y_{t}^{\ast}.
\end{align*}
Hence,%
\begin{multline*}
E_{t}\left(  y_{t+1}\right)  =\left(  \rho y_{t}+\rho^{\ast}D_{t}\left(
y_{t}^{\ast}-b\right)  \right)  \left(  1-\Phi\left(  \frac{b-\rho y_{t}%
-\rho^{\ast}D_{t}\left(  y_{t}^{\ast}-b\right)  }{\sigma}\right)  \right)
\\  +\sigma\phi\left(  \frac{b-\rho y_{t}-\rho^{\ast}D_{t}\left(  y_{t}^{\ast
}-b\right)  }{\sigma}\right)  .
\end{multline*}
In a linear model $\left(  \rho^{\ast}=0,\text{ }b=-\infty\right)  $, the
1-period ahead impulse response is $\rho,$ which coincides with the
coefficient on $y_{t}$ in the local projection $E_{t}\left(  y_{t+1}\right)
=\rho y_{t}.$ In that case, the coefficient $\rho$ corresponds to both
$\frac{\partial E_{t}\left(  y_{t+1}\right)  }{\partial u_{t}}=\frac{\partial
E_{t}\left(  y_{t+1}\right)  }{\partial y_{t}}$ and $E\left(  y_{t+1}%
|u_{t}=1,y_{t-1}\right)  -E\left(  y_{t+1}|u_{t}=0,y_{t-1}\right)  .$ None of
these properties hold in the dynamic Tobit model. For example, if we go with
$\frac{\partial E_{t}\left(  y_{t+1}\right)  }{\partial u_{t}}$ as our
definition of the impulse response, we will not be able to obtain it from the
slope of the conditional expectation function $E_{t}\left(  y_{t+1}\right)  $
with respect to $y_{t}$. One problem is that the function $E_{t}\left(  y_{t+1}\right)  $ is
non-differentiable at $u_{t}=b-\rho y_{t-1}+\rho^{\ast}D_{t-1}\left(
y_{t-1}^{\ast}-b\right)  $, i.e., exactly at the boundary. Another problem is that we
still need to rely on the parametric structure of the model to uncover
the impulse response from $E_{t}\left(  y_{t+1}\right)  $. For example, we
need to compute $\frac{\partial E_{t}\left(  y_{t+1}\right)  }{\partial u_{t}%
},$ which at all points $y_{t}^{\ast}\neq b$\ is given by:%
\[
\frac{\partial E_{t}\left(  y_{t+1}\right)  }{\partial u_{t}}=\left\{
\begin{array}
[c]{ll}%
\rho\left(  1-\Phi\left(  \frac{b-\rho y_{t}}{\sigma}\right)  +\frac{b}%
{\sigma}\phi\left(  \frac{b-\rho y_{t}}{\sigma}\right)  \right)  & \text{if
}D_{t}=0\\
\rho^{\ast}\left(  1-\Phi\left(  \frac{b-\rho b-\rho^{\ast}\left(  y_{t}%
^{\ast}-b\right)  }{\sigma}\right)  +\frac{b}{\sigma}\phi\left(  \frac{b-\rho
b-\rho^{\ast}\left(  y_{t}^{\ast}-b\right)  }{\sigma}\right)  \right)  &
\text{if }D_{t}=1.
\end{array}
\right.
\]
There is no clear way to obtain the above
impulse response from a local projection of $y_{t+1}$ on simple nonlinear
transformations of $y_{t}$, such as powers or interactions with the regime
indicator.

\end{appendix}

\end{document}